\pdfoutput=1
\documentclass[nolinenum]{jfp}

\newcommand\tomACMVersion0

\newcommand\ifTomACMVersion[2]{\ifthenelse{\equal\tomACMVersion1}{#1}{#2}}

\usepackage{mathtools}
\usepackage{booktabs}
\usepackage{subcaption}
\usepackage{hyperref} 
\usepackage{cleveref}
\usepackage{float}
\usepackage{stmaryrd}
\usepackage{xifthen}
\usepackage{tikz}
\usepackage{makecell}
\usepackage{xspace}
\usepackage{wrapfig} 
\usepackage[final]{pdfpages}
\usepackage[normalem]{ulem}
\usetikzlibrary{arrows.meta,tikzmark,calc}

\floatstyle{boxed}
\restylefloat{figure}

\newcommand\ID{\textrm{ID}}
\newcommand\JID{\textrm{JID}}
\newcommand\CID{\textrm{CID}}
\newcommand\fwdM{\operatorname*{\mathcal{M}}}
\newcommand\genID{\mathrm{genID}}
\newcommand\JobDescr{\mathrm{JobDescr}}
\newcommand\History{\mathrm{History}}
\newcommand\HStart{\mathrm{Start}}
\newcommand\HFork{\mathrm{Fork}}
\newcommand\R{\mathbb R}
\newcommand\lR{\underline\R}
\newcommand\Z{\mathbb Z}
\newcommand\N{\mathbb N}
\newcommand\ra\rightarrow
\newcommand\lra{\multimap }

\newcommand\fun{\lambda}
\newcommand\lfun{\underline\lambda}

\newcommand\inl{\textrm{inl}}
\newcommand\inr{\textrm{inr}}
\newcommand\case[1]{\textbf{case}\ {#1}\ \textbf{of}}
\newcommand\sign{\textrm{sign}}
\newcommand\fst{\textrm{fst}}
\newcommand\snd{\textrm{snd}}
\newcommand\id{\mathrm{id}}
\newcommand\Bool{\ensuremath{\mathrm{Bool}}}
\newcommand\True{\ensuremath{\mathrm{True}}}
\newcommand\Dual{\textrm{Dual}}

\newcommand\Map{\textrm{Map}}
\newcommand\coloneqqq{\mathbin{\raisebox{0.5pt}{::}{=}}}
\newcommand\Op{\mathrm{Op}}
\newcommand\trans[2]{\textbf{D}^{#1}_{#2}}

\newcommand\Interleave[1]{\textrm{Interleave}^{#1}}
\newcommand\Deinterleave[1]{\textrm{Deinterleave}^{#1}}
\newcommand\Wrap[1]{\textrm{Wrap}^{#1}}

\newcommand\Staged{\operatorname*{\textrm{Staged}}}
\newcommand\Expr{\operatorname*{\textrm{Expr}}}
\newcommand\ZeroStaged{0_{\text{Staged}}}
\newcommand\PlusStaged{\mathbin{+_{\text{Staged}}}}
\newcommand\StagedCall{\textrm{SCall}}
\newcommand\InitStaged{\textrm{SCotan}}
\newcommand\ResolveStagedWith[1]{#1{SResolve}}
\newcommand\ResolveStaged{\ResolveStagedWith{\textrm}}

\newcommand\Array{\operatorname*{\textrm{Array}}}
\newcommand\IArray{\operatorname*{\textrm{IArray}}}
\newcommand\StagedAlloc{\textrm{SAlloc}}
\newcommand\StagedOneHot{\textrm{SOneHot}}

\newcommand\arralloc{\texttt{alloc}}

\newcommand\arrget{\texttt{get}}

\newcommand\arrmodify{\texttt{modify}}
\newcommand\arrfreeze{\texttt{freeze}}
\newcommand\iarridx{\mathbin{@}}
\newcommand\Contrib{\textrm{Contrib}}
\newcommand\cost{\textrm{cost}}
\newcommand\size{\textrm{size}}
\newcommand\parCom{\star}
\newcommand\leqj{\leq_{\mathrm{j}}}
\newcommand\leqz{\leq_{\Z}}
\newcommand\leqc{\leq_{\mathrm{c}}}
\newcommand\ST{\texttt{ST}\xspace}
\newcommand\IO{\texttt{IO}\xspace}

\newcommand\tightunderbrace[2]{\parbox[t]{\widthof{\(#1\)}}{\(#1\)\vspace{-0.7em}\\\upbracefill\vspace{-0.4em}\\\centering{\scriptsize\(#2\)}}}

\newcommand\mbind{\mathbin{>\!\!>\!\!=}}

\newcommand\mseq{\mathbin{>\!\!>}}

\newcommand\figheading[1]{\textbf{#1}\hfill}

\newcommand\leadcomma{\mathrlap{\hspace{0.08em},}\phantom{(}}
\newcommand\leadcommai{\leadcomma\hspace{0.08em}}

\newenvironment{insight}{\textbf{Insight}:}{\par\vspace{0.2em}}

\newcommand\myparagraph[1]{\paragraph*{#1.}\;} 
 
\begin{document}

\journaltitle{JFP}
\cpr{Cambridge University Press}
\doival{10.1017/xxxxx}

\lefttitle{T.J.\ Smeding and M.I.L.\,Vákár}
\righttitle{Parallel Dual-Numbers Reverse AD}

\totalpg{\pageref{lastpage}}
\jnlDoiYr{2024}

\title{Parallel Dual-Numbers Reverse AD}

\begin{authgrp}
\author{Tom J.\ Smeding}
\affiliation{
  Utrecht University, The Netherlands \\
  (\email{t.j.smeding@uu.nl})
}

\author{Matthijs I.L. Vákár}
\affiliation{
  Utrecht University, The Netherlands \\
  (\email{m.i.l.vakar@uu.nl})
}
\end{authgrp}

\begin{abstract}
  Where dual-numbers forward-mode automatic differentiation (AD) pairs each scalar value with its tangent value, dual-numbers \emph{reverse-mode} AD attempts to achieve reverse AD using a similarly simple idea: by pairing each scalar value with a backpropagator function.
  Its correctness and efficiency on higher-order input languages have been analysed by Brunel, Mazza and Pagani, but this analysis used a custom operational semantics for which it is unclear whether it can be implemented efficiently.
  We
    take inspiration from their use of \emph{linear factoring} to optimise dual-numbers reverse-mode AD to an algorithm that has the correct complexity and enjoys an efficient implementation in a standard functional language with support for mutable arrays, such as Haskell.
  Aside from the linear factoring ingredient, our optimisation steps consist of well-known ideas from the functional programming community.
  We demonstrate the use of our technique by providing a practical implementation that differentiates most of Haskell98.
  Where previous work on dual numbers reverse AD has required sequentialisation to construct the reverse pass, we demonstrate that we can apply our technique to task-parallel source programs and generate a task-parallel derivative computation.
\end{abstract}

\maketitle

\section{Introduction}

An increasing number of applications requires computing derivatives of functions specified by a computer program.
The derivative of a function gives more qualitative information of its behaviour around a point (i.e.\ the local shape of the function's graph) than just the function value at that point.
This qualitative information is useful, for example, for optimising parameters along the function (because the derivative tells you how the function changes, locally) or inferring statistics about the function (e.g.\ an approximation of its integral).
These uses appear, respectively, in parameter optimisation in machine learning or numerical equation solving, and in Bayesian inference of probabilistic programs.
Both application areas are highly relevant today.

Automatic differentiation (AD) is the most effective technique for efficient computation of derivatives of programs, and it comes in two main flavours: forward AD and reverse AD.
In practice, by far the most common case is that functions have many input parameters and few, or even only one, output parameters; in this situation, forward AD is inefficient while reverse AD yields the desired computational complexity.
Indeed, reverse AD computes the gradient of a function implemented as a program in time\footnote{
  This refers to the number of primitive operations (including memory operations, etc.); naturally, the actual wall-clock time depends also on cache behaviour, memory locality, etc., which we consider out of scope.
} at most a constant factor more than the runtime of the original program, where forward AD has multiplicative overhead in the size of the program input.
However, reverse AD is also significantly more difficult to implement flexibly and correctly than forward AD.


Many approaches exist for doing reverse AD on flexible programming languages: using taping/tracing in an imperative language (e.g.~\citep{ad-2017-pytorch}) and in a functional language~\citep{ad-2021-kmett-hackage}, using linearisation and transposition code transformations~\citep{dex-2021-ad}, or sometimes specialised by taking advantage of common usage patterns in domain-specific languages~\citep{ad-2022-futhark-partial-recompute}.
In the programming language theory community, various algorithms have been described that apply to a wide variety of source languages, including approaches based on symbolic execution and tracing~\citep{ad-2020-dualnum-revad-linear-factoring,ad-2020-rev-ad-semantics} and on category theory~\citep{vakar-2022-chad}, as well as formalisations of existing implementations~\citep{ad-2021-krawiec-kmett-ad}.
Although all these source languages could, theoretically, be translated to a single generic higher-order functional language, each reverse AD algorithm takes a different approach to solve the same problem.
It is unclear how exactly these algorithms relate to each other, meaning that correctness proofs (if any) need to be rewritten for each individual algorithm.

This paper aims to improve on the situation by providing a link from the elegant, theoretical dual-numbers reverse AD algorithm analysed by~\cite{ad-2020-dualnum-revad-linear-factoring} to a practical functional taping approach as used in~\citep{ad-2021-kmett-hackage} and analysed by~\cite{ad-2021-krawiec-kmett-ad}.
Further, we ensure that the implementation exploits parallelism opportunities for the derivative computation that arise from parallelism present in the source program, rather than sequentialising the derivative computation, as was done in~\citep{ad-2021-kmett-hackage,ad-2021-krawiec-kmett-ad,ad-dualrev-th}.
The key point made by~\cite*{ad-2020-dualnum-revad-linear-factoring} is that one can attain the right computational complexity by starting from the very elegant dual-numbers reverse AD code transformation (\cref{sec:key-ideas,sec:naive}), and adding a \emph{linear factoring} rule to the operational semantics of the output language of the code transformation.
This linear factoring reduction rule states that for linear functions $f$, the expression $f\ x + f\ y$ should be reduced to $f\ (x + y)$.
We demonstrate how this factoring idea can motivate an efficient practical implementation that is also parallelism-preserving.

\myparagraph{Summary of contributions}
Our main contributions are the following:
\begin{itemize}
\item
  We show how the theoretical analysis based on the linear factoring rule can be used as a basis for an algorithm that assumes normal, call-by-value semantics.
  We do this by \emph{staging calls to backpropagators} in \cref{sec:staging}.
\item
  We show how this algorithm can be made complexity-efficient by using the standard functional programming techniques of Cayley transformation (\cref{sec:cayley}) and (e.g.\ linearly typed or monadic) functional in-place updates (\cref{sec:mutarrays}).
\item
  We explain how our algorithm relates to classical approaches based on taping (\cref{sec:improve}).
\item
  We demonstrate that, by contrast with previous similar approaches \citep{ad-2021-kmett-hackage,ad-2021-krawiec-kmett-ad,ad-dualrev-th},
  we do not need to sequentialise the derivative computation in case of a parallel source program, but instead can store the task parallelism structure during the primal pass and produce a task-parallel derivative (\cref{sec:parallelism}).
\item
  We give an implementation of the parallelism-ready algorithm of \cref{sec:parallelism} that can differentiate most of Haskell98 (but using call-by-value semantics), and that has the correct asymptotic complexity as well as decent constant-factor performance (\cref{sec:implementation}).
\item 
  We explain in detail how our technique relates to 
  the functional taping AD of \citep{ad-2021-kmett-hackage} and \citep{ad-2021-krawiec-kmett-ad} as well as \citep{ad-2019-fwd-ad-gradient-compiler-opts}'s approach of trying to optimise forward AD to reverse AD at compile time (\cref{sec:related-work}).
  We also briefly describe the broader relationship with related work.
\end{itemize}

\myparagraph{Relation to previous work}
This paper extends and develops our previous work \citep{ad-dualrev-th} presented
at the 50th ACM SIGPLAN Symposium on Principles of Programming Languages (POPL 2023). 
This version includes multiple elaborations.
Most notable is an analysis of how a variant of our dual numbers reverse AD algorithm can be applied to task-parallel source programs to produce task-parallel derivative code (\cref{sec:parallelism}), and an implementation of this parallelism-preserving AD method (\url{https://github.com/tomsmeding/ad-dualrev-th}).
Besides this novel parallel AD technique, we include extra sections to explain our method better: a detailed discussion of the desired type of reverse AD (\cref{sec:rev-ad-type}) including a comparison to CHAD~\citep{vakar-2022-chad,DBLP:journals/mscs/NunesV23,chad-efficient-popl} and a detailed description of how to use mutable arrays to eliminate the final log-factors from the complexity of our method (\cref{sec:mutarrays}).


\section{Key Ideas}\label{sec:key-ideas}


\myparagraph{Forward and reverse AD}
All modes of automatic differentiation exploit the idea that derivatives satisfy the chain rule:
\[
  D_x (f\circ g) = D_{g(x)}(f) \cdot D_x(g)
\] 
where we write the derivative of $h : \R^n \to \R^m$ at the point $x \in \R^n$ as $D_x (h)$, a linear map $\lR^n\lra \lR^m$ that computes directional derivatives.\footnote{
  These are known as Fr\'echet derivatives (we write $\lR$ for the Euclidian vector space on the set $\R$).
  The function $\lambda x.\ h(p) + D_p(h)(x - p)$ is the best linear approximation to $h$ around some point $p \in \R^n$.
  Equivalently: $D_p(h)$ describes how $h$'s output changes given a small perturbation to its input, i.e.\ $h(p) + D_p(h)(\Delta p)$ is close to $h(p + \Delta p)$ for small values of $\Delta p$.
}
(This linear map corresponds to the Jacobian matrix of partial derivatives of $h$. If $Jh(x)$ is the Jacobian of $h$ at $x$, then $D_x(h)(v) = Jh(x) \cdot v$.)
Using the chain rule, one can mechanically compute derivatives of a composition of functions.
The intermediate function values (such as $x$ and $g(x)$ in the example above) are called \emph{primals}; the (forward) derivative values (e.g.\ the input and output of $D_x(g)$ and $D_{g(x)}(f)$) are called \emph{tangents}.

Forward AD directly implements the chain rule above as a code transformation to compute derivatives.
Reverse AD instead computes the \emph{transposed derivative}, which satisfies this \emph{contravariant} chain rule instead:
\[
  D_x (f\circ g)^t = D_x(g)^t \cdot D_{g(x)}(f)^t
\] 
Here, we write $D_x (h)^t:\lR^m\lra \lR^n$ for the transpose of the linear map  $D_x(h):\lR^n\lra \lR^m$.
(In relation to the Jacobian of $h$, we have $D_x(h)^t(w) = w \cdot Jh(x)$.)
The values taken by and produced from transposed derivatives (or \emph{reverse derivatives}), such as $D_x(g)^t$ and $D_{g(x)}(f)^t$ above, are called \emph{cotangents} in this paper; other literature also uses the word \emph{adjoint} for this purpose.\footnote{
  If we write $\nabla_x f$ for the \emph{gradient} at $x$ of a scalar-valued function $f$, then we can rewrite the Jacobian equality to $D_x(h)^t(w) = w \cdot (\nabla_x h_1, \ldots, \nabla_x h_n)$.
  In particular, for $n = 1$ if $w = (1)$, one clearly recovers $h$'s gradient, which describes in what direction to move $h$'s input to make its (scalar) output change the fastest, and how fast it changes in that case.
}

Computing the full Jacobian of a function $f : \R^n \to \R^m$ at $x : \R^n$ requires $n$ evaluations of $D_x(f) : \lR^n \lra \lR^m$ (at the $n$ basis vectors of $\lR^n$; one such evaluation computes one column of the Jacobian), or alternatively $m$ evaluations of $D_x(f)^t : \lR^m \lra \lR^n$ (one such evaluation computes one row of the Jacobian).
Because many applications have $n \gg m$, reverse AD is typically of most interest, but forward AD is much simpler: for both modes (forward and reverse), we need the primals in the derivative computation, but in forward mode the derivatives are computed \emph{in the same order} as the original computation (and hence as the computation of the primals).
In reverse mode, the derivatives are computed in opposite order, requiring storage of all\footnote{Or nearly all, in any case, if optimal time complexity is desired. A trade-off between time and space complexity is provided by checkpointing \citep[e.g.][]{ad-2018-checkpointing-built-in}; a constant-factor improvement may be obtained by skipping some unnecessary primals~\citep[e.g.][]{ad-2005-to-be-recorded}.} primals during a \emph{forward pass} of program execution, after which the \emph{reverse pass} uses those stored primals to compute the derivatives.\footnote{
  For an survey of AD in general, see e.g.~\citep{ad-2018-survey-automatic-differentiation}.
}

This observation that, in forward AD, primals and tangents can be computed in tandem, leads to the idea of \emph{dual-numbers forward AD}: pair primals and derivatives explicitly to interleave the primal and derivative computations, and run the program with overloaded arithmetic operators to propagate these tangents forward using the chain rule.
The tangent of the output can be read from the tangents paired up with the output scalars of the program.
For example, transforming the program in \cref{subfig:key-example-orig} using dual-numbers forward AD yields \cref{subfig:key-example-fwd}.
As a result, we do not store the primals any longer than is necessary, and the resulting code transformation tends to be much simpler than a naive attempt that first computes and stores all primals before computing any derivatives.

\begin{figure}
  \begin{subfigure}[b]{0.16\textwidth}
    \( \begin{array}{@{}l@{}}
        \fun(x : \R, y : \R). \\
        \; \textbf{let}\ z = x + y \\
        \; \textbf{in}\ x \cdot z
    \end{array} \)
    \caption{\label{subfig:key-example-orig}
      Original
    }
  \end{subfigure}
  \hfill
  \begin{subfigure}[b]{0.32\textwidth}
    \( \begin{array}{@{}l@{}}
        \begin{array}{@{}r@{}l@{}}
          \fun( & (x : \R, dx : \lR) \\
          {,}\hspace{0.03cm}{} & (y : \R, dy : \lR)).
        \end{array} \\
        \; \textbf{let}\ (z, dz) = (x + y, dx + dy) \\
        \; \textbf{in}\ (x \cdot z, x \cdot dz + z \cdot dx)
    \end{array} \)
    \caption{\label{subfig:key-example-fwd}
      Dual-numbers forward AD
    }
  \end{subfigure}
  \hfill
  \begin{subfigure}[b]{0.47\textwidth}
    \( \begin{array}{@{}l@{}}
        \begin{array}{@{}r@{}l@{}}
          \fun( & (x : \R, dx : \lR\lra (\lR, \lR)) \\
          {,}\hspace{0.03cm}{} & (y : \R, dy : \lR\lra (\lR, \lR))).
        \end{array} \\
        \; \textbf{let}\ (z, dz) = (x + y , \lfun(d : \lR).\ dx\ d + dy\ d) \\
        \; \textbf{in}\ (x \cdot z , \lfun(d : \lR).\ dz\ (x \cdot d) + dx\ (z \cdot d))
    \end{array} \)
    \caption{\label{subfig:key-example-rev}
      Dual-numbers reverse AD
    }
  \end{subfigure}

  \caption{\label{fig:key-example}
    An example program together with its derivative, both using dual-numbers forward AD and using dual-numbers reverse AD.
    The original program is of type $(\R, \R) \ra \R$.
  }
\end{figure}

\myparagraph{Naive dual-numbers reverse AD}
For reverse AD, it is in general not possible to interleave the primal and derivative computations, as the reversal of derivatives generally requires us to have computed all primals before starting the derivative computation.
However, by choosing a clever encoding, it is possible to make reverse AD \emph{look} like a dual-numbers-style code transformation.
Even if the primal and derivative computations will not be performed in an interleaved fashion, we can interleave the primal computation with a computation that \emph{builds} a delayed reverse derivative function.
Afterwards, we still need to call the derivative function that has been built.
The advantage of such a ``dual-numbers-style'' approach to reverse AD is that the code transformation can be simple and widely applicable.

To make reverse AD in dual-numbers style possible, we have to encode the ``reversal'' in the tangent scalars that we called $dx$ and $dy$ in \cref{subfig:key-example-fwd}.
A solution is to replace those tangent scalars with \emph{linear functions} that take the \emph{cotangent} of the scalar it is paired with, and return the cotangent of the full input of the program.
Transforming the same example program \cref{subfig:key-example-orig} using this style of reverse AD yields \cref{subfig:key-example-rev}.
The linearity indicated by the $\lra$-arrow here is that of a monoid homomorphism (a function preserving 0 and (+)\footnote{And also scaling, making it a vector space homomorphism. The vector space structure of tangents and cotangents tends to be unused in AD.}); operationally, however, linear functions behave just like regular functions.

This naive dual-numbers reverse AD code transformation, shown in full in \cref{fig:algo-naive}, is simple and it is easy to see that it is correct via a logical relations argument~\citep{nunes-2024-dual-numbers}.
The idea of this argument is to prove via induction that a backpropagator $\mathit{dx} : \lR\lra c$ that is paired with an intermediate value $x : \R$ in the program, computes the reverse (i.e.\ transposed) derivative of the subcomputation that calculates $x : \R$ from the global input to the program.\footnote{That is to say: `$\mathit{dx}\ 1$' returns the gradient of $x$ with respect to the program input.}
$c$ is a type parameter of the code transformation, and represents the type of cotangents to the program input; if this input (of type $\tau$, say) is built from just scalars, discrete types and product and sum types, as we will assume in this paper, it suffices (for correctness, not efficiency) to set $c = \tau$.\footnote{
  In fact, for essentially all first-order $\tau$ (i.e.\ no function types), the conclusions in this paper continue to hold; some extensions are given in \cref{sec:source-language-extension}.
  For arrays (or other large product types), the complexity is correct but constant-factor performance is rather lacking.
  \emph{Efficient} support for arrays is left as future work.
}

Dual-numbers forward AD has the very useful property that it generalises over many types (e.g.\ products, coproducts, recursive types) and program constructs (e.g.\ recursion, higher-order functions), thereby being applicable to many functional languages; the same property is inherited by the style of dual-numbers reverse AD exemplified here.
However, unlike dual-numbers forward AD (which can propagate tangents through a program with only a constant-factor overhead over the original runtime), naive dual-numbers reverse AD is wildly inefficient: calling $dx_n$ returned by the differentiated program in \cref{fig:exponential-complexity-example} takes time \emph{exponential} in $n$.
\begin{figure}
  \begin{align*}
    \begin{array}{@{}l|l@{}}
      \begin{array}{@{}l@{}l@{}l@{}}
        \begin{array}{@{}l@{}l@{}}
          \fun&(x_0 : \R). \\
            &\textbf{let}\ x_1 = x_0 + x_0 \\
            &\textbf{in}\ \textbf{let}\ x_2 = x_1 + x_1 \\[-2pt]
            &\quad\vdots \\
            &\textbf{in}\ \textbf{let}\ x_{n+1} = x_{n} + x_{n} \\
            &\textbf{in}\ x_{n+1}    
        \end{array}
        & \rightsquigarrow &        
        \begin{array}{@{}l@{}l@{}}
          \fun&(x_0 : \R, dx_0 : \lR\lra \lR). \\
            &\textbf{let}\ (x_1, dx_1) = (x_0 + x_0, \lfun(d : \lR).\ dx_0\ d + dx_0\ d) \\
            &\textbf{in}\ \textbf{let}\ (x_2, dx_2) = (x_1 + x_1, \lfun(d : \lR).\ dx_1\ d + dx_1\ d) \\[-2pt]
            &\quad\vdots \\
            &\textbf{in}\ \textbf{let}\ (x_{n+1}, dx_{n+1}) = (x_{n} + x_{n}, \lfun(d : \lR).\ dx_{n}\ d + dx_{n}\ d) \\
            &\textbf{in}\ (x_{n+1}, dx_{n+1})
        \end{array}
      \end{array}
      \\[-8pt]
    &
      \smash{\raisebox{-4pt}{\scalebox{0.75}{%
        \begin{tikzpicture}
          \node (4) at (0, 0.9) {$dx_{n+1}$};
          \node (3) at (0, 1.8) {$dx_{n}$};
          \node (d) at (0, 2.7) {$\mathrlap{\smash{\raisebox{2pt}{\vdots}}}\phantom{\vdots}$};
          \node (2) at (0, 3.6) {$dx_1$};
          \node (1) at (0, 4.5) {$dx_0$};
          \draw[->, line width=1pt] (4) to [bend left=35] (3);
          \draw[->, line width=1pt] (4) to [bend right=35] (3);
          \draw[->, line width=1pt] (3) to [bend left=35] (d);
          \draw[->, line width=1pt] (3) to [bend right=35] (d);
          \draw[->, line width=1pt] (d) to [bend left=35] (2);
          \draw[->, line width=1pt] (d) to [bend right=35] (2);
          \draw[->, line width=1pt] (2) to [bend left=35] (1);
          \draw[->, line width=1pt] (2) to [bend right=35] (1);
        \end{tikzpicture}%
      }}}
    \end{array}
  \end{align*}
  \caption{\label{fig:exponential-complexity-example} 
    Left: an example showing how naive dual-numbers reverse AD can result in exponential blow-up when applied to a program with sharing.
    Right: the dependency graph of the backpropagators $dx_i$.
  }
\end{figure}
Such overhead would make reverse AD completely useless in practice---particularly because other (less flexible) reverse AD algorithms exist that indeed do a lot better.
(See e.g.\ \citep{adbook-2008-griewank-walther,ad-2018-survey-automatic-differentiation}.)

Fortunately, it turns out that this naive form of dual-numbers reverse AD can be \emph{optimised} to be as efficient (in terms of time complexity) as these other algorithms---and most of these optimisations are just applications of standard functional programming techniques.
This paper presents a sequence of changes to the code transformation (see the overview in \cref{fig:key-overview}) that fix all the complexity issues and, in the end, produce an algorithm with which the differentiated program has only a constant-factor overhead in runtime over the original program.
This complexity is as desired from a reverse AD algorithm, and is best possible, while nevertheless being applicable to a wide range of programming language features.
The last algorithm from \cref{fig:key-overview} can be enhanced to differentiate task-parallel source programs, and can also be further optimised to something essentially equivalent to classical taping techniques.

\begin{figure}
  \resizebox{\linewidth}{!}{
  \begin{tikzpicture}
    \node[draw] (1) at (0, 0) {\makecell[c]{
      $\trans1c$ (\cref{fig:algo-naive,fig:wrapper-naive}) \\
      Naive dual-numbers \\ reverse AD
    }};
    \node[draw] (2) at (5.6, 0) {\makecell[c]{
      (\cref{sec:staging}) \\
      Stage back- \\ propagator calls \\ with linear factoring${}^\dag$
    }};
    \node[draw] (3) at (10.5, 0) {\makecell[c]{
      $\trans2c$ (\cref{fig:algo-monadic}) \\
      Use ID generation \\ monad to \\ implement staging
    }};
    \node[draw] (4) at (0.2, -2.9) {\makecell[c]{
      $\trans3c$ (\cref{fig:algo-cayley}) \\
      Cheap zero and plus \\ on backpropagator \\ codomain
    }};
    \node[draw] (5) at (5.6, -2.9) {\makecell[c]{
      (\cref{sec:just-scalars}) \\
      Logarithmic-time \\ addition of sparse \\ one-hot vectors
    }};
    \node[draw] (6) at (10.7, -2.9) {\makecell[c]{
      (\cref{sec:mutarrays}) \\
      Final log-factors \\ removed from \\ complexity
    }};
    \draw[->] (1) -- (2)
      node [midway, above] {\footnotesize \vphantom{a}\smash{Replace top-level}}
      node [midway, below] {\footnotesize \vphantom{a}\smash{cotangent type}};
    \draw[->] (2) -- (3)
      node [midway, above] {\footnotesize \vphantom{a}\smash{To monadic}}
      node [midway, below] {\footnotesize \vphantom{a}\smash{code}};
    \draw[->] (3) .. controls (9.5, -2.4) and (1, -0.9) .. (4)
      node [midway, above] {\footnotesize Cayley transform};
    \draw[->] (4) -- (5)
      node [midway, above] {\footnotesize \vphantom{a}\smash{Use a $\Map{}{}$}}
      node [midway, below] {\footnotesize \vphantom{a}\smash{as collector}};
    \draw[->] (5) -- (6)
      node [midway, above] {\footnotesize \vphantom{a}\smash{Use mutable}}
      node [midway, below] {\footnotesize \vphantom{a}\smash{arrays}};
  \end{tikzpicture}}

  \caption{\label{fig:key-overview}
    Overview of the optimisations to dual-numbers reverse AD as a code transformation that are described in this paper.
    ($\dag$ = inspired by
    \protect\citep{ad-2020-dualnum-revad-linear-factoring})
  }
\end{figure}

\myparagraph{Optimisation steps}
The first step in \cref{fig:key-overview} is to apply \emph{linear factoring}: for a linear function $f$, such as a backpropagator, we have that $f\ x + f\ y = f\ (x + y)$.
Observing the form of the backpropagators in \cref{fig:algo-naive}, we see that in the end all we produce is a giant sum of applications of backpropagators to scalars; hence, in this giant sum, we should be able to contract applications of the same backpropagator using this linear factoring rule.
The hope is that we can avoid executing $f$ more often than is strictly necessary if we
represent and reorganise these applications at runtime in a sufficiently clever way.

We achieve this linear factoring by not returning a plain $c$ (presumably the type of (cotangents to) the program input) from our backpropagators, but instead a $c$ wrapped in an object that can delay calls to linear functions producing a $c$.
This object we call $\Staged$; aside from changing the monoid that we are mapping into from $(c, \underline0, (+))$ to $(\Staged c, \ZeroStaged, (\PlusStaged))$, the only material change is that the calls to argument backpropagators in $\trans1c[\textit{op}]$ are now wrapped using a new function $\StagedCall$, which delays the calls to $d_i$ by storing the relevant metadata in the returned $\Staged$ object.

However, it is not obvious how to implement this $\Staged$ type:
at the very least, we need decidable equality on linear functions
to be able to implement the linear factoring rule;
and if we want any hope of an efficient implementation,
we even need a linear order on such linear functions so that we can use them 
as keys in a tree map in the implementation of $\Staged$.
Furthermore, even if we can delay calls to backpropagators, we still need to \emph{call} them at some point, and it is unclear in what order we should do so (while this order turns out to be very important for efficiency).

We thus need two orders on our backpropagators.
It turns out that for \emph{sequential} input programs, it suffices to generate a unique, monotonically increasing identifier (ID) for each backpropagator that we create, and use that ID not only as a witness for backpropagator identity and as a key in the tree map, but also as a witness for the dependency order (see below) of the backpropagators.
These IDs are generated by letting the differentiated program run in an ID generation monad (a special case of a state monad).
The result is shown in \cref{fig:algo-monadic}, which is very similar to the previous version in \cref{fig:algo-naive} apart from threading through the next-ID-to-generate.
(On first glance the code looks very different, but this is only due to monadic bookkeeping.)

At this point, the code transformation already reaches a significant milestone: by staging (delaying) calls to backpropagators as long as possible, we can ensure that \emph{every backpropagator is called at most once}.
This milestone can be seen using the following observation: lambda functions in a pure functional program that do not take functions as arguments, can only call functions that appear in their closure.
Because backpropagators are never mutually recursive (that could only happen if their corresponding scalars are defined mutually recursively, which, being scalars, would never terminate anyway), this observation means that calling backpropagators (really \emph{unfolding}, as we delay evaluating subcalls) in their reverse dependency order achieves the goal of delaying the calls as long as possible.
Thus, for monotonically increasing IDs, a backpropagator will only call other backpropagators with lower IDs.
If we do not need parallelism in the differentiated programs, we can therefore suffice by simply resolving backpropagators from the highest to the lowest ID.

Effectively, one can think of the IDs as (labels of) the nodes in the computation graph of this run of the original program; from this perspective, the resolving process is nothing more than a choice of a reverse topological order of this graph, so that we can perform backpropagation (the reverse pass) in the correct order without redundant computation.

But we are not done yet.
The code transformation at this point ($\trans2c$ in \cref{fig:algo-monadic}) still has a glaring problem: orthogonal to the issue that backpropagators were called too many times (which we fixed), we are still creating one-hot input cotangents and adding those together.
This problem is somewhat more subtle, because it is not actually apparent in the program transformation itself; indeed, looking back at \cref{subfig:key-example-rev}, there are no one-hot values to be found.
However, the only way to \emph{use} the program in \cref{subfig:key-example-rev} to do something useful, namely to compute the cotangent (gradient) of the input, is to pass $(\lambda z.\ (z, 0))$ to $\textit{dx}$ and $(\lambda z.\ (0, z))$ to $\textit{dy}$; it is easy to see that generalising this to larger input structures results in input values like $(0, \ldots, 0, z, 0, \ldots, 0)$ that get added together.
(These are created in the \emph{wrapper} in \cref{fig:wrapper-naive}.)
Adding many zeros together can hardly be the most efficient way to go about things, and indeed this is a complexity issue in the algorithm.

The way we solve this problem of one-hots is less AD-specific: the most important optimisations that we perform are Cayley transformation (\cref{sec:cayley}) and using a better sparse vector representation ($\Map\ \Z\ \lR$ instead of a plain $c$ value; \cref{sec:just-scalars}).
Cayley transformation (also known in the Haskell community by a common use: \emph{difference lists}~\citep{fp-1986-difference-lists}) is a classic technique in functional programming that represents an element $m$ of a 
 monoid $M$ (written additively in this paper) by the function $m + - : M \ra M$ it induces through addition.
Cayley transformation helps us because the monoid $M \ra M$ has very cheap zero and plus operations: $\id$ and $(\circ)$.
Afterwards, using a better (sparse) representation for the value in which we collect the final gradient, we can ensure that adding a one-hot value to this gradient collector can be done in logarithmic time.

By now, the differentiated program can compute the gradient with a \emph{logarithmic} overhead over the original program.
If a logarithmic overhead is not acceptable, the log-factor in the complexity can be removed by using functional mutable arrays (\cref{sec:mutarrays}).
Such mutability can be safely accommodated in our code transformation by either swapping out the state monad for a resource-linear state monad, or by using mutable references in an \ST-like monad (\cref{sec:mutarrays-io}).
(The latter can be generalised to the parallel case; see below.)

And then we are done for sequential programs, because we have now obtained a code transformation with the right complexity: the differentiated program computes the gradient of the source program at some input with runtime proportional to the runtime of the source program at that input.

\myparagraph{Correctness}
Correctness of the resulting AD technique follows in two steps:
1. the naive dual-numbers reverse AD algorithm we start with is correct by a logical relations argument detailed by \cite{nunes-2024-dual-numbers};
2. we transform this into our final algorithm using a combination of (A) standard optimisations that are well-known to preserve semantics (and hence correctness)---notably sparse vectors via Cayley transformation\footnote{See \citep{fp-1986-difference-lists} for intuition, or \citep[\S3.3]{2018-yoneda-profunctor} for a relation to theory.} and using a mutable array to optimise a tree map---and (B) the custom optimisation of linear factoring, which is semantics-preserving because derivatives (backpropagators) are linear functions.

\myparagraph{Parallelism}
If the user is satisfied with a fully sequential (non-parallel) computation of the derivative,
it is enough to generate monotonically increasing integers and use them, together with their linear order, as IDs for backpropagators during the forward pass.
This is the approach described so far.

However, if the source program has (task) parallelism (we assume fork-join parallelism in this paper), we would prefer to preserve that parallelism when performing backpropagation on the (implied) computation graph.
The linear order (chronological, by comparing IDs) that we were using to witness the dependency order so far does not suffice any more: we need to be more frugal with adding spurious edges in the dependency graph that only exist because one backpropagator happened to be created after another on the clock.
However, we only need to make our order (i.e.\ dependency graph) precise enough that independent tasks are incomparable in the order (and hence independent in the dependency graph); recording more accurate (lack of) dependencies between individual scalar operations would even allow exploiting implicit parallelism within an a priori serial subcomputation, which is potentially interesting but beyond the scope of this paper.\footnote{It is also treacherous ground: it turns out to be very difficult to fulfil the promise of functional programming that all independent expressions are parallelisable, because thread management systems simply have too much overhead for that granularity of parallelism. (Interesting work here was done recently by \cite{westrick2024parallelism}.)}

Our solution is thus to switch from simple integer IDs to compound IDs, consisting of a job ID and a sequential ID within that job.\footnote{Being pairs of integers, these still have a natural linear order to use as a map key.}
We assume parallelism in the source program is expressed using a parallel pair constructor with the following typing rule:
\[
  \frac{\Gamma \vdash t : \tau \qquad
        \Gamma \vdash s : \sigma}
       {\Gamma \vdash t \parCom s : (\tau, \sigma)}
\]
(The method generalises readily to $n$-ary versions of this primitive.)
To differentiate code using this construct, we take out the ID generation monad that the target program ran in so far, and replace it with a monad in which we can also record the dependency graph between parallel jobs.
The derivative of $(\parCom)$ is the only place where we use the new methods of this monad: all other code transformation rules remain identical, save for writing the right-hand sides as a black-box monad instead of explicit state passing.
We can then make use of this additional recorded information in the backpropagator resolution phase to do so in parallel; in this process, the net effect is that \emph{forks from the primal become joins in the derivative computation}, and vice versa.

\myparagraph{Comparison to other algorithms}
We can relate the sequential version of our technique to that of \cite{ad-2021-krawiec-kmett-ad}
by noting that we can replace $\trans1c[\R] = (\R, \lR\lra c)$ with the isomorphic definition $\trans1c[\R] = (\R, c)$.
This turns the linear factoring rule into a \emph{distributive law} $v\cdot x+v\cdot y\leadsto v\cdot (x+y)$ that is effectively applied at runtime by using an intensional 
representation of the cotangent expressions of type $c$.
While their development is presented very differently and the equivalence is not at all clear at first sight, we explain the correspondence in \cref{sec:improve}.

This perspective also makes clear the relationship between our technique and that of \citep{ad-2019-fwd-ad-gradient-compiler-opts}.
Where they try to optimise vectorised forward AD to reverse AD at \emph{compile time} by using a distributive law (which sometimes succeeds for sufficiently simple programs), our technique proposes a clever way of efficiently applying the distributive law in the required places at \emph{runtime}, giving us the power to always achieve the desired reverse AD behaviour.

Finally, we are now in the position to note the similarity to (sequential\footnote{Traditional taping-based reverse AD methods are fundamentally sequential. They may have parallel primitive operations, such as matrix multiplications etc., but there is typically no general task parallelism.}) taping-based AD as in \citep{ad-2021-kmett-hackage}, older versions of PyTorch~\citep{ad-2017-pytorch}, etc.: the incrementing IDs that we attached to backpropagators earlier give a mapping from $\{0,\ldots,n\}$ to our backpropagators.
Furthermore, each backpropagator corresponds to either a primitive arithmetic operation performed in the source program, or to an input value; this already means that we have a tape, in a sense, of all performed primitive operations, albeit in the form of a chain of closures.
The optimisation using mutable arrays (\cref{sec:mutarrays}) which reifies this tape in a large array in the reverse pass, especially if one then proceeds to already use this array in the forward pass (\cref{sec:improve-taping}), eliminates also this last difference.

\section{Preliminaries: The Complexity of Reverse AD}\label{sec:rev-ad-complexity}

The only reason, in practice, for using reverse AD over forward AD (which is significantly easier to implement) is computational complexity.
Arguably, therefore, it is important that we fix precisely what the time complexity of reverse AD ought to be, and check that any proposed algorithm indeed conforms to this time complexity.

In this paper we discuss a code transformation, so we phrase the desired time complexity in terms of a code transformation $\mathcal R$ that takes a program $P$ of type $\R^n \ra \R^m$ to a program $\mathcal R[P]$ of type\footnote{We will discuss the type of reverse AD in more detail in \cref{sec:rev-ad-type}, generalised beyond vectors of reals.} $(\R^n, {\lR^m}) \ra {\lR^n}$ that computes the reverse derivative of $P$.
The classic result \citep[see][]{adbook-2008-griewank-walther} is that, for sequential first-order languages, $\mathcal R$ exists such that the following criterion is satisfied:
\[
  \begin{array}{@{}l@{}}
    \exists c > 0.\ \forall P : \mathrm{Programs}(\R^n \ra \R^m).\ \forall I : \R^n, A : {\lR^m}. \\
    \qquad \cost(\mathcal R[P]\ (I, A))
      \leq c \cdot (\cost(P\ I) + \size(I))
  \end{array}
\]
where we denote by $\cost(E)$ the amount of time required to evaluate the expression $E$, and by $\size(I)$ the amount of time required to read all of $I$ sequentially.
(Note that $\cost(\mathcal R[P]\ (I, A))$ does not measure the cost of evaluating the code transformation $\mathcal R$ itself; that is considered to be a compile-time cost.)
In particular, if $P$ reads its entire input (and does not ignore part of it), the second line can be simplified to $\cost(\mathcal R[P]\ (I, A))) \leq c \cdot \cost(P\ I)$.

The most important point of this criterion is that $c$ cannot depend on $P$: informally, the output program produced by reverse AD is not allowed to have more than a constant factor overhead in runtime over the original program, and this constant factor is uniform over all programs in the language.

A weaker form of the criterion is sometimes used where $c$ is dependent on the program in question but not on the size of the input to that program; for example, `$f = \lambda \textit{arr}.\ \text{sum}\ \textit{arr}$' is allowed to have a different $c$ than `$g = \lambda \textit{arr}.\ \text{sum}\ (\text{map}\ (\lambda x.\ x + 1)\ \textit{arr})$', but given $f$, the same constant $c$ will still apply regardless of the size of the \emph{input array}.
(Note that this can only make sense in a language that has variably sized arrays or similar structures.)
This criterion is used by e.g.~\cite{ad-2022-futhark-partial-recompute}, where $c$ is proportional to the largest scope depth in the program.
In this case, the criterion is expressed for a program family $\mathit{PF}$ that should be understood to be the same program for all $n$, just with different input sizes:
\[
  \begin{array}{@{}l@{}}
    \forall \mathit{PF} : (n : \N) \ra \mathrm{Programs}(\R^n \ra \R^m).\ \exists c > 0.\ \forall n \in \N.\ \forall I : \R^n, A : {\lR^m}. \\
    \qquad \mathrm{cost}(\mathcal R[\mathit{PF}_n]\ (I, A)) \leq c \cdot (\cost(\mathit{PF}_n\ I) + \size(I))
  \end{array}
\]
where $c$ is preferably at most linearly or sub-linearly dependent on the size of the program code of $\mathit{PF}$.

The final sequential version of the code transformation described in this paper (in \cref{sec:mutarrays}) satisfies the first (most stringent) criterion.
For the parallel version (\cref{sec:parallelism}), the criterion holds for the amount of work performed.

\section{Preliminaries: The Type of Reverse AD}\label{sec:rev-ad-type}
Before one can define an algorithm, one has to fix the type of that algorithm.
Similarly, before one can define a code transformation, one has to fix the domain and codomain of that transformation: the ``type'' of the transformation.

\myparagraph{Typing forward AD}
For forward AD on first-order programs (or at least, programs for which the input and output does not contain function values), the desired type seems quite evident: $\mathcal F : (\sigma \ra \tau) \rightsquigarrow ((\sigma, \underline{\sigma}) \ra (\tau, \underline{\tau}))$, where we write $T_1 \rightsquigarrow T_2$ for a (compiler) code transformation taking a program of type $T_1$ and returning a program of type $T_2$, and where $\underline{\tau}$ is the type of tangent vectors (derivatives) of values of type $\tau$.
This distinction between the type of values $\tau$ and the type of their derivatives $\underline{\tau}$ is important in some versions of AD, but will be mostly cosmetic in this paper; in an implementation one can take $\underline{\tau} = \tau$, but there is some freedom in this choice.\footnote{For example, $\underline{\R} = \R$, but for $\underline{\Z}$ one can choose the unit type $()$ and be perfectly sound and consistent.}
Given a program $f : \sigma \ra \tau$, $\mathcal F[f]$ is a program that takes, in addition to its regular argument, also a tangent at that argument; the output is then the regular result paired up with corresponding tangent at that result.

More specifically, for forward AD, we want the following in the case that $\sigma = \R^n$ and $\tau = \R^m$ (writing $\mathbf{x} = (x_1, \ldots, x_n)$):\footnote{This generalises to more complex (but still first-order) in/outputs by regarding those as collections of real values as well.}
\begin{align*}
\mathcal F[f]\left(\mathbf{x}, \left(\frac{\partial x_1}{\partial\alpha}, \ldots, \frac{\partial x_n}{\partial \alpha}\right)\right)
= \left(f(\mathbf{x}), \left(\frac{\partial f(\mathbf{x})_1}{\partial\alpha}, \ldots, \frac{\partial f(\mathbf{x})_m}{\partial\alpha}\right)\right)
\end{align*}
Setting $\alpha = x_i$ means passing $(0, \ldots, 1, \ldots, 0)$ as the argument of type $\underline{\sigma}$ and computing the partial derivative with respect to $x_i$ of $f(\mathbf{x})$.
In other words, $\snd(\mathcal F[f](x, \mathit{dx}))$ is the directional derivative of $f$ at $x$ in the direction $\mathit{dx}$.

\myparagraph{A first attempt at typing reverse AD}
For reverse AD, the desired type is less evident.
A first guess would be:
\begin{align*}
\mathcal R_1 : (\sigma \ra \tau) \rightsquigarrow ((\sigma, \underline{\tau}) \ra (\tau, \underline{\sigma}))
\end{align*}
with this intended meaning for $\sigma = \R^n$ and $\tau = \R^m$ (again writing $\mathbf{x} = (x_1, \ldots, x_n)$):
\begin{align*}
\mathcal R_1[f]\left(\mathbf{x}, \left(\frac{\partial\omega}{\partial f(\mathbf{x})_1}, \ldots, \frac{\partial\omega}{\partial f(\mathbf{x})_m}\right)\right)
= \left(f(\mathbf{x}), \left(\frac{\partial\omega}{\partial x_1}, \ldots, \frac{\partial\omega}{\partial x_n}\right)\right)
\end{align*}
In particular, if $\tau = \R$ and we pass 1 as its cotangent (also called adjoint) of type $\underline{\tau} = \lR$, the $\underline{\sigma}$-typed output contains the gradient with respect to the input.

\myparagraph{Dependent types}
However, $\mathcal R_1$ is not readily implementable for even moderately interesting languages.
One way to see this is to acknowledge the reality that the type $\underline{\tau}$ (of derivatives of values of type $\tau$) should really be dependent on the accompanying \emph{primal value} of type $\tau$.
Let us write the type of derivatives at $\tau$ not as $\underline\tau$ but as $\mathcal D[\tau](x)$, where $x : \tau$ is that primal value.
With just scalars and product types this dependence does not yet occur (e.g.\ $\mathcal D[\R](x) = \R$ independent of the primal value $x$), but when adding sum types (coproducts), the dependence becomes non-trivial: the only sensible derivatives for a value $\inl(x) : \sigma + \tau$ (for $x : \sigma$) are of type $\underline{\sigma}$.
Letting $\underline{\sigma + \tau} = \underline{\sigma} + \underline{\tau}$ would allow passing a derivative value of type $\underline{\tau}$ to $\inl(x) : \sigma + \tau$, which is nonsensical (and an implementation could do little else than return a bogus value like 0 or throw a runtime error).
The derivative of $\inl(2x) : \R + \Bool$ cannot be $\inr(\True)$; it should at least somehow contain a real value.

Similarly, the derivative for a dynamically sized array, if the input language supports those, must really be of the same size as the input array.
This, too, is a dependence of the type of the derivative on the \emph{value} of the input.

Therefore, the output type of forward AD which we wrote above as $(\sigma, \underline{\sigma}) \ra (\tau, \underline{\tau})$ should really\footnote{The notation `$\Sigma_{x : \sigma}\, \tau$' denotes a \emph{sigma type}: it is roughly equivalent to the pair type $(\sigma, \tau)$, but the type $\tau$ is allowed to refer to $x$, the value of the first component of the pair.} be $(\Sigma_{x : \sigma}\, \mathcal D[\sigma](x)) \ra (\Sigma_{y : \tau}\, \mathcal D[\tau](y))$, rendering what were originally pairs of value and tangent now as \emph{dependent} pairs of value and tangent.
This is a perfectly sensible type, and indeed correct for forward AD, but it does not translate at all well to reverse AD in the form of $\mathcal R_1$: the output type would be something like $(\Sigma_{x : \sigma}\, \mathcal D[\tau](y)) \ra (\Sigma_{y : \tau}\, \mathcal D[\sigma](x))$, which is nonsense because both $x$ and $y$ are out of scope.

\myparagraph{Let-bindings}
A different way to see that the type of $\mathcal R_1$ is unusable, is to note that one cannot even differentiate let-bindings using $\mathcal R_1$.
In order to apply to (an extension of) the lambda calculus, let us rewrite the types somewhat: where we previously put a function $f : \sigma \ra \tau$, we now put a term $x : \sigma \vdash t : \tau$ with its input in a free variable and producing its output as the returned value.
Making the modest generalisation to support any full environment as input (instead of just a single variable), we get $\mathcal R_1 : (\Gamma \vdash t : \tau) \rightsquigarrow (\Gamma, d : \underline{\tau} \vdash \mathcal R_1[t] : (\tau, \underline{\Gamma}))$, where $\underline{\Gamma}$ is a tuple containing the derivatives of all elements in the environment $\Gamma$.
(To be precise, we define $\underline{\varepsilon} = ()$ for the empty environment and $\underline{\Gamma, x : \tau} = (\underline{\Gamma}, \underline{\tau})$ inductively.)

Now, consider differentiating the following program using $\mathcal R_1$:
\begin{align*}
\Gamma \vdash (\mathbf{let}\ x = e_1\ \mathbf{in}\ e_2) : \tau
\end{align*}
where $\Gamma \vdash e_1 : \sigma$ and $\Gamma, x : \sigma \vdash e_2 : \tau$.
Substituting, we see that $\mathcal R_1$ needs to somehow build a program of this type:
\begin{align}
\Gamma, d : \underline{\tau} \vdash \mathcal R_1[\mathbf{let}\ x = e_1\ \mathbf{in}\ e_2] : (\tau, \underline{\Gamma})
\label{r1_deriv_letbind}
\end{align}
However, recursively applying $\mathcal R_1$ on $e_1$ and $e_2$ yields terms:
\begin{align*}
\Gamma, d : \underline{\sigma} &\vdash \mathcal R_1[e_1] : (\sigma, \underline{\Gamma}) \\
\Gamma, x : \sigma, d : \underline{\tau} &\vdash \mathcal R_1[e_2] : (\tau, (\underline{\Gamma}, \underline{\sigma}))
\end{align*}
To produce the program in \eqref{r1_deriv_letbind}, we cannot use $\mathcal R_1[e_2]$ because we do not yet have an $x : \sigma$ (which needs to come from $\mathcal R_1[e_1]$), and we cannot use $\mathcal R_1[e_1]$ because the $\underline{\sigma}$ needs to come from $\mathcal R_1[e_2]$!
The type of $\mathcal R_1$ demands the cotangent of the result \emph{too early}.

Of course, one might argue that we can just use $e_1$ to compute the $\sigma$, $\mathcal R_1[e_2]$ to get the $\underline{\sigma}$ and $e_2$'s contribution to $\underline{\Gamma}$, and finally $\mathcal R_1[e_1]$ to get $e_1$'s contribution to $\underline{\Gamma}$ based on its own cotangent of type $\underline{\sigma}$.
However, this would essentially compute $e_1$ twice (once directly and once as part of $\mathcal R_1[e_1]$), meaning that the time complexity becomes super-linear in the depth of let-bindings, which is quite disastrous for typical functional programs.

So in addition to not being precisely typeable, $\mathcal R_1$ is also not implementable in a compositional way.

\myparagraph{Fixing the type of reverse AD}
Both when looking at the dependent type of $\mathcal R_1$ and when looking at its implementation, we found that the cotangent $\mathit{dy} : \mathcal D[\tau](y)$ was required before the result $y : \tau$ was itself computed.
One way to solve this issue is to just postpone requiring the cotangent of $y$, i.e.\ to instead look at $\mathcal R_2$:\footnote{
  Where a $\Sigma$-type is a dependent pair, a $\Pi$-type is a dependent function: $\Pi_{x:\sigma}\, \tau$ means $\sigma \to \tau$ except that the type $\tau$ may depend on the argument value $x$.
}
\begin{align*}
\begin{array}{l@{\qquad}c}
  \mathcal R_2 : (\sigma \ra \tau) \rightsquigarrow (\sigma \ra (\tau, \underline{\tau} \ra \underline{\sigma}))
    & \text{(non-dependent version)} \\[0.2em]
  \mathcal R_2 : (\sigma \ra \tau) \rightsquigarrow (\Pi_{x : \sigma}\, \Sigma_{y : \tau}\, (\mathcal D[\tau](y) \ra \mathcal D[\sigma](x)))
    & \text{(dependent version)}
\end{array}
\end{align*}
Note that this type \emph{is} well-scoped.
Furthermore, this ``derivative function'' mapping the cotangent of the result to the cotangent of the argument is actually a linear function, in the sense of a vector space homomorphism: indeed, it is multiplication by the Jacobian matrix of $f$, the input function.
Thus we can write:
\begin{align*}
\mathcal R_2 : (\sigma \ra \tau) \rightsquigarrow (\Pi_{x : \sigma}\, \Sigma_{y : \tau}\, (\mathcal D[\tau](y) \lra \mathcal D[\sigma](x)))
\end{align*} 
which is the type of the reverse\footnote{
There is also a corresponding formulation of forward AD which would have type:
\[
  \mathcal F_2 : (\sigma \ra \tau) \rightsquigarrow (\Pi_{x : \sigma}\, \Sigma_{y : \tau}\, (\mathcal D[\sigma](x) \lra \mathcal D[\tau](y)))
\]
However, in the case of forward AD, there is no added value in using this more precise type, compared to our previous formulation $\mathcal F$.
In fact, there are downsides: as we are forced to consume tangents only after the primal computation has finished,
we can no longer interleave the primal and tangent computations, leading to larger memory use.
Moreover, the resulting code transformation is more complex than $\mathcal F$. 
} AD code transformation derived by \cite{adfp-2018-categories-ad} and in CHAD (e.g.\ \citep{vakar-2022-chad}; see also \cref{sec:related-work:pl}).\footnote{Actually, CHAD has non-identity type mappings for the primal types $x : \sigma$ and $y : \tau$ as well in order to compositionally support function values in a way that fits the type of $\mathcal R_2$. We consider only the top-level type in this discussion, and for first-order in- and output types, the two coincide.}

While this formulation of reverse AD admits a rich mathematical foundation \citep{DBLP:journals/mscs/NunesV23}
and has the correct complexity~\citep{chad-efficient-popl},
the required program transformation is more complex than the formulation $\mathcal F$ that we have for forward AD.
In particular, we need to compute for each programming language construct what its CHAD transformation is, which may be non-trivial (for example, for the case of function types).

This motivates us to pursue a reverse AD analogue of $\mathcal F$.

\myparagraph{Applying Yoneda/CPS}
An instance of the Yoneda lemma (or in this case: continuation-passing style; see also~\citep{2018-yoneda-profunctor}) is that $\sigma \lra \tau$ is equivalent to $\forall r.\, (\sigma \lra r) \to (\tau \lra r)$.
We can apply this to the $\lra$-arrow in $\mathcal R_2$ to obtain a type for reverse AD that is somewhat reminiscent of our formulation $\mathcal F$ of forward AD.
With just Yoneda, we get $\mathcal R_3'''$ below; we then weaken this type somewhat by enlarging the scope of the $\forall c$ quantifier, weaken some more by taking the $(\mathcal D[\tau](x) \lra c)$ argument before returning $y$, and finally we uncurry to arrive at $\mathcal R_3$:
\begin{align*}
\mathcal R_3''' &: (\sigma \ra \tau) \rightsquigarrow \Pi_{x : \sigma}\, \Sigma_{y : \tau}\, \forall c.\, ((\mathcal D[\sigma](x) \lra c) \ra (\mathcal D[\tau](y) \lra c)) \\
\mathcal R_3'' &: (\sigma \ra \tau) \rightsquigarrow \forall c.\, \Pi_{x : \sigma}\, \Sigma_{y : \tau}\, ((\mathcal D[\sigma](x) \lra c) \ra (\mathcal D[\tau](y) \lra c)) \\
\mathcal R_3' &: (\sigma \ra \tau) \rightsquigarrow \forall c.\, \Pi_{x : \sigma}\, ((\mathcal D[\sigma](x) \lra c) \ra \Sigma_{y : \tau}\, (\mathcal D[\tau](y) \lra c)) \\
\mathcal R_3 &: (\sigma \ra \tau) \rightsquigarrow \forall c.\, (\Sigma_{x : \sigma}\, (\mathcal D[\sigma](x) \lra c)) \ra \Sigma_{y : \tau}\, (\mathcal D[\tau](y) \lra c) \\
&\hphantom{: (\sigma \ra \tau) \;\rightsquigarrow\;} \textcolor{gray}{\forall c.\, (\mathrlap{\sigma,}\hphantom{\Sigma_{x : \sigma}\,(} \mathrlap{\underline{\sigma}}\hphantom{\mathcal D[\sigma](x)} \lra c\hphantom{)}) \ra \mathrlap{(\tau,}\hphantom{\Sigma_{y : \tau}\,(} \mathrlap{\underline{\tau}}\hphantom{\mathcal D[\tau](y)} \lra c)}
\end{align*}
We give both a dependently typed (black) and a simply typed (grey) signature for $\mathcal R_3$.

The $\lra$-arrows in these types, as well as the $c$ bound by the $\forall$-quantifier, live in the category of commutative monoids.
Indeed, $c$ will always have a commutative monoid structure in this paper; that is: it has a zero  $\underline 0$ as well as a commutative, associative addition operation $(+) : (c, c) \ra c$ for which $\underline 0$ is the unit.
(The $\lra$-arrows in these types are really vector space homomorphisms, but since we will only use the substructure of commutative monoids in this paper, and forget about scalar multiplication, we will always consider $\lra$-functions (commutative) monoid homomorphisms.)

Returning to the types in question, we see that we can convert $\mathcal R_2[t]$ to $\mathcal R_3[t]$:
\[
  \begin{array}{@{}l@{}}
    (\fun (x : \sigma, \mathit{dx} : \mathcal D[\sigma](x) \lra c). \\
    \hspace{3em} \mathbf{let}\ (y : \tau, \mathit{dy} : \mathcal D[\tau](y) \lra \mathcal D[\sigma](x)) = \mathcal R_2[t]\ x\ \mathbf{in}\ (y, \mathit{dx} \circ \mathit{dy})) \\
    \hspace{1em} : \forall c.\, (\Sigma_{x : \sigma}\, (\mathcal D[\sigma](x) \lra c)) \ra \Sigma_{y : \tau}\, (\mathcal D[\tau](y) \lra c)
  \end{array}
\]
where we write $\circ$ for the composition of linear functions.
We can also convert $\mathcal R_3[t]$ back to $\mathcal R_2[t]$, but due to how we weakened the types above, only in the non-dependent world:
\begin{align*}
&(\fun (x : \sigma).\ \mathcal R_3[t]\ (x, \lfun (z: \underline \sigma).\ z))
    : \sigma \ra (\tau, \underline \tau \lra \underline \sigma)
\end{align*}
So, in some sense, $\mathcal R_2$ and $\mathcal R_3$ compute the same thing, albeit with types that differ in how precisely they portray the dependencies.

In fact, $\mathcal R_3$ admits a very elegant implementation as a program transformation that is structurally recursive over all language elements except for the primitive operations in the leaves.
However, there are some issues with the computational complexity of this straightforward implementation of $\mathcal R_3$, one of which we will fix here immediately, and the other of which are the topic of the rest of this paper.

\myparagraph{Moving the pair to the leaves}
Let us return to forward AD for a moment.
Recall the type we gave for forward AD:\footnote{We revert to the non-dependent version for now because the dependencies are irrelevant for this point, and they clutter the presentation.}
\begin{align*}
\mathcal F : (\sigma \ra \tau) \rightsquigarrow ((\sigma, \underline{\sigma}) \ra (\tau, \underline{\tau}))
\end{align*}
Supposing we have a program $f : ((\R_1, \R_2), \R_3) \ra \R_4$, we get: (the subscripts are semantically meaningless and are just for tracking arguments)
\begin{align*}
\mathcal F[f] : (((\R_1, \R_2), \R_3), ((\underline{\R_1}, \underline{\R_2}), \underline{\R_3})) \ra (\R_4, \underline{\R_4})
\end{align*}
While this is perfectly implementable and correct and efficient, it is not the type that corresponds to what is by far the most popular implementation of forward AD, namely \emph{dual-numbers forward AD}, which has the following type:
\begin{gather*}
\mathcal F_{\text{dual}} : (\sigma \ra \tau) \rightsquigarrow (\Dual[\sigma] \ra \Dual[\tau]) \\
\Dual[\R] = (\R, \lR) \qquad
\Dual[()] = () \qquad
\Dual[(\sigma, \tau)] = (\Dual[\sigma], \Dual[\tau])
\end{gather*}
Intuitively, instead of putting the pair at the root like $\mathcal F$ does, $\mathcal F_{\text{dual}}$ puts the pair at the leaves---more specifically, at the scalars in the leaves, leaving non-$\R$ types like $()$ or $\Z$ alone.
For the given example program $f$, dual-numbers forward AD would yield the following derivative program type:
\begin{align*}
\mathcal F_{\text{dual}}[f] : (((\R_1, \underline{\R_1}), (\R_2, \underline{\R_2})), (\R_3, \underline{\R_3})) \ra (\R_4, \underline{\R_4})
\end{align*}
Of course, for any given types $\sigma, \tau$ the two versions are trivially inter-converted, and as stated, for forward AD both versions can be defined inductively equally well, resulting in efficient programs in terms of time complexity.

However, for reverse AD in the style of $\mathcal R_3$, the difference between $\mathcal R_3$ and its pair-at-the-leaves dual-numbers variant ($\mathcal R_{3\text{dual}}$ below) is more pronounced.
First note that indeed both styles (with the pair at the root and with the pair at the leaves) produce a sensible type for reverse AD: (again for $f : ((\R_1, \R_2), \R_3) \ra \R_4$)
\begin{align*}
\mathcal R_3[f] &: \forall c.\, (((\R_1, \R_2), \R_3), ((\underline{\R_1}, \underline{\R_2}), \underline{\R_3}) \lra c) \ra (\R_4, \underline{\R_4} \lra c) \\
\mathcal R_{3\text{dual}}[f] &: \forall c.\, (((\R_1, \underline{\R_1} \lra c), (\R_2, \underline{\R_2} \lra c)), (\R_3, \underline{\R_3} \lra c)) \ra (\R_4, \underline{\R_4} \lra c)
\end{align*}
The individual functions of type $\underline{\R} \lra c$ are usually called \emph{backpropagators} in literature, and we will adopt this terminology.

Indeed, these two programs are again easily inter-convertible, if one realises that:
\begin{enumerate}
\item $c$ is a commutative monoid and thus possesses an addition operation, which can be used to combine the three $c$ results into one for producing the input of $\mathcal R_3$ from the input of $\mathcal R_{3\text{dual}}$;
\item The function $g : ((\underline{\R_1}, \underline{\R_2}), \underline{\R_3}) \lra c$ is linear, and hence e.g.\ $\fun (x : \underline{\R_2}).\ g\ ((0, x), 0)$ suffices as value for $\underline{\R_2} \lra c$.
\end{enumerate}

However, the problem arises when defining $\mathcal R_3$ inductively as a program transformation.
To observe this difference between $\mathcal R_3$ and $\mathcal R_{3\text{dual}}$, consider the term $t = \fun(x : (\sigma, \tau)).\ \fst(x)$ of type $(\sigma, \tau) \ra \sigma$ and the types of its derivative using both methods:
\begin{gather*}
\begin{array}{r@{\ }l}
\mathcal R_3[t] &: \forall c.\ ((\sigma, \tau), (\underline{\sigma}, \underline{\tau}) \lra c) \ra (\sigma, \underline{\sigma} \lra c) \\[2pt]
\mathcal R_{3\text{dual}}[t] &: \forall c.\ (\Dual_c[\sigma], \Dual_c[\tau]) \ra \Dual_c[\sigma]
\end{array} \\
\Dual_c[\R] = (\R, \lR\lra c) \qquad
\Dual_c[()] = () \qquad
\Dual_c[(\sigma, \tau)] = (\Dual_c[\sigma], \Dual_c[\tau])
\end{gather*}

Their implementations look as follows:
\begin{align*}
\mathcal R_3[t] &= \fun (x : (\sigma, \tau), \mathit{dx} : (\underline{\sigma}, \underline{\tau}) \lra c).\ (\fst(x), \fun (d : \underline{\sigma}).\ \mathit{dx}\ (d, \underline{0}_{\underline{\tau}})) \\
\mathcal R_{3\text{dual}}[t] &= \fun (x : (\Dual_c[\sigma], \Dual_c[\tau])).\ \fst(x)
\end{align*}
where $\underline{0}_{\underline{\tau}}$ is the zero value of the cotangent type of $\tau$.
The issue with the first variant is that $\tau$ may be an arbitrarily complex type, perhaps even containing large arrays of scalars, and hence this zero value $\underline{0}_{\underline{\tau}}$ may also be large.
Having to construct this large zero value is not, in general, possible in constant time, whereas the primal operation ($\fst$) was a constant-time operation; this is anathema to getting a reverse AD code transformation with the correct time complexity.
Further, on our example program, we see that the variant $\mathcal R_{3}$ results in a more complex code transformation than $\mathcal R_{3\text{dual}}$, and this observation turns out to hold more generally.
$\mathcal R_{3}$ shares both these challenges with the CHAD formulation $\mathcal R_{2}$ of reverse AD.

As evidenced by the complexity analysis and optimisation of the CHAD reverse AD algorithm~\citep{chad-efficient-popl}, there are ways to avoid having to construct a non-constant-size zero value here.
In fact, we use one of those ways, in a different guise, later in this paper in \cref{sec:cayley}.
However, in this paper we choose the $\mathcal R_{3\text{dual}}$ approach.
We pursue the dual numbers approach not to avoid having to deal with the issue of large zeros---indeed, skipping the problem here just moves it somewhere else, namely to the implementation of the backpropagators ($\underline{\R} \lra c$).
Rather, we pursue this approach because $\mathcal R_{3\text{dual}}$ extends more easily to a variety of language features (see \cref{sec:source-language-extension,sec:parallelism}).

\section{Naive, Unoptimised Dual-Numbers Reverse AD}\label{sec:naive}

We first describe the naive  implementation of dual-numbers reverse AD: this algorithm is easy to define and prove correct compositionally, but it is wildly inefficient in terms of complexity.
Indeed, it tends to blow up to exponential overhead over the original function, whereas the desired complexity is to have only a constant factor overhead over the original function.
In subsequent sections, we will apply a number of optimisations to this algorithm that fix the complexity issues, to derive an algorithm that does have the desired complexity.

\subsection{Source and Target Languages}\label{sec:source-target-languages}

The reverse AD methods in this paper are code transformations, and hence have a source language (in which input programs may be written) and a target language (in which gradient programs are expressed).
While the source language will be identical for all versions of the transformation that we discuss, the target language will expand to support the optimisations that we perform.

\begin{figure}
  \figheading{Types: 
  }
  \begin{align*}
    \begin{array}{l@{\ \ }c@{\ \ }l}
    \sigma, \tau &\coloneqqq&
      \R \mid () \mid (\sigma, \tau) \mid \sigma \ra \tau \mid \Z
    \end{array}
  \end{align*}
  \figheading{Terms:}
  \begin{align*}
    \qquad\begin{array}{l@{\ \ }c@{\ \ }ll}
    s, t &\coloneqqq&
    x \mid \mathrlap{() \mid (s, t) \mid \fst(t) \mid \snd(t) \mid s\ t \mid \fun (x : \tau).\ t \mid \textbf{let}\ x : \tau = s\ \textbf{in}\ t} \\
      &\mid& r & \text{(literal $\R$ values)} \\
      &\mid& \textit{op}(t_1,\ldots,t_n) & \text{($\textit{op} \in \Op_n$, primitive operation application)} \\
    \end{array}
  \end{align*}
  \caption{\label{fig:source-language}
    The source language of all variants of this paper's reverse AD transformation.
    $\Z$, the type of integers, is added as an example of a type that AD does not act upon.
  }
\end{figure}

The source language is defined in \cref{fig:source-language}; the initial target language is given in \cref{fig:target-language-1}.
The typing of the source language is completely standard, so we omit typing rules here.
We assume call-by-value evaluation.
The only part that warrants explanation is the treatment of primitive operations: for all $n \in \Z_{>0}$ we presume the presence of a set $\Op_n$ containing $n$-ary primitive operations $\textit{op}$ on real numbers in the source language.
Concretely, given typed programs $\Gamma\vdash t_i : \R$ of type $\R$ in typing context $\Gamma$, for $1\leq i\leq n$, we have a program $\Gamma\vdash \textit{op}(t_1,\ldots,t_n):\R$.
The program transformation does not care what the contents of $\Op_n$ are, as long as the partial derivatives are available in the target language after differentiation.

\begin{figure}
  \figheading{Types:}
  \begin{align*}\qquad
    \begin{array}{l@{\ \ }c@{\ \ }ll}
    \overline\sigma, \overline\tau &\coloneqqq&
      \color{gray} \R \mid () \mid (\overline\sigma, \overline\tau) \mid \Z & \text{(types without functions)} \\
    \sigma, \tau &\coloneqqq& 
      \color{gray} \R \mid () \mid (\sigma, \tau) \mid \Z \mid \sigma \ra \tau \\
      &\mid& \overline\sigma \lra \overline\tau & \text{(linear functions)}
    \end{array}
  \end{align*}
  \figheading{Terms:}
  \begin{align*}
    \begin{array}{l@{\ \ }c@{\ \ }ll}
    s, t &\coloneqqq&
      \color{gray} \mathrlap{x \mid \textbf{let}\ x : \tau = s\ \textbf{in}\ t  \mid () \mid (s, t) \mid \fst(t) \mid \snd(t)\mid \fun (x : \tau).\ t  \mid s\ t  \mid r \mid \textit{op}(t_1,\ldots,t_n)} \\
      &\mid& \lfun(z : \tau).\ b & \text{(linear lambda abstraction ($\tau$ a type without function arrows))} {}
    \end{array}
  \end{align*}
  \figheading{Linear function bodies:
  }
  \begin{align*}
    \begin{array}{l@{\ \ }c@{\ \ }ll}
      b &\coloneqqq&
        () \mid (b, b') \mid \fst(b) \mid \snd(b) & \text{(tupling)} \\
        &\mid& z & \text{(reference to $\lfun$-bound variable)} \\
        &\mid& x\ b & \text{(linear function application; $x:\sigma\lra\tau$ an identifier)} \\
        &\mid& \partial_i\textit{op}(x_1,\ldots,x_n)(b) & \text{($\textit{op} \in \Op_n$, $i$'th partial derivative of $\textit{op}$ )} \\
        &\mid& b + b' & \text{(elementwise addition of results)} \\
        &\mid& \underline0 & \text{(zero of result type)}
      \end{array}
  \end{align*}
  \caption{\label{fig:target-language-1}
    The target language of the unoptimised variant of the reverse AD transformation.
    Components that are also in the source language (\cref{fig:source-language}) are set in \textcolor{gray}{grey}.
  }
\end{figure}

In the target language in \cref{fig:target-language-1}, we add linear functions with the type $\sigma \lra \tau$: these functions are linear in the sense of being monoid homomorphisms, meaning that $f(0) = 0$ and $f(x + y) = f(x) + f(y)$ if $f : \sigma \lra \tau$.
Because it is not well-defined what the derivative of a function value (in the input or output of a program) should be, we disallow function types on either side of the $\lra$-arrow.%
\footnote{In \cref{sec:cayley} we will, actually, put endomorphisms ($a \ra a$) on both sides of a $\lra$-arrow; for justification, see there.}
(Note that higher-order functions \emph{within} the program are fine; the full program should just have first-order input and output types.)
Operationally, however, linear functions are just regular functions: the operational meaning of all code in this paper remains identical if all $\lra$-arrows are replaced with $\ra$ (and partial derivative operations are allowed in regular terms).

On the term level, we add an introduction form for linear functions; because we disallowed linear function types from or to function spaces, neither $\tau$ nor the type of $b$ can contain function types in $\lfun(z : \tau).\ b$.
The body of such linear functions is given by the restricted term language under $b$, which adds application of linear functions (identified by a variable reference), partial derivative operators, and zero and plus operations, but removes variable binding and lambda abstraction.

Note that zero and plus will always be of a type that is (part of) the domain or codomain of a linear function, which therefore has the required commutative monoid structure.
The fact that these two operations are not constant-time will be addressed when we improve the complexity of our algorithm later.

Regarding the derivatives of primitive operations: in a linear function, we need to compute the linear (reverse) derivatives of the primitive operations.
For every $\textit{op} \in \Op_n$, we require chosen programs $\Gamma\vdash \partial_i\textit{op}(t_1,\ldots, t_n):\lR\lra \lR$, given $\Gamma\vdash t_i:\R$, for $1\leq i\leq n$.
We require that these implement the partial derivatives of $\textit{op}$ in the sense that they have semantics $\partial_i\textit{op}(x)(d) = d \cdot \frac{\partial (\textit{op}(x))}{\partial x_i}$.

\subsection{The Code Transformation}\label{ssec:naive-code-transformation}

The naive dual-numbers reverse AD algorithm acts homomorphically over all program constructs in the input program, except for those constructs that non-trivially manipulate real scalars.
The full program transformation is given in \cref{fig:algo-naive}.
We use some syntactic sugar: $\textbf{let}\ (x_1, x_2) = s\ \textbf{in}\ t$ should be read as $\textbf{let}\ y = s\ \textbf{in}\ \textbf{let}\ x_1 = \fst(y)\ \textbf{in}\ \textbf{let}\ x_2 = \snd(y)\ \textbf{in}\ t$, where $y$ is fresh.
\begin{figure}
  \figheading{On types:
  }
  \begin{gather*}
    \trans1c[\R] = (\R, \lR\lra c) \qquad
    \trans1c[()] = () \qquad
    \trans1c[(\sigma,\tau)] = (\trans1c[\sigma], \trans1c[\tau]) \\[-1pt]
    \trans1c[\sigma \ra \tau] = \trans1c[\sigma] \ra \trans1c[\tau] \qquad
    \trans1c[\Z] = \Z 
  \end{gather*}
  \figheading{On environments:}
  \begin{gather*}
  \trans1c[\varepsilon] = \varepsilon \qquad
  \trans1c[\Gamma, x : \tau] = \trans1c[\Gamma], x : \trans1c[\tau] 
  \end{gather*}
  \figheading{On terms:}
  \begin{align*}
    &\text{If}\ \Gamma \vdash t : \tau\ \text{then}\ \trans1c[\Gamma] \vdash \trans1c[t] : \trans1c[\tau] \\
    &\textcolor{gray}{\trans1c[x : \tau] = x : \trans1c[\tau]} 
    &&\textcolor{gray}{\begin{array}[t]{@{}l@{}}
      \trans1c[\textbf{let}\ x : \tau = s\ \textbf{in}\ t] = \\
      \qquad \textbf{let}\ x : \trans1c[\tau] = \trans1c[s]\ \textbf{in}\ \trans1c[t]
    \end{array}} \\
    &
    \textcolor{gray}{\trans1c[()] = ()} 
    &&\textcolor{gray}{\trans1c[\fst(t)] = \fst(\trans1c[t])}\\
    &
    \textcolor{gray}{\trans1c[(s,t)] = (\trans1c[s], \trans1c[t])} 
    && \textcolor{gray}{\trans1c[\snd(t)] = \snd(\trans1c[t])}\\
    &\textcolor{gray}{\trans1c[\fun(x : \tau).\ t] = \fun(x : \trans1c[\tau]).\ \trans1c[t]} 
    && \textcolor{gray}{\trans1c[s\ t] = \trans1c[s]\ \trans1c[t]} \\
    &\trans1c[r] = (r, \lfun(z : \lR).\ \underline0) \\
    &\mathrlap{\trans1c[\textit{op}(t_1,\ldots,t_n)] = \begin{array}[t]{@{}l@{}}
      \textbf{let}\ (x_1, d_1) = \trans1c[t_1]\ \textbf{in}\ \ldots\ \textbf{in}\ \textbf{let}\ (x_n, d_n) = \trans1c[t_n] \\
      \textbf{in}\ (\textit{op}(x_1, \ldots, x_n) \\
      \hphantom{\textbf{in}\ } , \lfun(z : \lR).\ d_1\ (\partial_1\textit{op}(x_1, \ldots, x_n)(z)) + \cdots + \\
      \hphantom{\textbf{in}\  , \lfun(z : \lR).\ } d_n\ (\partial_n\textit{op}(x_1, \ldots, x_n)(z)))
    \end{array}}
  \end{align*}
  \caption{\label{fig:algo-naive}
    The naive code transformation from the source (\cref{fig:source-language}) to the target (\cref{fig:target-language-1}) language.
    The cases where $\trans1c$ just maps homomorphically over the source language are set in \textcolor{gray}{gray}.
  }
\end{figure}

The transformation consists of a mapping $\trans1c[\tau]$ on types $\tau$ and a mapping $\trans1c[t]$ on terms $t$.\footnote{In this section we choose $c$ to be the domain type of the top-level program; later we will modify $c$ to support our optimisations.}
The mapping on types works homomorphically except on scalars, which it maps (in the style of dual-numbers AD) to a \emph{pair} of a scalar and a derivative of that scalar.
In constrast to forward AD, however, the derivative is not represented by another scalar (which in forward AD would contain the derivative of this scalar result with respect to a particular initial input value), but instead by a \emph{backpropagator}.
If a $\trans1c{}$-transformed program at some point computes a scalar--backpropagator pair $(x, d)$ from an top-level input $\mathit{input} : \sigma$, then given a $z : \lR$, $d(z) : \underline\sigma$ is equal to $z$ times the gradient of $x$ as a function of $\mathit{input}$.

Variable references, tuples, projections, function application, lambda abstraction and let-binding are mapped homomorphically, i.e., the code transformation simply recurses over the subterms of the current term.
However, note that for variable references, lambda abstractions and let-bindings, the types of the variables do change.

Scalar constants are transformed to a pair of that scalar constant and a backpropagator for that constant.
Because a constant clearly does not depend on the input at all, its gradient is zero, and hence the backpropagator is identically zero, thus $\lfun(z : \lR).\ \underline0$.

Finally, primitive scalar operations are the most important place where this code transformation does something non-trivial.
First, we compute the values and backpropagators of the (scalar) arguments to the operation, after which we can compute the original (scalar) result by applying the original operation to those argument values.
Now, writing $\alpha$ for the top-level program input, we have:
\[
  z \cdot \tfrac{\partial (\mathit{op}(x_1, \ldots, x_n))}{\partial \alpha}
  = z \cdot \sum_{i=1}^n \tfrac{\partial (\mathit{op}(x_1, \ldots, x_n))}{\partial x_i} \cdot \tfrac{\partial x_i}{\partial \alpha}
  = \sum_{i=1}^n \tfrac{\partial x_i}{\partial \alpha} \cdot \bigl(z \cdot \tfrac{\partial (\mathit{op}(x_1, \ldots, x_n))}{\partial x_i}\bigr)
\]
and because $d_i\ z = \frac{\partial x_i}{\partial \alpha}$ and $\partial_i\textit{op}(x_1,\ldots,x_n) = \frac{\partial(\textit{op}(x_1,\ldots,x_n))}{\partial x_i}$, the appropriate backpropagator to return is indeed
\(
  \lfun(z : \lR).\ \sum_{i=1}^n d_i\ (\partial_i\textit{op}(x_1,\ldots,x_n))
\)
as is written in \cref{fig:algo-naive}.
This sum is on values of type $c$, which is currently still the type of the top-level program input.

\myparagraph{Wrapper of the AD transformation}
We want the external API of the AD transformation to be like $\mathcal R_2$ from \cref{sec:rev-ad-type}:
\[
  \mathcal R_2[f] : \sigma \ra (\tau, \underline{\tau} \ra \underline{\sigma})
\]
given $f : \sigma \ra \tau$.
However, our compositional code transformation actually follows $R_{3\text{dual}}$:
\[
  \mathcal R_{3\text{dual}}[t] : \forall c.\ \trans1c[\sigma] \ra \trans1c[\tau]
\]
hence we need to convert from $\mathcal R_{3\text{dual}}$ form to the intermediate $\mathcal R_3$:
\[
  \mathcal R_3[t] : \forall c.\ (\sigma, \underline\sigma \lra c) \ra (\tau, \underline\tau \lra c)
\]
and from there to $\mathcal R_2$.
The conversion from $(\sigma, \sigma \lra c)$ to $\trans1c[\sigma]$, for first-order $\sigma$, consists of \emph{interleaving} the backpropagator into the data structure of type $\sigma$; the converse (for $\tau$) is a similar deinterleaving process.
These two conversions (back and forth) are implemented by $\Interleave1$ and $\Deinterleave1$ in \cref{fig:wrapper-naive}.
The final conversion from $\mathcal R_3$ to $\mathcal R_2$ is easy in the simply-typed world (as described in \cref{sec:rev-ad-type}); this conversion is implemented in the top-level wrapper, $\Wrap1$, also in \cref{fig:wrapper-naive}.

\begin{figure}
  \begin{align*}
    &\begin{array}{@{}l@{\ }c@{\ }l}
      \Interleave1_\tau &:& \forall c.\ (\tau, \underline \tau \lra c) \ra \trans1c[\tau] \\
      \Interleave1_{\R} &=& \fun(x, d).\ (x, d) \\
      \Interleave1_{()} &=& \fun((), d).\ () \\
      \Interleave1_{(\sigma, \tau)} &=& \fun((x, y), d).\ \begin{array}[t]{@{}l@{}}
        (\Interleave1_\sigma\ (x, \lfun(z : \sigma).\ d\ (z, \underline0)) \\
        \leadcomma \Interleave1_\tau\ (y, \lfun(z : \tau).\ d\ (\underline0, z)))
      \end{array} \\
      \Interleave1_{\Z} &=& \fun(n, d).\ n \\
      \Interleave1_{\sigma \ra \tau} &=& \text{not defined!}
    \end{array} \\[0.2cm]
    &\begin{array}{@{}l@{\ }c@{\ }l}
      \Deinterleave1_\tau &:& \forall c.\ \trans1c[\tau] \ra (\tau, \underline \tau \lra c) \\
      \Deinterleave1_{\R} &=& \fun(x, d).\ (x, d) \\
      \Deinterleave1_{()} &=& \fun().\ ((), \lfun(z : ()).\ \underline0) \\
      \Deinterleave1_{(\sigma, \tau)} &=& \fun(x, y).\ \begin{array}[t]{@{}l@{}}
        \textbf{let}\ (x_1, x_2) = \Deinterleave1_\sigma\ x \\
        \textbf{in}\ \textbf{let}\ (y_1, y_2) = \Deinterleave1_\tau\ y \\
        \textbf{in}\ ((x_1, y_1), \lfun(z : (\sigma, \tau)).\ x_2\ (\fst(z)) + y_2\ (\snd(z)))
      \end{array} \\
      \Deinterleave1_{\Z} &=& \fun n.\ (n, \lfun(z : \Z).\ \underline0) \\
      \Deinterleave1_{\sigma \ra \tau} &=& \text{not defined!}
    \end{array} \\[0.2cm]
    &\begin{array}{@{}l@{\ }l@{}}
      \Wrap1 : (\sigma \ra \tau) \rightsquigarrow (\sigma \ra (\tau,& \underline \tau \lra \underline \sigma)) \\
      \Wrap1[\fun(x : \sigma).\ t] = \fun(x : \sigma).\ &\textbf{let}\ x : \trans1\sigma[\sigma] = \Interleave1_\sigma\ (x, \lfun(z : \underline\sigma).\ z)\ \\
      &\textbf{in}\ \Deinterleave1_\tau\ (\trans1\sigma[t])
    \end{array}
  \end{align*}
  \caption{\label{fig:wrapper-naive} 
    Wrapper around $\trans1c$ of \cref{fig:algo-naive}.
  }
\end{figure}

\subsection{Running Example}\label{sec:running-example-naive}

Let us look at the simple example from \cref{subfig:key-example-orig} in \cref{sec:key-ideas}:
\begin{equation}
  \lambda(x : \R, y : \R).\ \tightunderbrace{\mathbf{let}\ z = x + y\ \mathbf{in}\ x \cdot z}{t}
  \label{eq:running-example-naive-1}
\end{equation}
We have $x : \R, y : \R \vdash t : \R$.
The code transformation $\trans1c$ from \cref{fig:algo-naive} maps $t$ to:
\[
  \trans1c[t]
  = \begin{array}[t]{@{}l@{}}
    \mathbf{let}\ z = \begin{array}[t]{@{}l@{}}
      \mathbf{let}\ (x_1, d_1) = x\ \mathbf{in}\ \mathbf{let}\ (x_2, d_2) = y \\
      \mathbf{in}\ (x_1 + x_2, \lfun(z' : \lR).\ d_1\ z' + d_2\ z')
    \end{array} \\
    \mathbf{in}\ \begin{array}[t]{@{}l@{}}
      \mathbf{let}\ (x_1, d_1) = x\ \mathbf{in}\ \mathbf{let}\ (x_2, d_2) = z \\
      \mathbf{in}\ (x_1 \cdot x_2, \lfun(z' : \lR).\ d_1\ (z \cdot z') + d_2\ (x \cdot z'))
    \end{array}
  \end{array}
\]
which satisfies $x : (\R, \lR \lra c), y : (\R, \lR \lra c) \vdash \trans1c[t] : (\R, \lR \lra c)$.
(We $\alpha$-renamed $z$ from \cref{fig:algo-naive} to $z'$ here.)
The wrapper $\Wrap1$ in \cref{fig:wrapper-naive} computes, given $x : (\R, \R)$:
\[
  \Interleave1_{(\R,\R)}\ (x, \lfun(z : (\lR,\lR)).\ z)
  = ((\fst(x), \lfun(z : \lR).\ (z, \underline0)), (\snd(x), \lfun(z : \lR).\ (\underline0, z)))
\]
The $x$ and $y$ in \cref{eq:running-example-naive-1} get bound to the first half and the second half of this pair, respectively.
$\Deinterleave1_\tau$ is the identity in this case, because $\tau = \R$.

In \cref{sec:running-example-staged,sec:running-example-cayley}, we will revisit this example to show how the outputs change.

\subsection{Complexity of the Naive Transformation}\label{sec:naive-complexity}

Reverse AD transformations like the one described in this section are well-known to be correct \citep[e.g.][]{ad-2020-dualnum-revad-linear-factoring,ad-2021-dual-revad-linear-factoring-pcf,ad-2020-sam-mathieu-matthijs,nunes-2024-dual-numbers}.
However, as given here, it does not at all have the right time complexity.

The forward pass is fine: calling the function $\Wrap1[\lambda(x : \sigma).\ t : \tau] : \sigma \ra (\tau, \tau \lra \sigma)$ at some input $x : \sigma$ takes time proportional to the original program $t$.
However, the problem arises when we call the top-level backpropagator returned by the wrapper.
When we do so, we start a tree of calls to the linear backpropagators of all scalars in the program, where the backpropagator corresponding to a particular scalar value will be invoked once for each usage of that scalar as an argument to a primitive operation.
This means that any sharing of scalars in the original program results in multiple calls to the same backpropagator in the derivative program.
\cref{fig:exponential-complexity-example} in \cref{sec:key-ideas} displays an example program $t$ with its naive  derivative $\trans1c[t]$, in which sharing of scalars thus results in exponential time complexity.

This overhead is unacceptable: we can do much better. For first-order programs, we understand well how to write a code transformation such that the output program computes the gradient in only a constant factor overhead over the original program~\citep{adbook-2008-griewank-walther}.
This is less immediately clear for higher-order programs, as we consider here, but it is nevertheless possible.

In~\citep{ad-2020-dualnum-revad-linear-factoring}, this problem of exponential complexity is addressed from a theoretical point of view by observing that calling a linear backpropagator multiple times is a waste of work: indeed, linearity of a backpropagator $f$ means that $f\ x + f\ y = f\ (x + y)$.
Hopefully, applying this \emph{linear factoring rule} from left to right (thereby taking together two calls into one) allows us to ensure that every backpropagator is executed at most once.

And indeed, should we achieve this, the complexity issue described above (the exponential blowup) is fixed: every created backpropagator corresponds to some computation in the original program (either a primitive operation, a scalar constant or an input value), so with maximal application of linear factoring, the number of backpropagator executions would become proportional to the runtime of the original program.
If we can further make the body of a single backpropagator (not counting its callees) constant-time,\footnote{Obstacles to this are e.g.\ $\underline0$ and $(+)$ on the type $c$; we will fix this in \cref{sec:cayley,sec:just-scalars,sec:mutarrays}.} the differentiated program will compute the gradient with only a constant-factor overhead over the original program---as it should be for reverse AD.

However, this argument crucially depends on us being able to ensure that every backpropagator gets invoked at most once.
The solution of \cite{ad-2020-dualnum-revad-linear-factoring} is to define a custom operational semantics that symbolically evaluates the output program of the transformation to a straight-line program with the input backpropagators still as symbolic variables, and afterwards symbolically reduces the obtained straight-line program in a very specific way, making use of the linear factoring rule ($f\ x + f\ y = f\ (x + y)$) in judicious places.

In this paper, we present an alternative way to achieve linear factoring in a standard, call-by-value semantics for the target language.
In doing so, we attain the correct computational complexity without any need for symbolic execution.
We achieve this by changing the type $c$ that the input backpropagators map to, to a more intelligent type than the space of cotangents of the input that we have considered so far.
Avoiding the need for a custom operational semantics allows the wrapper of our code transformation to be relatively small (though it will grow in subsequent sections), and the core of the differentiated program to run natively in the target language.

\section{Linear Factoring by Staging Function Calls}\label{sec:staging}

As observed above in \cref{sec:naive-complexity}, the most important complexity problem of the reverse AD algorithm is solved if we ensure that all backpropagators are invoked at most once, and for that we must use that linear functions $f$ satisfy $f\ x + f\ y = f\ (x + y)$.
We must find a way to ``merge'' all invocations of a single backpropagator with this linear factoring rule so that in the end only one invocation remains (or zero if it was never invoked in the first place).

\myparagraph{Evaluation order}
Ensuring this complete merging of linear function calls is really a question of choosing an order of evaluation for the tree of function calls created by the backpropagators.
Consider for example the (representative) situation where a program generates the following backpropagators:
\begin{center}
  \begin{tabular}{l@{\hspace{2cm}}l}
    \( \begin{array}{@{}c@{\ }c@{\ }l@{}}
      f_1 &=& \lfun(z : \lR).\ (0, (z, 0)) \\[0.2em]
      f_2 &=& \lfun(z : \lR).\ f_1\ (2 \cdot z) + f_1\ (3 \cdot z) \\[0.2em]
      f_3 &=& \lfun(z : \lR).\ f_2\ (4 \cdot z) + f_1\ (5 \cdot z) \\[0.2em]
      f_4 &=& \lfun(z : \lR).\ f_2\ z + f_3\ (2 \cdot z)
    \end{array} \)
    &
    \makecell[l]{
      \begin{tikzpicture}[scale=1.2]
        \node (4) at (0.2, -0.5) {$f_4$};
        \node (3) at (1, 0.5) {$f_3$};
        \node (2) at (1.7, -0.5) {$f_2$};
        \node (1) at (2.5, 0.5) {$f_1$};
        \draw[->] (4) -- (2);
        \draw[->] (4) -- (3);
        \draw[->] (3) -- (2);
        \draw[->] (3) -- (1);
        \draw[->] (2) to[bend left=15] (1);
        \draw[->] (2) to[bend right=15] (1);
      \end{tikzpicture}
    }
  \end{tabular}
\end{center}
Suppose $f_4$ is the (only) backpropagator contained in the result.
Normal call-by-value evaluation of $f_4$ would yield two invocations of $f_2$ and five invocations of $f_1$, following the call graph on the right.

However, taking inspiration from symbolic evaluation and moving away from standard call-by-value for a moment, we could also first invoke $f_3$ to expand the body of $f_4$ to $f_2\ z + f_2\ (4 \cdot (2 \cdot z)) + f_1\ (5 \cdot (2 \cdot z))$.
Now we can take the two invocations of $f_2$ together using linear factoring to produce $f_2\ (z + 4 \cdot (2 \cdot z)) + f_1\ (5 \cdot (2 \cdot z))$; then invoking $f_2$ first, producing two more calls to $f_1$, we are left with three calls to $f_1$ which we can take together to a single call using linear factoring, which we can then evaluate.
With this alternate evaluation order, we have indeed ensured that every linear function is invoked at most (in this case, exactly) once.

If we want to obtain something like this evaluation order, the first thing that we must achieve is to \emph{postpone} invocation of linear functions until we conclude that we have merged all calls to that function and that its time for evaluation has arrived.
To achieve this goal, we would like to change the representation of $c$ to a dictionary mapping linear functions to the argument at which we intend to later call them.\footnote{This is the intuition; it will not go through precisely as planned, but something similar will.}
Note that this uniform representation in a dictionary works because all backpropagators in the core transformed program (outside of the wrapper) have the same domain ($\lR$) and codomain ($c$).
The idea is that we replace what are now applications of linear functions with the creation of a dictionary containing one key-value (function-argument) pair, and to replace addition of values in $c$ with taking the union of dictionaries, where arguments for common keys are added together.

\myparagraph{Initial Staged object}
More concretely, we want to change $\trans1c[\R] = (\R, \lR\lra c)$ to instead read $\trans1c[\R] = (\R, \lR\lra \Staged c)$, where `$\Staged c$' is our ``dictionary''.\footnote{In the wrapper, we still instantiate $c$ to the domain type $\sigma$, meaning that $\Staged \sigma$ will actually appear in the derivative program.}
We define, or rather, would like to define $\Staged c$ as follows: (`$\Map \ k \ v$' is the usual type of persistent tree-maps with keys of type $k$ and values of type $v$)
\begin{align*}
  \Staged c = (c, \Map\ (\lR\lra \Staged c)\ \lR)
\end{align*}
Suspending disbelief about implementability, this type can represent both literal $c$ values (necessary for the one-hot vectors returned by the input backpropagators created in $\Interleave1$) and staged (delayed) calls to linear functions.
We use $\Map$ to denote a standard (persistent) tree-map as found in every functional language.
The intuitive semantics of a value $(x, \{f_1 \mapsto a_1, f_2 \mapsto a_2\}) $ of type $ \Staged c$ is its \emph{resolution}  $x + f_1\ a_1 + f_2\ a_2:c$.

To be able to replace $c$ with $\Staged c$ in $\trans1c$, we must support all operations that we perform on $c$ also on $\Staged c$.
We implement them as follows:
\begin{itemize}
\item $\underline0 : c$ becomes simply $\ZeroStaged \coloneqq (\underline0, \{\}) : \Staged c$.
\item $(+) : c \ra c \ra c$ becomes $(\PlusStaged)$, adding $c$ values using $(+)$ and taking the union of the two $\Map$s.
  \textbf{Here we apply linear factoring}: if the two $\Map$s both have a value for the same key (i.e.\ we have two staged invocations of the same linear function $f$), the resulting map will have \emph{one} value for that same key $f$: the sum of the arguments stored in the two separate $\Map$s.
  For example:
  \begin{center} 
  \( \begin{array}{l}
    (c_1, \{f_1 \mapsto a_1, f_2 \mapsto a_2\}) \PlusStaged (c_2, \{f_2 \mapsto a_3\}) \\
    \qquad = (c_1 + c_2, \{f_1 \mapsto a_1, f_2 \mapsto a_2 + a_3\})
    \end{array} \)
  \end{center}
\item The one-hot $c$ values created in the backpropagators from $\Interleave1$ are stored in the $c$ component of $\Staged c$.
\item An application $f\ x$ of a backpropagator $f : \lR\lra c$ to an argument $x : \lR$ now gets replaced with $\StagedCall\ f\ x \coloneqq (\underline0, \{f \mapsto x\}) : \Staged c$.
  This occurs in $\trans1c[\textit{op}(...)]$ and in $\Deinterleave1$.
\end{itemize}
Essentially, this step of replacing $c$ with $\Staged c$ can be seen as a clever partial defunctionalisation of our backpropagators.

What is missing from this list is how to ``resolve'' the final $\Staged c$ value produced by the derivative computation down to a plain $c$ value---we need this at the end of the wrapper.
This resolve algorithm:
\[
\ResolveStaged :  (\Staged c) \to c
\]
will need to call functions stored in the $\Staged c$ object in the correct order, ensuring that we only invoke a backpropagator when we are sure that we have collected all calls to it in the $\Map$.
For example, in the example at the beginning of this section, $f_4\ 1$ returns $(\underline0, \{f_2 \mapsto 1, f_3 \mapsto 2\})$.
At this point, ``resolving $f_3$'' means calling $f_3$ at 2, observing the return value $(\underline0, \{f_2 \mapsto 8, f_1 \mapsto 10\})$, and adding it to the remainder (i.e.\ without the $f_3$ entry) of the previous $\Staged c$ object to get $(\underline0, \{f_2 \mapsto 9, f_1 \mapsto 10\})$.

But as we observed above, the choice of which function to invoke first is vital to the complexity of the reverse AD algorithm: if we chose $f_2$ first instead of $f_3$, the later call to $f_3$ would produce another call to $f_2$, forcing us to evaluate $f_2$ twice---something that we must avoid.
There is currently no information in a $\Staged c$ object from which we can deduce the correct order of invocation, so we need something extra.

There is another problem with the current definition of $\Staged c$: it contains a $\Map$ keyed by functions, meaning that we need equality---actually, even an ordering---on functions!
This is nonsense in general.
Fortunately, both problems can be tackled with the same solution.

\myparagraph{Resolve order}
The backpropagators that occur in the derivative program (as produced by $\trans1c$ from \cref{fig:algo-naive}) are not just arbitrary functions.
Indeed, taking the target type $c$ of the input backpropagators to be equal to the input type $\sigma$ of the original program (of type $\sigma \ra \tau$), as we do in $\Wrap1$ in \cref{fig:wrapper-naive}, all backpropagators in the derivative program have one of the following three forms:
\begin{enumerate}
\item
  $(\lfun(z : \lR).\ t)$ where $t$ is a tuple (of type $\sigma$) filled with zero scalars except for one position, where it places $z$; we call such tuples \emph{one-hot tuples}.
  These backpropagators result, after trivial beta-reduction of the intermediate linear functions, from the way that $\Interleave1_\sigma$ (\cref{fig:wrapper-naive}) handles references to the global inputs of the program.
\item
  $(\lfun(z :  \lR).\ \underline0)$ occurs as the backpropagator of a scalar constant $r$.
  Note that since this $\underline0$ is of type $\sigma$, operationally it is equivalent to a tuple filled completely with zero scalars.
\item
  $(\lfun(z :  \lR).\ d_1\ (\partial_1\textit{op}(x_1,\ldots,x_n)(z)) + \cdots + d_n\ (\partial_n\textit{op}(x_1,\ldots,x_n)(z)))$ for an $\textit{op} \in \Op_n$ where $d_1,\ldots,d_n$ are other linear backpropagators: these occur as the back\-propagators generated for primitive operations.
\end{enumerate}

\begin{insight}
Hence, we observe that \emph{a backpropagator $f_r$ paired with a scalar $r$ will only ever call backpropagators $f_s$ that are paired with scalars $s$, such that $r$ already has a dependency on $s$ in the source program}.
In particular, $f_s$ must have been created (at runtime of the derivative program) before $f_r$ itself was created.
Furthermore, $f_r$ is not the same function as $f_s$ because that would mean that $r$ depends on itself in the source program.
Therefore, if, at runtime, we define a partial order on backpropagators with the property that 
$f_r \geq f_s$ if $r$ depends on $s$ (and $f_r > f_s$ if they are not syntactically equal), we obtain that a called backpropagator is always strictly \emph{lower} in the order than the backpropagator it was called from.
\end{insight}
In practice, we achieve this by giving unique IDs, of some form, to backpropagators and defining a partial order on those IDs at runtime, effectively building a computation graph.
This partial order tells us in which order to resolve backpropagators:
we walk the order from top to bottom, starting from the maximal IDs and repeatedly resolving the predecessors in the order after we finish resolving a particular backpropagator.
After all, any calls to other backpropagators that it produces in the returned $\Staged c$ value will have lower IDs, and so cannot be functions that we have already resolved (i.e.\ called) before.
And as promised, giving backpropagators IDs also solves the issue of using functions as keys in a $\Map$: we can use the ID as the $\Map$ key, which is perfectly valid and efficient as long as the IDs are chosen to be of some type that can be linearly ordered to perform binary search (such as tuples of integers).

We have still been rather vague about how precisely to assign the IDs and define their partial order.
In fact, there is some freedom in how to do that.
For the time being, we will simply work with \emph{sequentially incrementing integer IDs with their linear order}, which suffices for sequential programs.
Concretely, we number backpropagators with incrementing integer IDs at runtime, at the time of their creation by a $\lfun{}$.
We then resolve them from top to bottom, starting from the unique maximal ID. 
To support parallelism in \cref{sec:parallelism}, we will revisit this choice and work instead with \emph{pairs} of integers (a combination of a job ID and a sequentially increasing ID within that job) with a partial order that encodes the fork-join parallelism structure of the source program. 
That choice of non-linear partial order allows us to reflect the parallelism present in the source program in a parallel reverse pass to compute derivatives.
But because we can mostly separate the concerns of ID representation and differentiation, we will focus on simple, sequential integer IDs for now.

When we give backpropagators integer IDs, we can rewrite $\Staged c$ and $\StagedCall$:
\begin{align*}
  &\Staged c = (c, \Map\ \Z\ (\lR\lra \Staged c, \lR)) \\
  &\begin{array}{@{\,}l@{\ }c@{\ }l@{}}
    \StagedCall &:& (\Z, \lR\lra \Staged c) \ra \lR\lra \Staged c \\
    \StagedCall\ (i, f)\ x &=& (\underline0, \{i \mapsto (f, x)\})
  \end{array}
\end{align*}
We call the second component of a $\Staged c$ value, which has type $\Map\ \Z\ (\lR\lra \Staged c, \lR))$, the \emph{staging map}, after its function to stage (linear) function calls.

The only thing that remains is to actually generate the IDs for the backpropagators at runtime.
This we do using an ID generation monad (a state monad with a state of type $\Z$ to keep track of our integer IDs).
The resulting new program transformation, modified from \cref{fig:algo-naive,fig:wrapper-naive}, is shown in \cref{fig:algo-monadic}.

\newcommand\monstandout[1]{#1}
\begin{figure}
  \figheading{On types:}
  \[
    \begin{array}{l@{\qquad}l}
    \trans2c[\R] = (\R, \monstandout{(\Z, \lR\lra \Staged c)})
      & \textcolor{gray}{\trans2c[\Z] = \Z} \qquad \textcolor{gray}{\trans2c[()] = ()} \\
    \trans2c[\sigma \ra \tau] = \trans2c[\sigma] \ra \monstandout{\Z \ra (\trans2c[\tau], \Z)}
      & \textcolor{gray}{\trans2c[(\sigma,\tau)] = (\trans2c[\sigma], \trans2c[\tau])} \\
    \end{array}
  \]
  \figheading{On terms:}
  \begin{align*}
    &\text{If}\ \Gamma \vdash t : \tau\ \text{then}\ \trans2c[\Gamma] \vdash \trans2c[t] : \monstandout{\Z \ra (\trans2c[\tau], \Z)} \\
    &\textcolor{gray}{\trans2c[x : \tau] = \fun i.\ (x : \trans2c[\tau], i)} \\
    &\textcolor{gray}{\trans2c[(s,t)] = \fun i.\ \textbf{let}\ (x, i') = \trans2c[s]\ i\ \textbf{in}\ \textbf{let}\ (y, i'') = \trans2c[t]\ i'\ \textbf{in}\ ((x, y), i'')} \\
    &\textcolor{gray}{\trans2c[\textbf{let}\ x : \tau = s\ \textbf{in}\ t] = \fun i.\ \textbf{let}\ (x : \trans2c[\tau], i') = \trans2c[s]\ i\ \textbf{in}\ \trans2c[t]\ i'} \\
    &\textcolor{gray}{\textit{etc.}} \\
    &\trans2c[r] = \fun i.\ ((r, (\monstandout{i}, \lfun(z : \lR).\ \ZeroStaged)), \monstandout{i + 1}) \\
    &\begin{array}[t]{@{}l@{}}
      \trans2c[\textit{op}(t_1,\ldots,t_n)] = \\
      \qquad \begin{array}[t]{@{}l@{}}
        \fun i.\ \textbf{let}\ ((x_1, d_1), i_1) = \trans2c[t_1]\ i\ \textbf{in}\ \ldots\ \textbf{in}\ \textbf{let}\ ((x_n, d_n), i_n) = \trans2c[t_n]\ i_{n-1} \\
        \hphantom{\fun i.\ } \textbf{in}\ \begin{array}[t]{@{}l@{}}
          ((\textit{op}(x_1, \ldots, x_n), (\monstandout{i_n}, \lfun(z : \lR).\ \begin{array}[t]{@{}l@{}}
            \StagedCall\ d_1\ (\partial_1\textit{op}(x_1, \ldots, x_n)(z)) \PlusStaged \cdots \\
            \quad {} \PlusStaged \StagedCall\ d_n\ (\partial_n\textit{op}(x_1, \ldots, x_n)(z))))
          \end{array} \\
          \leadcommai \monstandout{i_n + 1})
        \end{array}
      \end{array}
    \end{array}
  \end{align*}
  \figheading{Changed wrapper:}
  \begin{align*}
    &\begin{array}{@{}l@{\ }c@{\ }l@{}}
      \Interleave2_\tau &:& \forall c.\ (\tau, \tau \lra \Staged c) \ra \monstandout{\Z \ra (\trans2c[\tau], \Z)} \\
      \Interleave2_\R &=& \fun(x, d).\ \fun i.\ ((x, (\monstandout{i}, d)), \monstandout{i + 1}) \\
      \textcolor{gray}{\Interleave2_{()}} &\textcolor{gray}{=}& \textcolor{gray}{\fun((), d).\ \fun i.\ ((), i)} \\
      \textcolor{gray}{\Interleave2_{(\sigma, \tau)}} &\textcolor{gray}{=}& \textcolor{gray}{\fun((x, y), d).\ \fun i.\ \begin{array}[t]{@{}l@{}}
        \textbf{let}\ (x', i') = \Interleave2_\sigma\ (x, \lfun(z : \sigma).\ d\ (z, \underline0))\ i \\
        \textbf{in}\ \textbf{let}\ (y', i'') = \Interleave2_\tau\ (y, \lfun(z : \tau).\ d\ (\underline0, z))\ i' \\
        \textbf{in}\ ((x', y'), i'')
      \end{array}} \\
      \textcolor{gray}{\Interleave2_\Z} &\textcolor{gray}{=}& \textcolor{gray}{\fun(n, d).\ \fun i.\ (n, i)}
    \end{array} \\[0.2em]
    &\begin{array}{@{}l@{}}
      \text{$\Deinterleave2_\tau$ gets type $\forall c.\ \trans2c[\tau] \ra (\tau, \tau \lra \Staged c)$ and ignores the new $\Z$ in $\trans2c[\R]$.} \\
      \text{$\underline0$ changes to $\ZeroStaged$ and $(+)$ changes to $(\PlusStaged)$.}
    \end{array} \\[0.2em]
    &\begin{array}{@{}l@{}}
      \Wrap2 : (\sigma \ra \tau) \rightsquigarrow (\sigma \ra (\tau, \tau \lra \sigma)) \\
      \Wrap2[\fun(x : \sigma).\ t] = \fun(x : \sigma).\ \begin{array}[t]{@{}l@{}}
        \textbf{let}\ (x : \trans2\sigma[\sigma], i) = \Interleave2_\sigma\ (x, \InitStaged)\ 0 \\
        \textbf{in}\ \textbf{let}\ (y, d) = \Deinterleave2_\tau\ (\fst (\trans2\sigma[t]\ i)) \\
        \textbf{in}\ (y, \lfun(z : \tau).\ \ResolveStaged\ (d\ z)) \vspace{-0.1em} \\
        \hspace{3cm}\text{\small \textit{\small --- see main text for \ResolveStagedWith{\textit}}}
        \vspace{-0.4em}
      \end{array}
    \end{array}
  \end{align*}
  \figheading{Staged interface:}
  \begin{align*}
    &\Staged c = (c, \Map\ \Z\ (\lR\lra \Staged c, \lR)) \\
    &\begin{array}{@{\,}l@{\ }c@{\ }l@{\quad}l@{\ }c@{\ }l@{\,}}
      \ZeroStaged &:& \Staged c
        & (\PlusStaged) &:& \Staged c \ra \Staged c \ra \Staged c \\
      \ZeroStaged &=& (\underline0, \{\})
        & (c, m) \PlusStaged (c', m') &=& (c + c', m \cup m') \quad \text{\small --- \textit{linear factoring}} \\[0.4em]
      \InitStaged &:& c \lra \Staged c
        & \StagedCall &:& (\Z, \lR\lra \Staged c) \ra \lR\lra \Staged c \\
      \InitStaged\ c &=& (c, \{\})
        & \StagedCall\ (i, f)\ x &=& (\underline0, \{i \mapsto (f, x)\})
    \end{array}
  \end{align*}
  \caption{\label{fig:algo-monadic}
    The monadically transformed code transformation (from \cref{fig:source-language} to \cref{fig:target-language-1} plus $\Staged$ operations), based on \cref{fig:algo-naive}.
    \textcolor{gray}{Grey} parts are unchanged or simply monadically lifted.
  }
\end{figure}

\myparagraph{New program transformation}
In \cref{fig:algo-monadic}, the term transformation now produces a term in the ID generation monad ($\Z \ra (-, \Z)$); therefore, all functions in the original program will also need to run in the same monad.
This gives the second change in the type transformation (aside from $\trans2c[\R]$, which now tags backpropagators with an ID): $\trans2c[\sigma \ra \tau]$ now produces a monadic function type instead of a plain function type.

On the term level, notice that the backpropagator for primitive operations (in $\trans2c[\textit{op}(...)]$) now no longer calls $d_1,\ldots,d_n$ (the backpropagators of the arguments to the operation) directly, but instead registers the calls as pairs of function and argument in the $\Staged c$ returned by the backpropagator.
The $\cup$ in the definition of $(\PlusStaged)$ refers to map union including linear factoring; for example:
\[ \{i_1 \mapsto (f_1, a_1), i_2 \mapsto (f_2, a_2)\} \cup \{i_2 \mapsto (f_2, a_3)\} = \{i_1 \mapsto (f_1, a_1), i_2 \mapsto (f_2, a_2 + a_3)\} \]
Note that the transformation assigns consistent IDs to backpropagators: it will never occur that two staging maps have an entry with the same key (ID) but with a different function in the value.
This invariant is quite essential in the algorithms in this paper.

In the wrapper, $\Interleave2$ is lifted into the monad and generates IDs for scalar backpropagators; $\Deinterleave2$ is essentially unchanged.
The initial backpropagator provided to $\Interleave2$ in $\Wrap2$, which was $(\lfun{(z:\underline\sigma)}.\ z) : \underline\sigma \lra \underline\sigma$ in \cref{fig:wrapper-naive}, has now become $\InitStaged : \underline\sigma \lra \Staged \underline\sigma$, which injects a cotangent into a $\Staged c$ object.
$\Interleave2$ ``splits'' this function up into individual $\lR\lra \Staged \underline\sigma$ backpropagators for each of the individual scalars in $\sigma$.

At the end of the wrapper, we apply the insight that we had earlier: by resolving (calling and eliminating) the backpropagators in the final $\Staged c$ returned by the differentiated program in order from the highest ID to the lowest ID, we ensure that every backpropagator is called at most once.\footnote{
  Technically, some backpropagators (namely, the ones that appear in the top-level function output), are invoked more than once because $\Deinterleave2$ indiscriminately calls all output backpropagators.
  If the function output contains $n$ scalars, this can lead to $O(n)$ overhead.
  (In particular, for a function $f : \tau \ra \R$, there is no such overhead.)
  The complexity of the algorithm is not in fact compromised, because the size of the output is at most the runtime of the original function.
  If desired, $\Deinterleave2$ can be modified to return a $\Staged c$ object that \emph{stages} calls to the output backpropagators, instead of directly calling them.
  We did not make this change for simplicity of presentation.
}
This is done in the following function:
\begin{align*}
  \begin{array}{l}
    \ResolveStaged : \Staged c \to c\\
    \ResolveStaged\ (\textit{grad} : c, m : \Map\ \Z\ (\lR\lra \Staged c, \lR)) \coloneqq \\
    \quad\textbf{if}\ m\ \text{is empty}\ \begin{array}[t]{@{}l@{}}
      \textbf{then}\ \textit{grad} \\
      \textbf{else}\ \begin{array}[t]{@{}l@{}}
        \textbf{let}\ i = \text{highest key in}\ m \\
        \textbf{in}\ \textbf{let}\ (f, a) = \text{lookup}\ i\ \text{in}\ m \\
        \textbf{in}\ \textbf{let}\ m' = \text{delete}\ i\ \text{from}\ m \\
        \textbf{in}\ \ResolveStaged\ (f\ a \PlusStaged (c', m'))
      \end{array}
    \end{array}
  \end{array}
  \hspace{0.5cm} 
\end{align*}
The three operations on $m$ are standard logarithmic-complexity tree-map operations.

\subsection{Running Example}\label{sec:running-example-staged}

In \cref{sec:running-example-naive}, we looked at the term $x : \R, y : \R \vdash t = \mathbf{let}\ z = x + y\ \mathbf{in}\ x \cdot z : \R$ from \cref{subfig:key-example-orig}.
With the updated transformation from \cref{fig:algo-monadic}, we now get (with lambda-application redexes already simplified for readability):
\[
  \trans2c[t]
  = \lambda i_1.\ 
    \begin{array}[t]{@{}l@{}}
      \mathbf{let}\ (z, i_4) = \begin{array}[t]{@{}l@{}}
        \mathbf{let}\ ((x_1, d_1), i_2) = (x, i_1)\ \mathbf{in}\ \mathbf{let}\ ((x_2, d_2), i_3) = (y, i_2) \\
        \mathbf{in}\ ((x_1 + x_2, (i_3, \lfun(z' : \lR).\ \StagedCall\ d_1\ z' \PlusStaged \StagedCall\ d_2\ z')), i_3 + 1)
      \end{array} \\
      \mathbf{in}\ \begin{array}[t]{@{}l@{}}
        \mathbf{let}\ ((x_1, d_1), i_5) = (x, i_4)\ \mathbf{in}\ \mathbf{let}\ ((x_2, d_2), i_6) = (z, i_5) \\
        \mathbf{in}\ ((x_1 \cdot x_2, (i_6, \lfun(z' : \lR).\ \StagedCall\ d_1\ (z \cdot z') \PlusStaged \StagedCall\ d_2\ (x \cdot z'))), i_6 + 1)
      \end{array}
    \end{array}
\]
The result of $\Interleave2_{(\R,\R)}\ (x, \InitStaged)\ 0$, as called from $\Wrap2$ in \cref{fig:algo-monadic}, is:
\[
  (((\fst(x), (0, d_{\text{in},0})), (\snd(x), (1, d_{\text{in},1}))), 2)
\]
where we abbreviated the input backpropagators as $d_{\text{in},0} = \lfun(z:\lR).\ ((z, \underline0), \{\})$ and $d_{\text{in},1} = \lfun(z:\lR).\ ((\underline0, z), \{\})$.
Now, assuming that the input $x$ is, say, $(12, 13)$ and that the initial cotangent is $1$, the $\Staged\ (\R, \R)$ object that gets passed to $\ResolveStaged$ in $\Wrap2$ (i.e.\ the result of $d\ z = d\ 1$ there) looks as follows:
\[
  (\underline0 + \underline0, \begin{array}[t]{@{}l@{}}
    \{0 \mapsto (d_{\text{in},0}, 25) \\
    \leadcomma 2 \mapsto (\lfun(z':\lR).\ (\underline0 + \underline0, \{0 \mapsto (d_{\text{in},0}, z'), 1 \mapsto (d_{\text{in},1}, z')\}), 12) \})
  \end{array}
\]
where $25 = (\fst(x) + \snd(x)) \cdot 1$ and $12 = \fst(x) \cdot 1$.
This result makes sense because the last expression in the term $t$ is `$x \cdot z$', so its backpropagator directly contributes cotangents to (1.) the input $x$ with a partial derivative of $25$, and (2.) the intermediate value $z = x + y$ with a partial derivative of $12$.
The sums $\underline0 + \underline0$, of course, directly reduce to $\underline0$ instead of staying unevaluated, but we left them as-is to show what computation is happening in $(\PlusStaged)$.

\subsection{Remaining Complexity Challenges}\label{sec:staged-complexity}

\begin{tabular}{@{}ll@{}}
  \makecell[>p{\textwidth - 1.47cm}]{
    We have gained a lot with the function call staging so far: where the naive algorithm from \cref{sec:naive} easily ran into exponential blowup of computation time if the results of primitive operations were used in multiple places, the updated algorithm from \cref{fig:algo-monadic} completely solves this issue.
    For example, the program of \cref{fig:exponential-complexity-example} now results in the call graph displayed on the right: each backpropagator is called exactly once.
    However, some other complexity problems still remain.\footnotemark{} \\
    \hspace{10pt}As discussed in \cref{sec:rev-ad-complexity}, for a reverse AD algorithm to have the right complexity, we want the produced derivative program $P'$ to compute the gradient\vspace{1.2pt}
  }
&
\makecell[>p{1.04cm}]{
  \scalebox{0.75}{ \framebox{\begin{tikzpicture}
      \node (4) at (0, 0.9) {$ dx_n$};
      \node (3) at (0, 1.8) {$ dx_{n-1}$};
      \node (d) at (0, 2.7) {$\mathrlap{\smash{\raisebox{2pt}{\vdots}}}\phantom{\vdots}$};
      \node (2) at (0, 3.6) {$ dx_1$};
      \node (1) at (0, 4.5) {$ dx_0$};
      \draw[->, line width=1pt] (4) to  (3);
      \draw[->, line width=1pt] (3) to (d);
      \draw[->, line width=1pt] (d) to (2);
      \draw[->, line width=1pt] (2) to  (1);
    \end{tikzpicture}}}
}
\end{tabular}
\noindent 
of the original program $P$ at a given input $x$ with runtime only a constant factor times the runtime of $P$ itself on $x$---and this constant factor should work for all programs $P$.
To account for programs that ignore their input, we add an additional term that $P'$ may also read the full input regardless of whether $P$ did so: the program $P = (\fun{(x:\tau)}.\ x) : \tau \ra \tau$ always takes constant time whereas its gradient program must at the very least construct the value of $P$'s full gradient, which has size $\size(x)$.
Hence, we require that:\footnotetext{This section does not provide a proof that $\Wrap2{}$ does \emph{not} have the correct complexity; rather, it argues that the expected complexity analysis does not go through. The same complexity analysis \emph{will} go through for $\Wrap4$ after the improvements of \cref{sec:cayley,sec:just-scalars,sec:mutarrays}.}
\begin{align*}
  \begin{array}{l}
    \exists c > 0.
    \ \forall P \in \textrm{Programs}(\sigma \ra \tau).
    \ \forall x : \sigma, \textit{dy} : \underline\tau.
     \\
    \qquad \cost(\snd (\Wrap{}[P]\ x)\ \textit{dy}) \leq c \cdot (\cost(P\ x) + \size(x))
  \end{array}
\end{align*}
where $\cost(E)$ is the time taken to evaluate $E$, and $\size(x)$ is the time taken to read all of $x$ sequentially.

So, what is $\cost(\snd(\Wrap{}[P]\ x)\ \textit{dy})$?
First, the primal pass ($\Wrap{}[P]\ x$) consists of interleaving, running the differentiated program, and deinterleaving.
\begin{itemize}
\item
  $\Interleave2$ itself runs in $O(\size(x))$.
  (The backpropagators it creates are more expensive, but those are not called just yet.)
\item
  For the differentiated program, $\trans2\sigma[P]$, we can see that in all cases of the transformation $\trans2c$, the right-hand side does the work that $P$ would have done, plus threading of the next ID to generate, as well as creation of backpropagators.
  Since this additional work is a constant amount per program construct, $\trans2\sigma[P]$ runs in $O(\cost(P\ x))$.
\item
  $\Deinterleave2$ runs in $O(\size(P\ x))$, i.e.\ the size of the program output; this is certainly in $O(\cost(P\ x) + \size(x))$ but likely much less.
\end{itemize}
Summarising, the primal pass as a whole runs in $O(\cost(P\ x) + \size(x))$, which is precisely as required.

Then, the dual pass ($f\ \textit{dy}$, where $f$ is the linear function returned by $\Wrap2$) first calls the backpropagator returned by $\Deinterleave2$ on the output cotangent, and then passes the result through $\ResolveStaged$ to produce the final gradient.
Let $t$ be the function body of $P$ (i.e.\ $P = \fun(x : \sigma).\ t$).
\begin{itemize}
\item
  Because the number of scalars in the output is potentially as large as $O(\cost(P\ x) + \size(x))$, the backpropagator returned by $\Deinterleave2$ is only allowed to perform a constant-time operation for each scalar.
  However, looking back at \cref{fig:wrapper-naive}, we see that this function calls all scalar backpropagators contained in the result of $\trans2\sigma[t]$ once, and adds the results using $(\PlusStaged)$.
  Assuming that the scalar backpropagators run in constant time (not yet---see below), we are left with the many uses of $(\PlusStaged)$; if these are constant-time, we are still within our complexity budget.
  However: \\
  \textbf{Problem}: $(\PlusStaged)$ (see \cref{fig:algo-monadic}) is not constant-time: it adds values of type $c$ and takes the union of staging maps, both of which may be expensive.
\item
  Afterwards, we use $\ResolveStaged$ on the resulting $\Staged \sigma$ to call every scalar backpropagator in the program (created in $\trans2\sigma[r]$, $\trans2\sigma[\textit{op}(...)]$ and $\Interleave2$) at most once; this is accomplished using three $\Map$ operations and one call to $(\PlusStaged)$ per backpropagator.
  However, each of the scalar backpropagators corresponds to either a constant-time operation\footnote{Assuming primitive operations all have bounded arity and are constant-time. A more precise analysis, omitted here, lifts these restrictions---as long as the gradient of a primitive operation can be computed in the same time as the original.} in the original program $P$ or to a scalar in the input $x$; therefore, in order to stay within the time budget of $O(\cost(P\ x) + \size(x))$, we are only allowed a constant-time overhead per backpropagator here.
  Since $(\PlusStaged)$ was covered already, we are left with: \\
  \textbf{Problem}: The $\Map$ operations in $\ResolveStaged$ are not constant-time.
\item
  While we have arranged to invoke each scalar backpropagator at most once, we still need those backpropagators to individually run in constant-time too: our time budget is $O(\cost(P\ x) + \size(x))$, but there could be $O(\cost(P\ x) + \size(x))$ distinct backpropagators.
  Recall from earlier that we have three kinds of scalar backpropagators:
  \begin{enumerate}
  \item\hspace{-6.4pt}%
    $(\lfun(z : \lR).\ \InitStaged\ (0, \ldots, 0, z, 0, \ldots, 0))$ created in $\Interleave2$ (with $\InitStaged$ from $\Wrap2$). \\
    \textbf{Problem}: The interleave backpropagators take time $O(\size(x))$, not $O(1)$.
  \item\hspace{-6.4pt}%
    $(\lfun(z : \lR).\ \ZeroStaged)$ created in $\trans2\sigma[r]$. \\
    \textbf{Problem}: $\ZeroStaged$ takes time $O(\size(x))$, not $O(1)$.
  \item\hspace{-6.4pt}%
    $(\lfun(z : \lR).\ \StagedCall\ d_1\ (\partial_1\textit{op}(...)(z)) \PlusStaged \cdots \PlusStaged \StagedCall\ d_n\ (\partial_n\textit{op}(...)(z)))$, as created in $\trans2\sigma[\textit{op}(...)]$.
    Assuming that primitive operation arity is bounded, we are allowed a constant-time operation for each argument to \textit{op}. \\
    \textbf{Problem}: $\StagedCall$ creates a $\underline0 : c$ and therefore runs in $O(\size(x))$, not $O(1)$.
    (The problem with $(\PlusStaged)$ was already covered above.)
  \end{enumerate}
\end{itemize}
Summarising again, we see that we have three categories of complexity problems to solve:
\begin{itemize}
\item[(A)]
  We are not allowed to perform monoid operations on $c$ as often as we do.
  (This affects $\ZeroStaged$, $(\PlusStaged)$ and $\StagedCall$.)
  Our fix for this (in \cref{sec:cayley}) will be to Cayley-transform the $\Staged c$ object, including the contained $c$ value, turning zero into the identity function `$\id$' and plus into function composition $(\circ)$ on the type $\Staged c \ra \Staged c$.
\item[(B)]
  The $\Interleave{}$ backpropagators that create a one-hot $c$ value should avoid touching parts of $c$ that they are zero on.
  After Cayley-transforming $\Staged c$ in \cref{sec:cayley}, this problem becomes less pronounced: the backpropagators now \emph{update} a $\Staged c$ value, where they can keep untouched subtrees of $c$ fully as-is.
  However, the one-hot backpropagators will still do work proportional to the \emph{depth} of the program input type $c$.
  We will turn this issue into a simple log-factor in the complexity in \cref{sec:just-scalars} by replacing the $c$ in $\Staged c$ with a more efficient structure (namely, $\Map\ \Z\ \lR$).
  This log-factor can optionally be further eliminated using mutable arrays as described in \cref{sec:mutarrays}.
\item[(C)]
  The $\Map$ operations in $\ResolveStaged$ are logarithmic in the size of the staging map.
  Like in the previous point, mutable arrays (\cref{sec:mutarrays}) can eliminate this final log-factor in the complexity.
\end{itemize}
From the analysis above, we can conclude that after we have solved each of these issues, the algorithm attains the correct complexity for reverse AD.

\section{Cayley-Transforming the Cotangent Collector}\label{sec:cayley}

For any monoid $(M, 0, +)$ we have a function $C_M:M\ra (M\ra M)$, given by $m\mapsto (m'\mapsto m+m')$.
In fact, due to the associativity and unitality laws for monoids, this function defines a monoid homomorphism from $(M, 0, +)$ to the endomorphism monoid $(M \ra M, \id, \circ)$ on $M$.
By observing that $C_M$ has a left-inverse ($C_M(m)(0)=m$), we see that it even defines an isomorphism of monoids to its image, a fact commonly referred to as Cayley's theorem.
This trick of realising a monoid $M$ as a submonoid of its endomorphism monoid $M\ra M$
is surprisingly useful in functional programming, as the operations of $(M \ra M, \id, \circ)$ 
may have more desirable operational/complexity characteristics than the operations of  $(M, 0, +)$.
The optimisation discussed in this section is to switch to this \emph{Cayley-transformed} representation for cotangents.

This trick is often known as the ``difference list'' trick in functional programming, due to its original application to improving the performance of repeated application of the list-append operation~\citep{fp-1986-difference-lists}.
The intent of moving from $[\tau]$ (i.e.\ lists of values of type $\tau$) to $[\tau] \ra [\tau]$ was to ensure that the list-append operations are consistently associated to the right.
In our case, however, the primary remaining complexity issues are not due to operator associativity, but instead because our monoid has very expensive $0$ and $+$ operations (namely, $\ZeroStaged$ and $(\PlusStaged)$).
If we Cayley-transform $\Staged c$, i.e.\ if we replace $\Staged c$ with $\Staged c \ra \Staged c$, all occurrences of $\ZeroStaged$ in the code transformation turn into $\id$ and all occurrences of $(\PlusStaged)$ turn into $(\circ)$.
Since $\id$ is a value (of type $\Staged c\ra \Staged c$) and the composition of two functions can be constructed in constant time, this makes the monoid operations on the codomain of backpropagators (which now becomes $\Staged c \ra \Staged c$) constant-time.

Of course, a priori this just moves the work to the primitive monoid values, which now have to update an existing value instead of directly returning a small value themselves.
Because $M \ra M$ is essentially a kind of sparse representation, however, this can sometimes be done more efficiently than with a separate addition operation.

After the Cayley transform, all non-trivial work with $\Staged c$ objects that we still perform is limited to: 1.\ the single $\ZeroStaged$ value that the full composition is in the end applied to (to undo the Cayley transform), and 2.\ the primitive $\Staged c$ values, that is to say: the implementation of $\StagedCall$, $\InitStaged$ and $\ResolveStaged$.
We do not have to worry about one single zero of type $c$, hence we focus only on $\StagedCall$, $\InitStaged$ and $\ResolveStaged$, which get the following updated types after the Cayley transform:\footnote{%
  Despite the fact that we forbade it in \cref{sec:source-target-languages}, we are putting function types on both sides of a $\lra$-arrow here.
  The monoid structure here is the one from the Cayley transform (i.e.\ with $\id$ and $(\circ)$).
  Notice that this monoid structure is indeed the one we want in this context: the ``sum'' (composition) of two values of type $(\Staged c \ra \Staged c)$ corresponds with the sum (with $(\PlusStaged)$) of the $\Staged c$ values that they represent.
  (This is the Cayley isomorphism described above.)
} (the changed parts are shown in \textcolor{red}{red})
\begin{align*}
  \begin{array}{l@{\ }c@{\ }l}
    \StagedCall &:& (\Z, \lR\lra \textcolor{red}{(\Staged c \ra \Staged c)}) \ra \lR\lra \textcolor{red}{(\Staged c \ra \Staged c)} \\
    \InitStaged &:& c \lra \textcolor{red}{(\Staged c \ra \Staged c)} \\
    \ResolveStaged &:& \textcolor{red}{(\Staged c \ra \Staged c)} \lra c
  \end{array}
\end{align*}
The definition of $\Staged c$ itself also gets changed accordingly:
\begin{align*}
  \Staged c = (c, \Map\ \Z\ (\lR\lra \textcolor{red}{(\Staged c \ra \Staged c)}, \lR))
\end{align*}
The new definition of $\StagedCall$ arises from simplifying the composition of the old $\StagedCall$ with $(\PlusStaged)$:
\begin{align*}
  \begin{array}{@{}l@{}l@{}l@{}}
    \StagedCall\ (i, f)\ x\ (c, m) = (c, \text{if}\ i \not\in m
    & \ \text{then} & \ \text{insert}\ i \mapsto (f, x)\ \text{into}\ m \\
    & \ \text{else} & \ \text{update}\ m\ \text{at}\ i\ \text{with}\ (\fun(\mathunderscore, x').\ (f, x + x')))
  \end{array}
\end{align*}
Note that $(\PlusStaged)$ has been eliminated, and we do not use $(+)$ on $c$ here anymore.
For $\InitStaged$ we have to modify the type further (Cayley-transforming its $c$ argument as well) to lose all $(+)$ operations on $c$:
\begin{align*}
  \begin{array}{l}
    \InitStaged : \textcolor{red}{(c \ra c)} \lra (\Staged c \ra \Staged c) \\
    \InitStaged\ f\ (c, m) = (f\ c, m)
  \end{array}
\end{align*}
Before calling $\ResolveStaged$, we simply apply the $(\Staged c \ra \Staged c)$ function to $\ZeroStaged$ (undoing the Cayley transform by using its left-inverse as discussed---this is now the only remaining $\ZeroStaged$); $\ResolveStaged$ is then as it was in \cref{sec:staging}, only changing $f\ a \PlusStaged (c, m')$ to $f\ a\ (c, m')$ on the last line: $f$ from the $\Map$ now has type $\lR\lra (\Staged c \ra \Staged c)$.

\subsection{Code Transformation}

\begin{figure}
  \figheading{On types:}
  \[
    \begin{array}{@{}l@{\qquad}l@{}}
    \trans3c[\R] = (\R, (\Z, \lR\lra (\Staged c \ra \Staged c))) &
    \textcolor{gray}{\trans3c[\Z] = \Z} \qquad \textcolor{gray}{\trans3c[()] = ()} \\
    \textcolor{gray}{\trans3c[\sigma \ra \tau] = \trans3c[\sigma] \ra \Z \ra (\trans3c[\tau], \Z)} &
    \textcolor{gray}{\trans3c[(\sigma,\tau)] = (\trans3c[\sigma], \trans3c[\tau])}
    \end{array}
  \]
  \figheading{On terms:}
  \begin{align*}
    &\textcolor{gray}{\text{If}\ \Gamma \vdash t : \tau\ \text{then}\ \trans3c[\Gamma] \vdash \trans3c[t] : \Z \ra (\trans3c[\tau], \Z)} \\
    &\text{Same as $\trans2c$, except with `$\id$' in place of $\ZeroStaged$ and `$\circ$' in place of $(\PlusStaged)$.}
  \end{align*}
  \figheading{Changed wrapper:}
  \begin{align*}
    &\begin{array}{@{}l@{\ }c@{\ }l@{}}
      \textcolor{gray}{\Interleave3_\tau} &\textcolor{gray}{:}& \textcolor{gray}{\forall c.\ (\tau,}\, (\tau \ra \tau) \lra (\Staged c \ra \Staged c)\textcolor{gray}{) \ra \Z \ra (\trans3c[\tau], \Z)} \\
      \textcolor{gray}{\Interleave3_{\R}} &\textcolor{gray}{=}& \textcolor{gray}{\fun(x, d).\ \fun i.}\ \begin{array}[t]{@{}l@{}}
        \textcolor{gray}{((x, (i,}\, \lfun(z : \lR).\ d\ (\fun(a : \R).\ z + a)\textcolor{gray}{))} \\
        \textcolor{gray}{\leadcommai i + 1)}
      \end{array}\\
      \textcolor{gray}{\Interleave3_{()}} &\textcolor{gray}{=}& \textcolor{gray}{\fun((), d).\ \fun i.\ ((), i)} \\
      \textcolor{gray}{\Interleave3_{(\sigma, \tau)}} &\textcolor{gray}{=}& \textcolor{gray}{\fun((x, y), d).\ \fun i.} \\
      &&\hspace{-1.2cm} \textcolor{gray}{\textbf{let}\ (x', i') = \Interleave3_\sigma\ (x,}\, \fun(f : \sigma \ra \sigma).\ \textcolor{gray}{d\ (}\fun((v, w) : (\sigma, \tau)).\ (f\ v, w)\textcolor{gray}{))\ i} \\
      &&\hspace{-1.2cm} \textcolor{gray}{\textbf{in}\ \textbf{let}\ (y', i'') = \Interleave3_\tau\ (y,}\, \fun(f : \tau \ra \tau).\ \textcolor{gray}{d\ (}\fun((v, w) : (\sigma, \tau)).\ (v, f\ w)\textcolor{gray}{))\ i'} \\
      &&\hspace{-1.2cm} \textcolor{gray}{\textbf{in}\ ((x', y'), i'')} \\
      \textcolor{gray}{\Interleave3_{\Z}} &\textcolor{gray}{=}& \textcolor{gray}{\fun(n, d).\ \fun i.\ (n, i)}
    \end{array} \\[0.2cm]
    &\begin{array}{@{}l@{\ }c@{\ }l}
      \Deinterleave3_\tau &:& \forall c.\ \trans3c[\tau] \ra (\tau, \tau \lra (\Staged c \ra \Staged c)) \\
      \mathrlap{\text{(Same as $\Deinterleave2$ in \cref{fig:algo-monadic}, except with $\id$ and $(\circ)$ in place of $\ZeroStaged$ and $(\PlusStaged)$)}}
    \end{array} \\[0.2cm]
    &\begin{array}{@{}l}
      \Wrap3 : (\sigma \ra \tau) \rightsquigarrow (\sigma \ra (\tau, \tau \lra \sigma)) \\
      \Wrap3[\lambda(x : \sigma).\ t] = \fun(x : \sigma).\ \begin{array}[t]{@{}l@{}}
        \textbf{let}\ (x : \trans3\sigma[\sigma], i) = \Interleave3_\sigma\ (x, \InitStaged)\ 0 \\
        \textbf{in}\ \textbf{let}\ (y, d) = \Deinterleave3_\tau\ (\fst (\trans3\sigma[t]\ i)) \\
        \textbf{in}\ (y, \lfun(z : \tau).\ \ResolveStaged\ (d\ z\ \ZeroStaged))
      \end{array}
    \end{array}
  \end{align*}
  \caption{\label{fig:algo-cayley}
    The Cayley-transformed code transformation, based on \cref{fig:algo-monadic}.
    Grey parts are unchanged.
  }
\end{figure}

We show the new code transformation in \cref{fig:algo-cayley}.
Aside from the changes to types and to the target monoid of the backpropagators, the only additional change is in $\Interleave3$, which is adapted to accomodate the additional Cayley transform on the $c$ argument of $\InitStaged$.
Note that the backpropagators in $\Interleave3$ do not create any $\underline0$ values for untouched parts of the collected cotangent of type $c$, as promised, and that the new type of $\InitStaged$ has indeed eliminated all uses of $(+)$ on $c$, not just moved them around.

\subsection{Running example}\label{sec:running-example-cayley}

Apart from types, the only change in $\trans3c$ since \cref{sec:running-example-staged} is replacement of $(\PlusStaged)$ by $(\circ)$.
The result of $\Interleave3_{(\R,\R)}\ (x, \InitStaged)\ 0$ is the same as before, except that the input backpropagators now update a $\Staged\ (\R,\R)$ pair instead of constructing one: $d_{\text{in},0} = \lfun(z:\lR).\ \fun((\mathit{dx},\mathit{dy}),m).\ ((z + \mathit{dx}, \mathit{dy}), m)$ and $d_{\text{in},1} = \lfun(z:\lR).\ \fun((\mathit{dx},\mathit{dy}),m).\ ((\mathit{dx}, z + \mathit{dy}), m)$.
The $\Staged\ (\R,\R)$ object passed to $\ResolveStaged$ is now:
\[
  (\underline0, \{0 \mapsto (d_{\text{in},0}, 25), 2 \mapsto (\lfun(z':\lR).\ \fun(c,m).\ (c, f\ m), 12)\})
\]
where $f$ is a function that adds the key-value pairs $0 \mapsto (d_{\text{in},0}, z')$ and $1 \mapsto (d_{\text{in},1}, z')$ into its argument $m$, inserting if not yet present and adding the second components of the values if they are.
In this case, because the item at ID 2 (the backpropagator for $z = x + y$) is the first to be resolved by $\ResolveStaged$, this $f$ will be passed an empty map, so the two pairs will be inserted.

\subsection{Remaining Complexity Challenges}\label{sec:cayley-complexity}

In \cref{sec:staged-complexity}, we pinpointed the three remaining complexity issues with the reverse AD algorithm after function call staging: costly monoid operations on $c$, costly one-hot backpropagators from $\Interleave{}$, and logarithmic $\Map$ operations in $\ResolveStaged$.
The first issue has been solved by Cayley-transforming $\Staged c$: only $\underline0 : c$ is still used, and that only once (in $\Wrap3$).
For the second issue, although performance of the one-hot backpropagators has improved in most cases, it is still unsatisfactory; for example, given the input type $\sigma = ((\R, \Z), (\R, \R))$, the backpropagator for the second scalar looks as follows before and after the Cayley transform:
\begin{center}
  \begin{tabular}{c@{\quad}|@{\quad}c}
    \textit{Before Cayley}
    &
    \textit{After Cayley}
    \\
    \begin{tikzpicture}[scale=0.9]
      \node at (-0.23, 0) {$\lfun(d : \lR).\ \InitStaged\ \biggl($};
      \node at (4, 0) {$\biggr)$};

      \node (1) at (2.6, 0.8) {\small$(,)$};
      \node (2a) at (1.9, 0) {\small$(,)$};
      \node (2b) at (3.3, 0) {\small$(,)$};
      \node (3a) at (1.5, -0.8) {$0$};
      \node (3b) at (2.3, -0.8) {$0$};
      \node (3c) at (2.9, -0.8) {\rlap{\smash{\raisebox{0.3pt}{$d$}}}\phantom{$d$}};
      \node (3d) at (3.7, -0.8) {$0$};
      \draw (1) -- (2a);
      \draw (1) -- (2b);
      \draw (2a) -- (3a);
      \draw (2a) -- (3b);
      \draw (2b) -- (3c);
      \draw (2b) -- (3d);
    \end{tikzpicture}
    &
    \begin{tikzpicture}[scale=0.9]
      \node at (-0.23, -2.3) {$\lfun(d : \lR).\ \InitStaged\ \biggl($};
      \node at (3.8, -2.3) {$\biggr)$};

      \node (c1) at (2.2, -1.5) {\small$(,)$};
      \node (c2a) at (1.5, -2.3) {$\vphantom{()}\id$};
      \node (c2b) at (2.9, -2.3) {\small$(,)$};
      \node (c3c) at (2.5, -3.1) {$(+d)$};
      \node (c3d) at (3.4, -3.1) {$\id$};
      \draw (c1) -- (c2a);
      \draw (c1) -- (c2b);
      \draw (c2b) -- (c3c);
      \draw (c2b) -- (c3d);
    \end{tikzpicture}
  \end{tabular}
\end{center}
This yields complexity logarithmic in the size of the input if the input is balanced, but can degrade to linear in the size of the input in the worst case---which is no better than the previous version.
We will make these backpropagators properly logarithmic in the size of the input in \cref{sec:just-scalars}, essentially reducing issue two to issue three.
Afterwards, in \cref{sec:mutarrays}, we finally properly resolve issue three (logarithmic overhead from Map operations) by introducing mutable arrays.
Doing so removes the   final log-factors from the algorithm's complexity.

\section{Keeping Just the Scalars: Efficient Gradient Updates}\label{sec:just-scalars}

The codomain of the backpropagators is currently $\Staged c \ra \Staged c$, where $\Staged c$ is defined as:
\[ \Staged c = (c, \Map\ \Z\ (\lR\lra (\Staged c \ra \Staged c), \lR)) \]
The final cotangent of the input to the program is collected in the first component of the pair, of type $c$.
This collector is initialised with $\underline0 : c$ in $\ZeroStaged$, and added to by the one-hot input backpropagators from $\Interleave{}$, called in $\ResolveStaged$.
The task of these input backpropagators is to add the cotangent (of type $\lR$) that they receive in their argument, to a particular scalar in the collector.

Hence, all we need of $c$ in $\Staged c$ is really the collection of its scalars: the rest simply came from $\underline0 : c$ and is never changed.\footnote{If $c$ contains coproducts (sum types) as we discuss in \cref{sec:source-language-extension}, this $\underline0 : c$ becomes dependent on the actual input to the program, copying the structure from there.}
Furthermore, the reason why the one-hot input backpropagators currently do not finish in logarithmic time is that $c$ may not be a balanced tree of its scalars.
But if we are interested only in the scalars anyway, we can \emph{make} the collector balanced---by replacing it with $\Map\ \Z\ \lR$:
\[ \Staged c = (\textcolor{red}{\Map\ \Z\ \lR}, \Map\ \Z\ (\lR\lra (\Staged c \ra \Staged c), \lR)) \]
To accomodate this, $\Interleave{}$ changes to number all the scalars in the input with distinct IDs (conveniently with the same IDs as their corresponding input backpropagators, but this is no fundamental requirement); the cotangent of the input scalar with ID $i$ is stored in the $\Map$ at key $i$.
The input backpropagators can then modify the correct scalar in the collector (now of type $\Map\ \Z\ \lR$) in time logarithmic in the size of the input.
To be able to construct the final gradient from this collection of just its scalars, $\Interleave{}_\tau$ additionally builds a \emph{reconstruction} function of type $(\Z \ra \lR) \ra \tau$, which we pass a function that looks up the ID in the final collector $\Map$ to compute the actual gradient value.

\myparagraph{Complexity}
Now that we have fixed (in \cref{sec:cayley}) the first complexity problem identified in \cref{sec:staged-complexity} (expensive monoid operations) and reduced the second (expensive input backpropagators) to a logarithmic overhead over the original program, we have reached the point where we satisfy the complexity requirement stated in \cref{sec:rev-ad-complexity,sec:staged-complexity} apart from log-factors.
More precisely, if we set $T = \cost(P\ x)$ and $I = \size(x)$, then $\Wrap{}[P]$ computes the gradient of $P$ at $x$ not in the desired time $O(T + I)$ but instead in time $O\bigl(T \max(\log(T), \log(I)) + I \log(I)\bigr)$.\footnote{
  There are $O(T)$ backpropagators to resolve, each of which could either modify the staging map (of size $O(T)$) or the gradient collector map (of size $O(I)$).
  (De)interleaving then does $O(n \log(n))$ work on the input of size $O(I)$.
}
If we accept logarithmic overhead, we could choose to stop here.
However, if we wish to strictly conform to the required complexity, or if we desire to lose the non-negligible constant-factor overhead of dealing with a persistent tree-map, we need to make the input backpropagators and $\Map$ operations in $\ResolveStaged$ constant-time; we do this using mutable arrays in \cref{sec:mutarrays}.

\section{Using Mutable Arrays to Shave Off Log Factors}\label{sec:mutarrays}

The analysis in \cref{sec:staged-complexity} showed that after the Cayley transform in \cref{sec:cayley}, the strict complexity requirements are met if we make the input backpropagators constant-time and make $\ResolveStaged$ have only constant overhead for each backpropagator that it calls.
Luckily, in both cases the only component that is not constant-time is the interaction with one of the $\Map$s in $\Staged c$:
\[ \Staged c = (\Map\ \Z\ \lR, \Map\ \Z\ (\lR\lra (\Staged c \ra \Staged c), \lR)) \]
The input backpropagators perform (logarithmic-time) updates to the first $\Map$ (the cotangent collector), and $\ResolveStaged$ reads, deletes and updates entries in the second $\Map$ (the staging map for recording delayed backpropagator calls).
Both of these $\Map$s are keyed by increasing, consecutive integers starting from 0, and are thus ideal candidates to be replaced by an array:
\[ \Staged c = (\mathrlap{\textcolor{red}{\Array}}\hphantom{\Map\ \Z}\ \lR, \mathrlap{\textcolor{red}{\Array}}\hphantom{\Map\ \Z}\ (\lR\lra (\Staged c \ra \Staged c), \lR)) \]

To allocate an array, one must know how large it should be.
Fortunately, at the time when we allocate the initial $\Staged c$ value using $\ZeroStaged$ in $\Wrap{}$, the primal pass has already been executed and we know (from the output ID of $\Interleave{}$) how many input scalars there are, and (from the output ID of the transformed program) how many backpropagators there are.
Hence, the size of these arrays is indeed known when they are allocated; and while these arrays are large, the resulting space complexity is equal to the worst case for reverse AD in general.\footnote{
  For worst-case programs, the space complexity of reverse AD is equal to the time complexity of the original program.~\citep{adbook-2008-griewank-walther}
  Reducing this space complexity comes at a trade-off to time complexity, using checkpointing \citep[e.g.][]{ad-2018-checkpointing-built-in}.
}

To get any complexity improvements from replacing a $\Map$ with an $\Array$ (indeed, to not pessimise the algorithm completely!),\ the write operations to the arrays need to be done \emph{mutably}.
These write operations occur in two places: in the updater functions (of type $\Staged c \ra \Staged c$)  produced by backpropagators, and in $\ResolveStaged$.
Hence, in these two places we need an effectful function type that allows us to encapsulate the mutability and ensure that it is not visible from the outside; options include a resource-linear function type and a monad for local side-effects such as the \ST monad in Haskell.
In this paper, we use a side-effectful monad; for a presentation of the sequential algorithm in terms of resource-linear types, see \citep[Appendix A]{ad-dualrev-th-arxivv2}.

\myparagraph{Time complexity}
We now satisfy all the requirements of the analysis in \cref{sec:staged-complexity}, and hence have the correct time complexity for reverse AD.
In particular, let $I$ denote the size of the input and $T$ the runtime of the original program.
Let $\trans4{}$, $\Interleave4$, $\Deinterleave4$, etc.\ be the definitions that use arrays as described above (see \cref{sec:mutarrays-io} for details).
Then we can observe the following:
\begin{itemize}
\item
  The number of operations performed by $\trans4c[t]$ (with the improvements from \cref{sec:just-scalars,sec:mutarrays}) is only a constant factor times the number of operations performed by $t$, and hence in $O(T)$.
  This was already observed for $\trans2c[t]$ in \cref{sec:staged-complexity}, and still holds.

\item
  The number of backpropagators created while executing $\trans4c[t]$ is clearly also in $O(T)$.

\item
  The number of operations performed in any one backpropagator is constant.
  This is new, and only true because $\id$ (replacing $\ZeroStaged$), $(\circ)$ (replacing $(\PlusStaged)$), $\InitStaged$ (with a constant-time mutable array updater as argument) and $\StagedCall$ are now all constant-time.

\item
  Hence, because every backpropagator is invoked at most once thanks to our staging, and because the overhead of $\ResolveStaged$ is constant per invoked backpropagator, the amount of work performed by calling the top-level input backpropagator is again in $O(T)$.

\item
  Finally, the (non-constant-time) extra work performed in $\Wrap4$ is interleaving ($O(I)$), deinterleaving ($O(\text{size of output})$ and hence $O(T + I)$), resolving ($O(T)$) and reconstructing the gradient from the scalars in the $\Array \lR$ in $\Staged c$ ($O(I)$); all this work is in $O(T + I)$.
\end{itemize}
Hence, calling $\Wrap4[t]$ with an argument and calling its returned top-level derivative once takes time $O(T + I)$, i.e.\ at most proportional to the runtime of calling $t$ with the same argument, plus the size of the argument itself.
This is indeed the correct time complexity for an efficient reverse AD algorithm, as discussed in \cref{sec:rev-ad-complexity}.


\subsection{Implementation Using Mutable References in a Monad}\label{sec:mutarrays-io}

As a purely functional language, Haskell chooses to disallow, in most parts of a program, any behavior that breaks referential transparency, including computational effects like mutable state.
Its workaround is to force the programmer to encapsulate such ``dangerous'' effectful behavior, when it is truly desired, in a monad, thus using the type system to isolate it from the pure code.
The result is that the compiler can aggressively optimize the pure parts of the code while mostly leaving the effectful code, where correctness of optimizations is much more subtle, as is.

In particular, a typical design for mutable arrays in a purely functional language like Haskell is to use mutable references inside some monad.
In Haskell, one popular solution is to use the \ST monad~\citep{fp-1994-st-monad} together with a mutable array library that exposes an API using \ST, such as \texttt{STVector} in the \texttt{vector}\footnote{\url{https://hackage.haskell.org/package/vector-0.13.1.0/docs/Data-Vector-Mutable.html}} library.
Because the \ST monad is designed to be deterministic, it has a pure handler:
\[ \mathrm{runST} : (\forall s.\ \ST\ s\ \alpha) \ra \alpha \]
allowing the use of local mutability without compromising referential transparency of the rest of the program.\footnote{
  The $s$ parameter is informationless and only there to ensure correct scoping of mutable references in \ST.
  For more info, see~\citep[\S2.4]{fp-1994-st-monad}, or \citep{fp-2022-st-monad-proof} for a formalised proof.
}

However, precisely because of this design, \ST does not support parallelism.
(Parallelism in the presence of mutable references trivially allows non-deterministic behaviour.)
For this reason we will write the definitions from this point on in terms of \IO, Haskell's catch-all monad for impurity.
Fortunately, the only functionality we use from \IO is parallelism and mutable arrays and references, and furthermore the design of our algorithm is such that the result is, in fact, deterministic even when the source program includes parallelism.\footnote{
  From a theoretical perspective, this determinism follows from the fact that we do not actually use unrestricted mutation, but only \emph{accumulation}---and accumulation is a commutative effect.
  In practice, however, the claim is technically untrue, because floating-point arithmetic is not associative.
  Given the nature of the computations involved, however, we still think getting parallelism is worth this caveat.}
Thus we can justify using $\texttt{unsafePerformIO} :: \IO\ \alpha \ra \alpha$ around the differentiated program, making the interface to the differentiated program pure again.

Letting the updater functions run in \IO changes $\Staged c$ as follows:
\[ \Staged c = (\Array \lR, \Array {(\lR \lra (\textcolor{red}{\Staged c \ra \IO\ ()}), \lR)}) \]
for a suitable definition of $\Array$, such as \texttt{IOVector} from \texttt{vector}.
To write this definition, we need to give a monoid structure on $\Staged c \ra \IO\ ()$; fortunately, the only reasonable one ($f + g = \fun x.\ f\ x \mseq g\ x$) corresponds to the monoid structure on $\Staged c$ and is therefore precisely the one we want.\footnote{$(\mseq) :: \text{Monad}\ m \Rightarrow m\ \alpha \ra m\ \beta \ra m\ \beta$. $m_1 \mseq m_2$ runs both computations, discarding the result of $m_1$.}

Note that this definition now no longer structurally depends on $c$!
This is to be expected, because the information about the structure of $c$ is now contained in the \emph{length} of the first array in a $\Staged c$.
For uniformity of notation, however, we will continue to write the $c$ parameter to $\Staged$.

\myparagraph{Mutable arrays interface}
We assume an interface to mutable arrays that is similar to that for \texttt{IOVector} in the Haskell \texttt{vector} library, cited earlier.
In summary, we assume the following functions:
\[
  \begin{array}{@{}l@{\ }c@{\ }l@{}}
    \arralloc &:& \Z \ra \alpha \ra \IO\ (\Array \alpha) \\
    \arrget &:& \Z \ra \Array \alpha \ra \IO\ \alpha \\
    \arrmodify &:& \Z \ra (\alpha \ra \alpha) \ra \Array \alpha \ra \IO\ () \\
    \arrfreeze &:& \Array \alpha \ra \IO\ (\IArray \alpha) \\
    (@) &:& \IArray \alpha \ra \Z \ra \alpha
  \end{array}
\]
where $\IArray$ is an immutable array type.
$\arralloc$ and $\arrfreeze$ are linear in the length of the array; $\arrget$, $\arrmodify$ and $(@)$ are constant-time (in addition to calling the function once, of course, for $\arrmodify$).

Let us see how the code transformation changes with mutable arrays.

\myparagraph{Code transformation}
Using the new definition of $\Staged$, we change the code transformation once more, this time from \cref{fig:algo-cayley} to the version given in \cref{fig:algo-mutarrays}.
The transformation on types and terms simply sees the type of scalar backpropagators change to use effectful updating instead of functional updating, so they do not materially change: we just give yet another interpretation of $\ZeroStaged$ and $(\PlusStaged)$, using the monoid structure on $\Staged c \ra \IO\ ()$ defined above in terms of $(\mseq)$.
However, some important changes occur in the $\Staged c$ interface and the wrapper.
Let us first look at the algorithm from the top, by starting with $\Wrap4$; after understanding the high-level idea, we explain how the other components work.

\begin{figure}
  \figheading{On types:}
  \[\begin{array}{l@{\qquad}l}
    \trans4c[\R] = (\R, (\Z, \lR\lra (\Staged c \ra \IO\ ())))
      & \textcolor{gray}{\trans4c[\Z] = \Z} \qquad \textcolor{gray}{\trans4c[()] = ()} \\
    \textcolor{gray}{\trans4c[\sigma \ra \tau] = \trans4c[\sigma] \ra \Z \ra (\trans4c[\tau], \Z)} 
      & \textcolor{gray}{\trans4c[(\sigma,\tau)] = (\trans4c[\sigma], \trans4c[\tau])} \\
  \end{array}
  \]
  \figheading{On terms:}
  \begin{align*}
    &\textcolor{gray}{\text{If}\ \Gamma \vdash t : \tau\ \text{then}\ \trans4{c}[\Gamma] \vdash \trans4c[t] : \Z \ra (\trans4c[\tau], \Z)} \\
    &\begin{array}[t]{@{}l@{}}
      \text{Same as $\trans2c$, except with `$\fun\_.\ \mathbf{return}\ ()$' in place of $\ZeroStaged$ and `$\fun f\ g\ x.\ f\ x \mseq g\ x$' in} \\
      \text{place of $(\PlusStaged)$.}
    \end{array}
  \end{align*}
  \figheading{New $\Staged$ interface:}
  \begin{align*}
    &\begin{array}{@{}l@{}}
      \Staged c = (\Array \lR, \Array{} (\lR\lra (\Staged c \ra \IO\ ()), \lR)) \\[2mm]
      \StagedAlloc : \Z \ra \Z \ra \IO\ (\Staged c) \\
      \StagedAlloc\ i_{\text{inp}}\ i_{\text{out}} =
        \texttt{alloc}\ i_{\text{inp}}\ 0 \mbind \fun c.\ \texttt{alloc}\ i_{\text{out}}\ (\lfun(z : \lR).\ \id, 0) \mbind \fun m.\ \mathbf{return}\ (c, m) \\[2mm]
      %
      \StagedCall : (\Z, \lR\lra (\Staged c \ra \IO\ ())) \ra \lR\lra (\Staged c \ra \IO\ ()) \\
      \StagedCall\ (i, f)\ a\ (c, m) = \arrmodify\ i\ (\fun (\mathunderscore, a').\ (f, a + a'))\ m \\[2mm]
      \StagedOneHot : \Z \ra \lR \lra (\Staged c \ra \IO\ ()) \\
      \StagedOneHot\ i\ a\ (c, m) = \arrmodify\ i\ (\fun(a' : \R).\ a + a')\ c \\[2mm]
      \ResolveStaged : \Z \ra \Staged c \ra \IO\ (\IArray \lR) \\
      \ResolveStaged\ i_{\text{out}}\ (c, m) = \textit{loop}\ (i_{\text{out}} - 1) \mseq \arrfreeze\ c \\
      \quad\text{where}\ \begin{array}[t]{@{}l@{\ }c@{\ }l@{}}
        \textit{loop}\ (-1) &=& \mathbf{return}\ () \\
        \textit{loop}\ i &=& \texttt{get}\ i\ m \mbind \fun(f, a).\ f\ a\ (c, m) \mseq \textit{loop}\ (i - 1)
      \end{array}
    \end{array}
  \end{align*}
  \figheading{Wrapper:}
  \begin{align*}
    &\begin{array}{@{}l@{\;}c@{\;}l@{}}
      \Interleave4_\tau &:& \tau \ra \Z \ra ((\trans4{c}[\tau], \IArray \lR \ra \tau), \Z) \\
      \Interleave4_\R &=& \fun x.\ \fun i.\ (((x, (i, \StagedOneHot\ i)), \fun a.\ a \iarridx i), i + 1) \\
      \Interleave4_{()} &=& \fun ().\ \fun i.\ (((), \fun a.\ ()), i) \\
      \Interleave4_{(\sigma, \tau)} &=& \fun (x, y).\ \fun i.\ \begin{array}[t]{@{}l@{}}
        \textbf{let}\ ((x', f_1), i') = \Interleave4_\sigma\ x\ i \\
        \textbf{in}\ \textbf{let}\ ((y', f_2), i'') = \Interleave4_\tau\ y\ i' \\
        \textbf{in}\ (((x', y'), \fun a.\ (f_1\ a, f_2\ a)), i'')
      \end{array} \\
      \Interleave4_\Z &=& \fun n.\ \fun i.\ ((n, \fun a.\ n), i)
    \end{array} \\[1.5mm]
    &\begin{array}{@{}l}
      \Deinterleave4_\tau : \trans4{c}[\tau] \ra (\tau, \tau \lra (\Staged c \ra \IO\ ())) \\
      \text{(Same as $\Deinterleave2$, except with the same monoid changes as $\trans4c$ above)}
    \end{array} \\[1.5mm]
    &\begin{array}{@{}l}
      \Wrap4 : (\sigma \ra \tau) \rightsquigarrow (\sigma \ra (\tau, \tau \lra \sigma)) \\
      \Wrap4[\lambda(x : \sigma).\ t] = \fun(x : \sigma). \\
      \qquad \textbf{let}\ ((x : \trans4{\sigma}[\sigma], \textit{rebuild} : \IArray \lR \ra \sigma), i) = \Interleave4_\sigma\ x\ 0 \\
      \qquad \textbf{in}\ \textbf{let}\ (y', i') = \trans4{\sigma}[t]\ i \\
      \qquad \textbf{in}\ \textbf{let}\ (y, d : \tau \lra (\Staged \sigma \ra \IO\ ())) = \Deinterleave4_\tau\ y' \\
      \qquad \textbf{in}\ (y, \lfun(z : \tau).\ \textit{rebuild}\ (\texttt{unsafePerformIO} \\
      \hphantom{\qquad \textbf{in}\ (y, \lfun(z : \tau).\ \textit{rebuild}\ (} \qquad (\StagedAlloc\ i\ i' \mbind \fun s.\ d\ z\ s \mseq \ResolveStaged\ i'\ s)))
    \end{array}
  \end{align*}
  \caption{\label{fig:algo-mutarrays}
    Code transformation plus wrapper using mutable arrays, modified from \cref{fig:algo-cayley}.
    Grey parts are unchanged.
  }
\end{figure}

In basis, $\Wrap4$ does the same as $\Wrap3$ from \cref{fig:algo-cayley}: interleave injector backpropagators with the input of type $\sigma$, execute the transformed function body using the interleaved input, and then deinterleave the result.
However, because we (since \cref{sec:just-scalars}) represent the final cotangent not directly as a value of type $\sigma$ in a $\Staged \sigma$ but instead as an array of only the embedded scalars ($\Array \lR$), some more work needs to be done.

Firstly, $\Interleave4_\sigma$ (monadically, in the ID generation monad that we are writing out explicitly) produces, in addition to the interleaved input, also a \emph{reconstruction} function\footnote{Implementing the $(\Z \ra \lR)$ in $(\Z \ra \lR) \ra \tau$ from \cref{sec:just-scalars} as $\IArray \lR$.} of type $\IArray \lR \ra \sigma$.
This rebuilder takes an array with precisely as many scalars as were in the input, and produces a value of type $\sigma$ with the structure (and discrete-typed values) of the input, but the scalars from the array.
The mapping between locations in $\sigma$ and indices in the array is the same as the numbering performed by $\Interleave4$.

Having $x$, $\textit{rebuild}$ and $i$ (the next available ID), we execute the transformed term $\trans4\sigma[t]$ monadically (with $x$ in scope), resulting in an output $y' : \trans4{\sigma}[\tau]$.
This output we deinterleave to $y : \tau$ and $d : \tau \lra (\Staged \sigma \ra \IO\ ())$.

The final result then consists of the regular function result ($y$) as well as the top-level derivative function of type $\tau \lra \sigma$.
In the derivative function, we allocate two arrays to initialise an empty $\Staged \sigma$ (note that the given sizes are indeed precisely large enough), and apply $d$ to the incoming $\tau$ cotangent.
This gives us an updater function that (because of how $\Deinterleave4$ works) calls the top-level backpropagators contained in $y'$ in the $\Staged$ arrays, and we apply this function to the just-allocated $\Staged$ object.
Then we use the new $\ResolveStaged$ to propagate the cotangent contributions backwards, by invoking each backpropagator in turn in descending order of IDs.
Like before in the Cayley-transformed version of our AD technique, those backpropagators update the state (now mutably) to record their own contributions to (i.e.\ invocations of) other backpropagators.
At the end of $\ResolveStaged$, the backpropagator staging array is dropped and the cotangent collection array is frozen and returned as an $\IArray \lR$ (corresponding to the $c$ value in a $\Staged c$ for the Cayley-transformed version in \cref{fig:algo-cayley}).

This whole derivative computation is, in the end, pure (in that it is deterministic and has no side-effects), so we can safely evaluate the \IO using \texttt{unsafePerformIO} to get a pure $\IArray \lR$, which contains the scalar cotangents that $\textit{rebuild}$ from $\Interleave4$ needs to put in the correct locations in the input, thus constructing the final gradient.

\myparagraph{Implementation of the components}
Having discussed the high-level sequence of operations, let us briefly discuss the implementation of the $\Staged c$ interface and (de)interleaving.
In $\Interleave4$, instead of passing structure information down in the form of a setter ($(\tau \ra \tau) \lra (\Staged c \ra \Staged c)$) like we did in $\Interleave3$ in \cref{fig:algo-cayley}, we build structure information up in the form of a getter ($\IArray \lR \ra \tau$).
This results in a somewhat more compact presentation, but in some sense the same information is still communicated.

The program text of $\Deinterleave4$ is again unchanged, because it is agnostic about the codomain of the backpropagators, as long as it is a monoid, which it remains.

On the $\Staged$ interface the transition to mutable arrays had a significant effect.
The $\ZeroStaged$ created by $\Wrap3$ in \cref{sec:cayley} is now essentially in $\StagedAlloc$, which uses $\arralloc$ to allocate a zero-filled cotangent collection array of size $i_{\text{inp}}$, and the backpropagator staging array of size $i_{\text{out}}$ filled with zero-backpropagators with an accumulated argument of zero.

$\StagedCall$ has essentially the same type, but its implementation differs because it now performs a constant-time mutable update on the backpropagator staging array instead of a logarithmic-complexity immutable $\Map$ update.
Note that, unlike in \cref{sec:cayley}, there is no special case if $i$ is not yet in the array, because unused positions are already filled with zeros.

$\StagedOneHot$ takes the place of $\InitStaged$, with the difference that we have specialised it using the knowledge that all relevant $c \ra c$ functions add a particular scalar to a particular index in the input, and that these functions can hence be defunctionalised to a pair $(\Z, \lR)$.
The monoid-linearity here is in the real scalar, as it was before, hence the placement of the $\lra$-arrow.

Finally, $\ResolveStaged$ takes an additional $\Z$ argument that should contain the output ID of $\trans4{c}[t]$, i.e.\ one more than the largest ID generated.
$\textit{loop}$ then does what the original $\ResolveStaged$ did directly, iterating over all IDs in descending order and applying the state updaters in the backpropagator staging array one-by-one to the state.
After the loop is complete, we freeze and return just the cotangent collection array, because we have no need for the staging array any more.
This frozen collection array will then be used to build the final gradient in $\Wrap4$.

\section{Was It Taping All Along?}\label{sec:improve}

In this section we first apply one more optimisation to our algorithm 
to make it slightly more efficient (\cref{sec:improve-one-array}).
Afterwards, we show that defunctionalising the backpropagators (\cref{sec:improve-defunctionalisation}) essentially reduces the technique to 
classical taping approaches (\cref{sec:improve-taping}).

\subsection{Dropping the Cotangent Collection Array}\label{sec:improve-one-array}

Recall that the final transformation of \cref{sec:mutarrays} used two mutable arrays threaded through the backpropagators in the $\Staged c$ pair: a cotangent collection array of type $\Array \lR$ and a backpropagator call staging array of type $\Array{} (\lR\lra (\Staged c \ra \IO\ ()), \lR)$, using the monad-based implementation from \cref{sec:mutarrays-io}.
The first array is modified by $\Interleave{}_\R$ and the second by $\StagedCall$.
No other functions modify these arrays.

Looking at the function of $\Interleave{}_\R$ in the algorithm,
all it does is produce input backpropagators with some ID $i$, which act by adding their  argument to index $i$ in the cotangent collection array.
This means that we have $c[i] = \snd (m[i])$ for all $i$ for which $c[i]$ is defined, if $(c, m)$ is the input to $\ResolveStaged$ for which the recursion terminates.
Therefore, the cotangent collection array is actually unnecessary: its information can be read off directly from the backpropagator staging array.

With this knowledge, we can instead use $\Staged c = \Array {(\lR \lra (\Staged c \ra \IO\ ()), \lR)}$ as our definition. The reconstruction functions of \cref{sec:just-scalars} simply take the second projection of the correponding array element.


\subsection{Defunctionalisation of Backpropagators}\label{sec:improve-defunctionalisation}

In the core code transformation ($\trans{}{c}$, excluding the wrapper), all backpropagators are (now) of type $\lR\lra (\Staged c \ra \IO\ ())$, and, as observed earlier in \cref{sec:staging}, these backpropagators come in only a limited number of forms:
\begin{enumerate}
\item\label{item:backprop-form-inputcot}
  The input backpropagators created in $\Interleave{}_\R$, reduced to $(\lfun(z : \lR).\ \mathbf{return}\ ())$ in \cref{sec:improve-one-array};
\item\label{item:backprop-form-scalar}
  $(\lfun(z : \lR).\ \mathbf{return}\ ())$ created in $\trans4{c}[r]$ for scalar constants $r$;
\item\label{item:backprop-form-operator}
  $(\lfun(z : \lR).\ \StagedCall\ d_1\ (\partial_1 \textit{op}(x_1,\ldots,x_n)(z)) \circ \cdots \circ \StagedCall\ d_n\ (\partial_n \textit{op}(x_1,\ldots,x_n)(z)))$ created in $\trans{}{c}[\textit{op}(x_1,\ldots,x_n)]$ for primitive operations $\textit{op}$.
\end{enumerate}
Furthermore, the information contained in an operator backpropagator of form (3) can actually be described without reference to the value of its argument $z$: 
because our operators return a single scalar (as opposed to e.g.\ a vector), we have
\[
  \partial_i \textit{op}(x_1,\ldots,x_n)(z) = z \cdot \partial_i \textit{op}(x_1,\ldots,x_n)(1)
\]
Hence, we can defunctionalise~\citep{fp-1998-defunctionalisation} and change all occurrences of the type $\lR\lra (\Staged c \ra \IO\ ())$ to $\Contrib$, where $\Contrib = [(\lR, (\Z, \Contrib))]$: a list of triples of a scalar, an integer ID, and a recursive $\Contrib$ structure.
The ID is the ID of the $\Contrib$ structure (i.e.\ the backpropagator) that it is adjacent to.
(As before, these IDs make sharing observable.)
In this representation, we think of $[(a_1, (i_1, \textit{cb}_1)), \ldots,(a_n, (i_n, \textit{cb}_n))]$ of type $\Contrib$ as the backpropagator
\begin{equation*}
\lfun(z : \lR).\ \StagedCall\ (i_1, \textit{cb}_1)\ (z \cdot a_1) \circ \cdots \circ \StagedCall\ (i_n, \textit{cb}_n)\ (z \cdot a_n)
\end{equation*}

For example, suppose we differentiate the running example program:
\begin{align*}
  \fun(x, y).\ \textbf{let}\ z = x + y\ \textbf{in}\ x \cdot z
\end{align*}
using the final algorithm of \cref{sec:mutarrays-io}.
The return value from the $\trans{}{(\R,\R)}$-transformed code (when applied to the output from $\Interleave{}_{(\R,\R)}$) has the sharing structure shown in \cref{subfig:sharing-structure-before}.
This shows how the backpropagators refer to each other in their closures.

If we perform the type replacement (\cref{sec:improve-one-array}) and the defunctionalisation (this subsection), $\Interleave{}$ simplifies and $\StagedCall$ disappears; backpropagators of forms (\ref{item:backprop-form-inputcot}) and (\ref{item:backprop-form-scalar}) become $[]$ (the empty list) and those of form (\ref{item:backprop-form-operator}) become:
\[
  [(\partial_1 \textit{op}(x_1,\ldots,x_n)(1), d_1), \ldots, (\partial_n \textit{op}(x_1,\ldots,x_n)(1), d_n)]
\]
$\ResolveStaged$ then interprets a list of such $(a, (i, \textit{cb}))$ by iterating over the list and for each such triple, replacing $(\textit{cb}', a')$ at index $i$ in the staging array with $(\textit{cb}, a' + a \cdot d)$, where $d$ is the cotangent recorded in the array cell where the list was found.

\begin{figure}
  \begin{center}
  \begin{subfigure}[b]{0.43\textwidth}
    \begin{tikzpicture}[remember picture]
      \tikzset{tikzmark prefix=defunc-a}
      \node at (0, 0) {$\textcolor{gray}{(x, (1,\,} \subnode[inner sep=1pt]{fun1}{\lfun} d.\ ... \textcolor{gray}{))}$};
      \node at (2.5, 0) {$\textcolor{gray}{(y, (2,\,} \subnode[inner sep=0.8pt]{fun2}{\lfun} d.\ ... \textcolor{gray}{))}$};
      \node at (2.2, -1) {$\textcolor{gray}{(z, (3,\,} \subnode[inner sep=1pt]{fun3}{\lfun} d.\ \subnode{box3-1}{\Box}\ d \circ \subnode{box3-2}{\Box}\ d\textcolor{gray}{))}$};
      \node at (1, -2) {$(x\cdot z, (4, \lfun d.\ \subnode{box4-1}{\Box}\ (z\cdot d) \circ \subnode{box4-3}{\Box}\ (x \cdot d)))$};
      \draw[->] (box3-1.mid) .. controls (2.4, -0.4) and (0.3, -0.9) .. (fun1);
      \draw[->] (box3-2.mid) .. controls (3.1, -0.7) and (2.8, -0.4) .. (fun2.south east);
      \draw[->] (box4-3.mid) .. controls (2.1, -1.8) and (2.0, -1.4) .. (fun3);
      \draw[->] (box4-1.mid) .. controls (0.51, -1.64) and (0.193, -0.56) .. (fun1);
    \end{tikzpicture}
    \caption{\label{subfig:sharing-structure-before}
      Before defunctionalisation
    }
  \end{subfigure}
  \hfill
  \begin{subfigure}[b]{0.55\textwidth}
    \begin{tikzpicture}[remember picture]
      \tikzset{tikzmark prefix=defunc-b}
      \node at (-0.3, 0) {$\textcolor{gray}{(x, (1,\,} \subnode[inner sep=1pt]{cb1}{[]}\textcolor{gray}{))}$};
      \node at (2.43, 0) {$\textcolor{gray}{(y, (2,\,} \subnode[inner sep=1pt]{cb2}{[]}\textcolor{gray}{))}$};
      \node at (2.8, -1) {$\textcolor{gray}{(z, (3,\,} \subnode[inner sep=1pt]{cb3}{[}(1.0, (1, \subnode{box3-1}{\Box})), (1.0, (2, \subnode{box3-2}{\Box}))]\textcolor{gray}{))}$};
      \node at (1.13, -2) {$(x\cdot z, (4, [(z, (1, \subnode{box4-1}{\Box})), (x, (3, \subnode{box4-3}{\Box}))]))$};
      \draw[->] (box3-1.mid) .. controls (2.12, -0.3) and (0.03, -0.9) .. (cb1.south);
      \draw[->] (box3-2.mid) .. controls (4.046, -0.4) and (3.1, -0.7) .. (cb2);
      \draw[->] (box4-3.mid) .. controls (2.2, -1.5) and (1.5, -1.7) .. (cb3);
      \draw[->] (box4-1.mid) .. controls (0.8, -1.3) and (0.0, -1.8) .. (cb1.south);
    \end{tikzpicture}
    \caption{\label{subfig:sharing-structure-after}
      After defunctionalisation
    }
  \end{subfigure}
  \end{center}
  \caption{\label{fig:sharing-structure}
    The sharing structure before and after defunctionalisation.
    $\StagedCall$ is elided here; in \cref{subfig:sharing-structure-before}, the backpropagator calls are depicted as if they are still normal calls.
    Boxes ($\Box$) are the same in-memory value as the value their arrow points to; two boxes pointing to the same value indicates that this value is \emph{shared}: referenced in two places.
  }
\end{figure}

\subsection{Was It Taping All Along?}\label{sec:improve-taping}
After the improvements from \cref{sec:improve-one-array,sec:improve-defunctionalisation}, what previously was a tree of (staged) calls to backpropagator functions is now a tree of $\Contrib$ values with attached IDs\footnote{Note that we now have $\trans{}{}[\R] = (\R, (\Z, \Contrib))$, the integer being the ID of the $\Contrib$ value.} that are interpreted by $\ResolveStaged$.
This interpretation (eventually) writes the $\Contrib$ value with ID $i$ to index $i$ in the staging array (possibly multiple times), and furthermore accumulates argument cotangents in the second component of the pairs in the staging array.
While the argument cotangents must be accumulated in reverse order of program execution (indeed, that is the whole point of \emph{reverse} AD), the mapping from ID to $\Contrib$ value can be fully known in the forward pass: the partial derivatives of operators, $\partial_i \textit{op}(x_1,\ldots,x_n)(1)$, can be computed in the forward pass already.

This means that we can already compute the $\Contrib$ lists and write them to the array in the forward pass, if we change the ID generation monad that the differentiated code already lives in (which is a state monad with a single $\Z$ as state) to additionally carry the staging array, and furthermore change the monad to thread its state in a way that allows mutation, again using the techniques from \cref{sec:mutarrays}, but now in the forward pass too.
All that $\ResolveStaged$ then has to do is loop over the array in reverse order (as it already does) and add cotangent contributions to the correct positions in the array according to the $\Contrib$ lists that it finds there.

At this point, there is no meaningful difference any more between this algorithm and what is classically known as taping: we have a tape (the staging array) to which we write the performed operations in the forward pass (automatically growing the array as necessary)---although the tape entries are the already-differentiated operations in this case, and not the original ones.
In this way, we have related the naive version of dual-numbers reverse AD, which admits neat correctness proofs, to the classical, very imperative approach to reverse AD based on taping, which is used in industry-standard implementations of reverse AD (e.g. PyTorch~\citep{ad-2017-pytorch}).


\section{Extending the Source Language}\label{sec:source-language-extension}

The source language (\cref{fig:source-language}) that the algorithm discussed so far works on, is a higher-order functional language including product types and primitive operations on scalars.
However, dual-numbers reverse AD generalises to much richer languages in a very natural way, because most of the interesting work happens in the scalar primitive operations.
The correctness proof for the algorithm can be extended to many expressive language constructs in the source language, such as coproducts and recursive types by using suitable logical relations arguments \citep{nunes-2024-dual-numbers}.
Further, the efficiency of the algorithm is independent of the language constructs in the source language.
Indeed, in the forward pass, the code transformation is fully structure-preserving outside of the scalar constant and primitive operation cases; and in the reverse pass (in $\ResolveStaged$), all program structure is forgotten anyway, because the computation is flattened to the (reverse of the) linear sequence of primitive operations on scalars that was performed in the particular execution of the forward pass.

\myparagraph{(Mutual) recursion}
For example, we can allow recursive functions in our source language by adding recursive let-bindings with the following typing rule:
\[
  \frac{\Gamma, f : \sigma \ra \tau, x : \sigma \vdash s : \tau \qquad
        \Gamma, f : \sigma \ra \tau \vdash t : \rho}
       {\Gamma \vdash \textbf{letrec}\ f\ x = s\ \textbf{in}\ f\ t : \rho}
\]
The code transformation $\trans i{}$ for all $i$ then treats $\textbf{letrec}$ exactly the same as $\textbf{let}$---note that the only syntactic difference between $\textbf{letrec}$ and $\textbf{let}$ is the scoping of $f$---and the algorithm remains both correct and efficient.
Note that due to the assumed call-by-value semantics, recursive non-function definitions would make little sense.

Recursion introduces the possibility of non-termination; because the reverse pass is nothing more than a loop over the primitive scalar operations performed in the forward execution, the derivative program terminates exactly if the original program terminates (on a machine with sufficient memory).

\myparagraph{Coproducts}
To support dynamic control flow (necessary to make recursion useful), we can easily add coproducts to the source language.
First add coproducts to the syntax for types ($\sigma,\tau \coloneqqq \cdots \mid \sigma \sqcup \tau$) both in the source language and in the target language, and add constructors and eliminators to all term languages (both linear and non-linear):
\begin{align*}
s,t \coloneqqq \cdots \mid \inl(t) \mid \inr(t) \mid \case s\ \{ \inl(x) \ra t_1 ; \inr(y) \ra t_2 \}
\end{align*}
where $x$ is in scope in $t_1$ and $y$ is in scope in $t_2$.
Then the type and code transformations extend in the unique structure-preserving manner:
\begin{gather*}
  \trans1c[\sigma \sqcup \tau] = \trans1c[\sigma] \sqcup \trans1c[\tau] \\
  \trans1c[\inl(t)] = \inl(\trans1c[t]) \qquad
    \trans1c[\inr(t)] = \inr(\trans1c[t]) \\
  \trans1c[\case s\ \{ \inl(x) \ra t_1 ; \inr(x) \ra t_2 \}] =
    \case{\trans1c[s]}\ \{ \begin{array}[t]{@{}l@{\;\ra\;}l@{}}
      \inl(x) & \trans1c[t_1] ; \\
      \inr(x) & \trans1c[t_2] \}
    \end{array}
\end{gather*}

To create an interesting interaction between control flow and differentiation, we can add a construct `$\sign$' with the unsurprising typing rule
\[
  \frac{\Gamma \vdash t : \R}
       {\Gamma \vdash \sign(t) : \Bool}
\]
where $\Bool = () \sqcup ()$, which allows us to perform a case distinction on the sign of a real number.
For differentiation of this construct it suffices to define
$\trans1c[\sign(t)] = \sign(\fst(\trans1c[t]))$; if one wants to give up the structure-preserving aspect of the transformation, it is also possible to prevent redundant differentiation of $t$ and just make clever substitutions in $t$'s free variables to convert back from dual-numbers form to the plain data representation.\footnote{
  The idea is to define functions $\varphi_\tau : \trans{}c[\tau] \ra \tau$ and $\psi_\tau : \tau \ra \trans{}c[\tau]$ by induction on $\tau$, where $\varphi_\tau$ projects out the value from a dual number and $\psi_\tau$ injects scalars into a dual number as constants (i.e.\ with a zero backpropagator).
  $\varphi_\tau$ and $\psi_\tau$ are mutually recursive at function types: e.g.\ $\varphi_{\sigma \ra \tau}$ converts a function that operates on dual numbers to a plain function by upgrading the plain argument to a dual argument (using $\psi_\sigma$), running that through the dual function, and finally downgrading the result back to a plain value using a recursive call to $\varphi_\tau$.
  Then one can define a non-differentiating transformation $\trans{\text{plain}}c$ on terms that uses $\varphi$ to convert free variables to plain values, and otherwise keeps all code plain.
  We then get $\trans1c[\sign(t)] = \sign(\trans{\text{plain}}c[t])$.
}

The type transformation stays unchanged when moving to $\trans4c$, and the only change for the term definitions is to transition to monadic code in $\trans2c$.
Lifting a computation to monadic code is a well-understood process.
The corresponding cases in $\Interleave{}$ and $\Deinterleave{}$ are the only reasonable definitions that type-check.

The introduction of dynamic control flow complicates the correctness story for any AD algorithm.
The approach presented here has the usual behaviour: derivatives are correct in the interior of domains leading to a particular execution path (if `$\sign$' is the only ``continuous conditional'', this is for inputs where none of the `$\sign$' operations receive zero as their arguments), but may be unexpected at the points of branching.
For discussion see e.g.~\citep[\S3.3]{ad-2023-pitfalls}; for proofs see e.g.~\citep[\S11]{nunes-2024-dual-numbers} or~\citep{ad-2021-dual-revad-linear-factoring-pcf}.

\myparagraph{Polymorphic and (mutually) recursive types}
In Haskell one can define (mutually) recursive data types e.g.\ as follows:
\begin{align*}
  &\textbf{data}\ T_1\ \alpha = C_1\ \alpha\ (T_2\ \alpha) \mid C_2\ \R \\
  &\textbf{data}\ T_2\ \alpha = C_3\ \Z\ (T_1\ \alpha)\ (T_2\ \alpha)
\end{align*}
If the user has defined some data types, then we can allow these data types in the code transformation.
We generate new data type declarations that simply apply $\trans1c[-]$ to all parameter types of all constructors:
\begin{align*}
  &\textbf{data}\ DT_1\ \alpha = DC_1\ \alpha\ (DT_2\ \alpha) \mid DC_2\ (\R,\lR\lra c) \\
  &\textbf{data}\ DT_2\ \alpha = DC_3\ \Z\ (DT_1\ \alpha)\ (DT_2\ \alpha)
\end{align*}
and we add one rule for each data type that simply maps:
\begin{gather*}
  \trans1c[T_1\ \tau] = DT_1\ \trans1c[\tau] \qquad
  \trans1c[T_2\ \tau] = DT_2\ \trans1c[\tau]
\end{gather*}
Furthermore, for plain type variables, we set $\trans1c[\alpha] = \alpha$.\footnote{As declaring new data types is inconvenient in Template Haskell, our current implementation only handles recursive data types that do not contain explicit scalar values.
As we can pass all required scalar types by instantiating their type parameters with a type containing $\R$, this is not a real restriction.
}

The code transformation on terms is completely analogous to a combination of coproducts (given above in this section, where we take care to match up constructors as one would expect: $C_i$ gets sent to $DC_i$) and products (given already in \cref{fig:algo-naive}).
The wrapper also changes analogously: $\Interleave{}$ and $\Deinterleave{}$ get clauses for $\Interleave{}_{(T_i\,\tau)}$ and $\Deinterleave{}_{(T_i\,\tau)}$.

Finally, we note that with the mentioned additional rule that $\trans1c[\alpha] = \alpha$, polymorphic functions can also be differentiated transparently, similarly to how the above handles polymorphic data types.

\section{Parallelism}\label{sec:parallelism}

So far, we have assigned sequentially increasing IDs to backpropagators in the forward pass and resolved them in their linear order from top to bottom during the reverse pass.
As long as the source program is executed sequentially, such ID generation is appropriate.
However, if the source program uses parallelism in its execution,
such linear ID assignment discards this parallelism structure and prevents us from exploiting it for computing the derivative in parallel.



In this section, we explore how to perform dual numbers reverse AD on source programs that contain fork-join task parallelism using a simple, but representative, parallel combination construct $(\parCom)$ that has the semantics that $s \parCom t$ computes $s$ and $t$ in parallel and returns the pair $(s, t)$.
We discuss a different ID assignment scheme for backpropagators that takes parallelism into account, and we show that we can take advantage of these new IDs to resolve backpropagators in parallel during the reverse pass.

\subsection{Fork-Join Parallelism}\label{sec:parallel-combinator}

We work with a parallel operational model in the fork-join style.
Roughly speaking: to run two subcomputations in parallel, a task \emph{forks} into two sub-tasks; this pauses the parent task.
After the sub-tasks are done, they \emph{join} back into the parent task, which then resumes execution.
A task may fork as many times as it likes.
Each individual sequential section of execution (i.e.\ the part of a task before its first fork, between its join and the next fork, etc., and finally the part after the last join) we call a \emph{job}.
Each job in the program gets a fresh job ID that is its unique identifier.
The intent is that independent jobs can execute in parallel on different CPU cores.
A typical runtime for this model is a thread pool together with a job queue to distribute work over the operating system threads: new jobs are submitted to the queue when they are created, and threads from the pool pick up waiting jobs from the queue when idle.

Concretely, we extend our source language syntax with a parallel pairing construct $(\parCom)$ with the following typing rule:
\[
  \frac{\Gamma \vdash s : \sigma \qquad \Gamma \vdash t : \tau}
       {\Gamma \vdash s \parCom t : (\sigma, \tau)}
\]
Operationally, when we encounter the term $t_1 \parCom t_2$ while evaluating some term $t$ in job $\alpha$, two new jobs are created with fresh job IDs $\beta$ and $\gamma$.
The term $t_1$ starts evaluating in job $\beta$, potentially does some forks and joins, and finishes in a (potentially) different job $\beta'$, returning a result $v$; $t_2$ starts evaluating in job $\gamma$ and finishes in $\gamma'$, returning a result $w$.
When $\beta'$ and $\gamma'$ terminate, evaluation of $t$ continues in a new job $\delta$ with the result $(v,w)$ for $t_1 \parCom t_2$.
We say that $\alpha$ \emph{forks} into $\beta$ and $\gamma$ and that $\beta'$ and $\gamma'$ \emph{join} into $\delta$.
The two parallel subgraphs, one from $\beta$ to $\beta'$ and one from $\gamma$ to $\gamma'$, we call \emph{tasks}.
The operational model does not know about tasks (it only knows about jobs), but we will use the concept of tasks in \cref{sec:parallel-monad} as a compositional building block for the job graph.
\Cref{fig:parallel-operational-model} contains a diagrammatic representation of the preceding paragraph.

\begin{figure}
  \begin{center}
  \vspace{0.1em}
  \begin{tikzpicture}[scale=0.5]
    \tikzset{every node/.style={font=\footnotesize, inner sep=0.4mm}}

    \draw[fill, color=gray!30] (2.65, 7.65) rectangle (3.35, 8.35);
    \draw[fill, color=gray!30] (1.65, 6.65) rectangle (2.35, 7.35);
    \draw[fill, color=gray!30] (3.4, 7.3) rectangle (4.6, 3.2);

    \node (a) at (3, 8) {$\alpha$};
    \node[inner sep=0.1mm] (b) at (2.0, 7) {$\beta$};
    \node (g) at (4.0, 7) {$\gamma$};
    \node (bl) at (1.6, 6) {};
    \node (br) at (2.4, 6) {};
    \node (bpl) at (1.6, 4.5) {};
    \node (bpr) at (2.4, 4.5) {};
    \node (bp) at (2.0, 3.4) {$\beta\mathrlap{\smash{'}}$};
    \node (gl) at (3.6, 6) {};
    \node (gr) at (4.4, 6) {};
    \node (gpl) at (3.6, 4.5) {};
    \node (gpr) at (4.4, 4.5) {};
    \node (gp) at (4.0, 3.5) {$\gamma\mathrlap{\smash{'}}$};
    \node (d) at (3, 2.5) {$\hspace{-1pt}\delta\hspace{1pt}$};

    \draw[->, -{Stealth[length=1.4mm]}] (a) -- ($(b.north east) + (0.05, 0)$);
    \draw[->, -{Stealth[length=1.4mm]}] (a) -- (g);
    \draw[->, -{Stealth[length=1.4mm]}] (b) -- (bl);
    \draw[->, -{Stealth[length=1.4mm]}] (b) -- (br);
    \draw[densely dashed] (bl) -- (bpl);
    \draw[densely dashed] (br) -- (bpr);
    \draw[->, -{Stealth[length=1.4mm]}] (bpl) -- (bp);
    \draw[->, -{Stealth[length=1.4mm]}] (bpr) -- (bp);
    \draw[->, -{Stealth[length=1.4mm]}] (g) -- (gl);
    \draw[->, -{Stealth[length=1.4mm]}] (g) -- (gr);
    \draw[densely dashed] (gl) -- (gpl);
    \draw[densely dashed] (gr) -- (gpr);
    \draw[->, -{Stealth[length=1.4mm]}] (gpl) -- (gp);
    \draw[->, -{Stealth[length=1.4mm]}] (gpr) -- (gp);
    \draw[->, -{Stealth[length=1.4mm]}] ($(bp.south east) + (-0.05, 0.13)$) -- (d);
    \draw[->, -{Stealth[length=1.4mm]}] (gp) -- (d);

    \node (jobtext) at (-0.3, 7.5) {jobs};
    \draw[->] (jobtext) .. controls (1.0, 8.0) and (1.9, 8.0) .. (2.65, 7.95);
    \draw[->] (jobtext) .. controls (0.6, 7.16) and (0.9, 7.1) .. (1.65, 7.05);

    \node (tasktext) at (6.5, 5.0) {task};
    \draw[->] (tasktext) .. controls (5.45, 5.25) and (5.2, 5.3) .. (4.6, 5.33);
  \end{tikzpicture}
  \end{center}
  \vspace{-1.1em}
  \caption{\label{fig:parallel-operational-model}
    Schematic view of the operational model underlying $(\parCom)$.
  }
\end{figure}

The algorithm in this section extends readily to $n$-ary parallel tupling constructs:
\[
  \frac{\Gamma \vdash t_1 : \tau_1 \quad \cdots \quad \Gamma \vdash t_n : \tau_n}
       {\Gamma \vdash (t_1, \ldots, t_n)^\parCom : (\tau_1, \ldots, \tau_n)}
\]
or even parallel combination constructs of dynamic arity, where $[\tau]$ denotes lists of $\tau$:
\[
  \frac{\Gamma \vdash t : [() \to \tau]}
       {\Gamma \vdash \parCom t : [\tau]}
\]
but for simplicity of presentation, we restrict ourselves here to binary forking.

\subsection{Parallel IDs and Their Partial Order}\label{sec:parallel-ids}

\newcommand\footnoteExplicitParallelism{
  Because our source language is pure, one could in principle detect and exploit implicit parallelism.
  We focus on explicit parallelism here because automatic parallelism extraction is (difficult and) orthogonal to this work.
  In effect, the dependency graph that we construct is a weakening of the perfectly accurate one: computations within a job are assumed sequentially dependent.
  The reverse pass in \cref{sec:parallelism-wrapper} simply walks our constructed dependency graph, exploiting all apparent parallelism; $\ResolveStaged$ there only inherits the concept of tasks and jobs because we encode the graph in a particular way that makes use of that structure (\cref{sec:parallel-monad}).
}
To avoid discarding the (explicit,\footnote{\footnoteExplicitParallelism} using $(\parCom)$) parallelism structure in the source program, we have to somehow record the dependency graph of the backpropagators (the same graph as the computation graph of scalars in the source program) in a way that is more precise than the chronological linear order used for the sequential algorithm.
Specifically, we want backpropagators that were created in parallel jobs in the forward pass to not depend on each other in the dependency graph, not even transitively.
In other words, they should be incomparable in the partial order that informs the reverse pass ($\ResolveStaged$) what backpropagator to resolve next.

To support the recording of this additional dependency information, we switch to \emph{compound IDs}, consisting of two integers instead of one:
\begin{itemize}
  \item
    The \emph{job ID} that uniquely identifies the job a backpropagator is created from.
    This requires that we have a way to generate a unique ID in parallel every time a job forks or two jobs join.
    Job IDs do \emph{not} carry a special partial order; see below.
  \item
    The \emph{ID within the job}, which we assign sequentially (starting from 0) to all backpropagators created in a job.
    The operations within one forward-pass job are sequential (because of our fork-join model where a fork ends a job; see the previous subsection); this ID-within-job simply witnesses this.
\end{itemize}
Compound IDs have lexicographic order: $(\alpha,i) \leqc (\beta, i')$ iff $\alpha \leqj \beta \land (\alpha \not= \beta \lor i \leqz i')$.
The order on sequential IDs (within a job) is simply the standard linear order $\leqz$, but the partial order on job IDs is different and is instead defined as the transitive closure of the following three cases:
\begin{enumerate}
\item $\alpha \leqj \alpha$;
\item If $\alpha$ forks into $\beta$ and $\gamma$, then $\alpha \leqj \beta$ and $\alpha \leqj \gamma$;
\item If $\alpha$ and $\beta$ join into $\gamma$, then $\alpha \leqj \gamma$ and $\beta \leqj \gamma$.
\end{enumerate}

In \cref{fig:example:partial-order}, we give an example term together with graphs showing the (generators of the) partial orders $\leqj$ on job IDs and $\leqc$ on compound IDs.
Both graphs have arrows pointing to the successors of each node: $n_2$ is a successor of $n_1$ (and $n_1$ a predecessor of $n_2$) if $n_1 < n_2$ (i.e.\ $n_1 \leq n_2$ and $n_1 \not= n_2$) and there is no $m$ such that $n_1 < m < n_2$.
The graphs generate their respective partial orders if one takes the transitive closure and includes trivial self-loops.
For \cref{fig:example:partial-order:job} ($\leqj$), the arrows thus show the fork/join relationships; for \cref{fig:example:partial-order:compound} ($\leqc$), this is refined with the linear order on sequential IDs within each job.
Note that these compound IDs replace the integer IDs of the sequential algorithm of \crefrange{sec:naive}{sec:improve}; that is to say: the result of every primitive operation gets a unique ID.

\begin{figure}
  \centering
  \begin{subfigure}[b]{0.49\textwidth}
    \(
      \mathbf{let}\ (z_1, z_2)=\\
      \phantom{....}\Big(\mathbf{let}\ x=a*b+\sin(b)\\ 
      \phantom{....\Big(}\mathbf{in} \ \mathbf{let}\ (y_1,y_2)= (x*a+a)\parCom \cos(x)\\
      \phantom{....\Big(}\mathbf{in}\ x + y_1*y_2\Big) \parCom \bigl(\exp(a) + b\bigr)\\
      \mathbf{in}\ z_1 * z_2
    \)
    \vspace{0.95cm}
    \caption{\label{fig:example:partial-order:term}
      Example program
    }
  \end{subfigure}
  \begin{subfigure}[b]{0.18\textwidth}
    \begin{tikzpicture}[scale=0.7]
      \tikzset{every node/.style={font=\small, inner sep=0.4mm}}
      \node (al) at (5.74, 6) {$\alpha$};
      \node (be) at (5, 5) {$\beta$};
      \node (de) at (4.4, 4) {$\delta$};
      \node (ep) at (5.6, 4) {$\varepsilon$};
      \node (ze) at (5, 3) {$\zeta$};
      \node (ga) at (6.5, 4) {$\gamma$};
      \node (he) at (5.8, 2) {$\eta$};
      \draw[->, -{Stealth[length=1.4mm]}] (al) -- (be);
      \draw[->, -{Stealth[length=1.4mm]}] (al) -- (ga);
      \draw[->, -{Stealth[length=1.4mm]}] (be) -- (de);
      \draw[->, -{Stealth[length=1.4mm]}] (be) -- (ep);
      \draw[->, -{Stealth[length=1.4mm]}] (de) -- (ze);
      \draw[->, -{Stealth[length=1.4mm]}] (ep) -- (ze);
      \draw[->, -{Stealth[length=1.4mm]}] (ze) -- (he);
      \draw[->, -{Stealth[length=1.4mm]}] (ga) -- (he);
    \end{tikzpicture}
    \vspace{0.9cm}
    \caption{\label{fig:example:partial-order:job}
      Job IDs ($\leqj$)
    }
  \end{subfigure}
  \begin{subfigure}[b]{0.28\textwidth}
    \begin{tikzpicture}[scale=0.5]
      \tikzset{every node/.style={font=\footnotesize, inner sep=0.4mm}}

      \node (b0) at (2.5,16.75) {$(\beta, 0)$};
      \node (b1) at (2.5,15.5) {$(\beta, 1)$};
      \node (b2) at (2.5,14.25) {$(\beta, 2)$};
      \node (d0) at (1.25,13) {$(\delta, 0)$};
      \node (d1) at (1.25,11.75) {$(\delta, 1)$};
      \node (e0) at (3.75,12.25) {$(\varepsilon, 0)$};
      \node (z0) at (2.5,10.5) {$(\zeta, 0)$};
      \node (z1) at (2.5,9.25) {$(\zeta, 1)$};
      \node (g0) at (5.7,16.75) {$(\gamma, 0)$};
      \node (g1) at (5.7,9.25) {$(\gamma, 1)$};
      \node (h0) at (4.1,8) {$(\eta, 0)$};

      \draw[->, -{Stealth[length=1.4mm]}] (b0) -- (b1);
      \draw[->, -{Stealth[length=1.4mm]}] (b1) -- (b2);
      \draw[->, -{Stealth[length=1.4mm]}] (b2) -- (d0);
      \draw[->, -{Stealth[length=1.4mm]}] (d0) -- (d1);
      \draw[->, -{Stealth[length=1.4mm]}] (b2) -- (e0);
      \draw[->, -{Stealth[length=1.4mm]}] (d1) -- (z0);
      \draw[->, -{Stealth[length=1.4mm]}] (e0) -- (z0);
      \draw[->, -{Stealth[length=1.4mm]}] (z0) -- (z1);
      \draw[->, -{Stealth[length=1.4mm]}] (z1) -- (h0);
      \draw[->, -{Stealth[length=1.4mm]}] (g0) -- (g1);
      \draw[->, -{Stealth[length=1.4mm]}] (g1) -- (h0);
    \end{tikzpicture}
    \caption{\label{fig:example:partial-order:compound}
      Compound IDs ($\leqc$)
    }
  \end{subfigure}
  \caption{\label{fig:example:partial-order}
    An example program.
    Note that the program starts by forking, before performing any primitive operations, hence job $\alpha$ is empty and the partial order on compound IDs happens to have multiple minimal elements.
  }
\end{figure}

The reverse pass ($\ResolveStaged$) needs to traverse the dependency graph on compound IDs (\cref{fig:example:partial-order:compound}) in reverse dependency order (taking advantage of task parallelism), but it is actually unnecessary to construct the full graph at runtime.
It is sufficient to construct the dependency graph on \emph{job IDs} (\cref{fig:example:partial-order:job}) together with, for each job $\alpha$, the number $n_\alpha$ of sequential IDs generated in that job (note that this number may be zero if no scalar computation was done while running in that job).
With this information, $\ResolveStaged$ can walk the job graph in reverse topological order, resolving parallel tasks in parallel, and for each job $\alpha$ sequentially resolve the individual backpropagators from $n_\alpha - 1$ to $0$.
We will collect this additional information during the forward pass by extending the monad that the forward pass runs in.

\subsection{Extending the Monad}\label{sec:parallel-monad}

In the first optimisation that we applied to the (sequential) algorithm, namely linear factoring via staging of backpropagators (\cref{sec:staging}, \cref{fig:algo-monadic}), we modified the code transformation to produce code that runs in an state monad $\fwdM \tau = \ID \ra (\tau, \ID)$, for $\ID = \Z$:
\[
  \text{If}\ \Gamma \vdash t : \tau
  \ \text{then}\ \trans2c[\Gamma] \vdash \trans2c[t] : \fwdM \trans2c[\tau]
\]
In fact, this state monad was simply the natural implementation of an \emph{ID generation monad} with one method:
\[
  \begin{array}{@{}l@{\ }c@{\ }l@{}}
    \genID &:& \fwdM \ID \\
    \genID &=& \fun(i : \Z).\ (i+1, i)
  \end{array}
\]
We saw above in \cref{sec:parallel-ids} that we need to extend this monad to be able to do two things: (1) generate compound IDs, not just sequential IDs, and (2) record the job graph resulting from parallel execution using $(\parCom)$.
Write $\JID$ for the type of job IDs (in an implementation we can simply set $\JID \coloneqq \Z$) and write $\CID \coloneqq (\JID, \Z)$ for the type of compound IDs.
The extended monad needs two methods:
\[
  \begin{array}{@{}r@{\ }c@{\ }l@{}}
    \genID &:& \fwdM \CID \\
    (\parCom') &:& \fwdM \sigma \to \fwdM \tau \to \fwdM {(\sigma, \tau)}
  \end{array}
\]
The updated $\genID$ reads the current job ID from reader context inside the monad and pairs that with a sequential ID generated in the standard fashion with monadic state.
The monadic parallel combination method, $(\parCom')$, generates some fresh job IDs by incrementing an atomic mutable cell in the monad, runs the two monadic computations \emph{in parallel} by spawning jobs as described in \cref{sec:parallel-combinator}, and records the structure of the job graph thus created in some state inside the monad.

In an implementation, one can take the definitions in \cref{fig:fwdm-implementation}.
Working in Haskell, we write \texttt{IORef} for a mutable cell and use \IO as the base monad in which we can access that mutable cell as well as spawn and join parallel threads.
(We do not use \IO in any other way in $\fwdM$, although $\ResolveStaged$, which runs after the forward pass, will also use mutable arrays as in \cref{sec:mutarrays-io}.)
The given implementation of $\fwdM$ has the atomic mutable cell in a reader context, and is a state monad in `$\JobDescr$': a description of a job that contains the \emph{history} of a job together with its job ID and the number of sequential IDs generated in that job (numbered $0, \ldots, n - 1$).
The history of a job $\alpha$ is the subgraph of the job graph given by all jobs $\beta$ satisfying $\beta < \alpha$ in the smallest \emph{task} (recall \cref{fig:parallel-operational-model}) containing the job.
For the special case of the (unique) last job of a task, its history is precisely the whole task excluding itself.
This definition of a ``history'' ensures that $(\parCom')$ has precisely the parts of the job graph that it needs to build up the job graph of the task that it itself is running in, which makes everything compose.

\begin{figure}
  \newcommand\oversetJD[1]{\mathrlap{\smash{\overset{(\text{#1})}{\JobDescr}}}\hphantom{\JobDescr}}
  \[
    \begin{array}{@{}c@{}}
      \fwdM \alpha = \texttt{IORef}\ \JID \ra \JobDescr \ra \IO\ (\JobDescr, \alpha) \\[0.5em]
      \begin{array}{@{}l@{\qquad}l@{}}
        \begin{array}{@{}c@{}}
          \JobDescr = (\History, \JID, \Z) \\[0.5em]
          \mathbf{data}\ \History
            \begin{array}[t]{@{\;}c@{\;}l@{}}
              =& \HStart \\
              \mid& \HFork\ \oversetJD{A}\ \oversetJD{B}\ \oversetJD{C}
            \end{array}
          \end{array}
        &
        \mathrlap{\raisebox{-0.62cm}{\begin{tikzpicture}[scale=0.5]
          \tikzset{every node/.style={font=\footnotesize, inner sep=0.4mm}}
          \node (A) at (0, 2) {A};
          \node (B) at (-0.5, 1) {B};
          \node (C) at (0.5, 1) {C};
          \node[inner sep=0mm] (cur) at (0, 0) {current job};
          \draw[->, -{Stealth[length=1.4mm]}] (A) -- (B.north);
          \draw[->, -{Stealth[length=1.4mm]}] (A) -- (C.north);
          \draw[->, -{Stealth[length=1.4mm]}] (B.south) -- (cur);
          \draw[->, -{Stealth[length=1.4mm]}] (C.south) -- (cur);
        \end{tikzpicture}}}
      \end{array} \\[2em]
      \begin{array}{@{}l@{}}
        f \parCom' g = \fun \mathit{ref}\ \mathit{jd}_0.\ \mathbf{do} \\
        \qquad \begin{array}[t]{@{}l@{}}
          (j_1, j_2, j_3) \leftarrow \texttt{atomicModifyIORef'}\ \mathit{ref}\ (\fun j.\ (j + 3, (j, j+1, j+2))) \\
          ((\mathit{jd}_1, x), (\mathit{jd}_2, y)) \leftarrow
            \texttt{inParallel}\ (f\ \mathit{ref}\ (\HStart, j_1, 0))\ (\mathrlap{g}\hphantom{f}\ \mathit{ref}\ (\HStart, j_2, 0)) \\
          \mathbf{return}\ ((\HFork\ \mathit{jd}_0\ \mathit{jd}_1\ \mathit{jd}_2, j_3, 0), (x, y))
        \end{array} \\
        \quad \text{where:}\ \begin{array}[t]{@{}l@{}}
          \texttt{atomicModifyIORef'} :: \texttt{IORef}\ \alpha \ra (\alpha \ra (\alpha, \beta)) \ra \IO\ \beta \\
          \texttt{inParallel} :: \IO\ \alpha \ra \IO\ \beta \ra \IO\ (\alpha, \beta)
        \end{array}
      \end{array}
    \end{array}
  \]
  \caption{\label{fig:fwdm-implementation}
    Sketch of the implementation of the monad $\fwdM$.
    The diagram shows the meaning of the job descriptions in `$\HFork$': the first field (labeled `$A$') contains the history up to the last fork in this task (excluding subtasks), and the fields labeled $B$ and $C$ describe the subtasks spawned by that fork.
    The first job in a task has no history, indicated with `$\HStart$'.
  }
\end{figure}

\myparagraph{Differentiation}
We keep the differentiation rules for all existing language constructs the same, except for changing the type of IDs to $\CID$ and using the monad $\fwdM$ instead of doing manual state passing of the next ID to generate.
A representative rule showing how this looks is: (compare \cref{fig:algo-monadic})
\[
  \trans5c[(s, t)] = \trans5c[s] \mbind \fun x.\ \trans5c[t] \mbind \fun y.\ \mathbf{return}\ (x, y)
\]
Because we now generate fresh compound IDs rather than plain integer IDs when executing primitive operations, we change $\Z$ to $\CID$ in $\trans4c[\R]$:\footnote{Here we work from the transformation $\trans4{}$ as described in \cref{sec:mutarrays-io} on monadic mutable arrays. With the defunctionalisation described in \cref{sec:improve-defunctionalisation}, the backpropagator would simply read `$\Contrib$' instead.}
\[
  \trans5c[\R] = (\R, (\textcolor{red}{\CID}, \lR \lra (\Staged c \ra \IO\ ())))
\]
The new rule for the parallel pairing construct is simply:
\[
  \trans5c[t\parCom s] = \trans5c[t] \parCom' \trans5c[s]
\]
This is the only place where we use the operation $(\parCom')$, and thus the only place in the forward pass where we directly use the extended functionality of our monad $\fwdM$.

\subsection{Updating the Wrapper}\label{sec:parallelism-wrapper}

The wrapper is there to glue the various components together and to provide the (now parallel) definition of $\ResolveStaged$.
We discuss the main ideas behind the parallel wrapper implementation in this section, skipping over some implementation details.

In \cref{sec:mutarrays}, the backpropagators staged their backpropagator calls in a single array, indexed by their ID.
With compound IDs, we still need one array slot for each backpropagator, meaning that we need a \emph{nested} staging array:
\[
  \Staged c = (\Array \lR, \Array {(\Array {(\lR \lra (\Staged c \ra \IO\ ()), \lR)})})
\]
where the outer array is indexed by the job ID and the inner by the sequential ID within that job.
Because jobs have differing lengths, the nested arrays also have differing lengths and we cannot use a rectangular two-dimensional array.
The implementation of $\StagedOneHot$ must be changed to update the cotangent collection array atomically, because backpropagators, and hence $\StagedOneHot$, will now be called from multiple threads.
$\StagedCall$ needs a simple modification to (atomically) modify the correct element in the now-nested staging array.

To construct the initial $\Staged$ object, $\StagedAlloc$ needs to know the correct length of all the nested staging arrays; it can get this information from the job graph (in the $\History$ structure) that $\Wrap{}$ receives from the monadic evaluation of the forward pass.

This leaves the parallel implementation of $\ResolveStaged$.
The idea here is to have two functions: one that resolves a \emph{task} (this one is new and handles all parallelism), and one that resolves a \emph{job} (and is essentially identical to the last version of $\ResolveStaged$, from \cref{fig:algo-mutarrays}).
An example implementation is given in \cref{fig:resolve-parallel}, where the first function is called $\mathit{resolveTask}$ and the second $\mathit{resolveJob}$.
In this way, $\ResolveStaged$ traverses the job graph (\cref{fig:example:partial-order:job}) from the terminal job backwards to the initial job, doing the usual sequential resolving process for each sequential job in the graph.

\begin{figure}
  \[
    \begin{array}{@{}l@{}}
      \ResolveStaged : \JobDescr \ra \Staged c \ra \IO\ (\IArray \lR) \\
      \ResolveStaged\ \mathit{jd}\ (c, m) = \mathit{resolveTask}\ \mathit{jd} \\
      \quad \mathrm{where}\ \begin{array}[t]{@{}l@{}}
        \mathit{resolveTask}\ (\mathit{history}, \mathit{jid}, i) = \mathbf{do} \\
        \quad \begin{array}[t]{@{}l@{}}
          \mathit{jobarr} \leftarrow \arrget\ \mathit{jid}\ m \\
          \mathit{resolveJob}\ (i - 1)\ \mathit{jobarr} \\
          \mathbf{case}\ \mathit{history}\ \mathbf{of}\ \begin{array}[t]{@{}l@{}}
            \HStart \ra \mathbf{return}\ () \\
            \HFork\ \mathit{jd}_0\ \mathit{jd}_1\ \mathit{jd}_2 \ra \mathbf{do} \\
            \quad \texttt{inParallel}\ (\mathit{resolveTask}\ \mathit{jd}_1)\ (\mathit{resolveTask}\ \mathit{jd}_2) \\
            \quad \mathit{resolveTask}\ \mathit{jd}_0
          \end{array} \\
        \end{array} \\
        ~ \\[-0.8em]
        \mathit{resolveJob}\ (-1)\ \mathit{arr} = \mathbf{return}\ () \\
        \mathit{resolveJob}\ i\ \mathit{arr} = \mathbf{do} \\
        \quad \begin{array}[t]{@{}l@{}}
          (f, a) \leftarrow \arrget\ i\ \mathit{arr} \\
          f\ a\ (c, m) \\
          \mathit{resolveJob}\ (i - 1)\ \mathit{arr}
        \end{array}
      \end{array}
    \end{array}
  \]
  \caption{\label{fig:resolve-parallel}
    Implementation of $\ResolveStaged$ for the parallel-ready dual-numbers reverse AD algorithm.
    The \texttt{inParallel} function is as in \cref{fig:fwdm-implementation}.
  }
\end{figure}

\myparagraph{Duality}
The usual mantra in reverse-mode AD is that ``sharing in the primal becomes addition in the dual''.
When parallelism comes into play, not only do we have this duality in the data flow graph, we also get an interesting duality in the control-flow graph: $\ResolveStaged$ forks where the primal joined, and joins where the primal forked.
Perhaps we can add a mantra for task-parallel reverse AD: ``forks in the primal become joins in the dual''.

\section{Implementation}\label{sec:implementation}

To show the practicality of our method, we provide a prototype implementation\footnote{The code is available at \url{https://github.com/tomsmeding/ad-dualrev-th}.} of the parallel algorithm of \cref{sec:parallelism}, together with the improvements from \cref{sec:improve-one-array,sec:improve-defunctionalisation}, that differentiates a sizeable fragment of Haskell98 including recursive types (reinterpreted as a strict, call-by-value language) using Template Haskell.
As described in \cref{sec:mutarrays-io}, we realise the mutable arrays using \texttt{IOVector}s.
The implementation does not incorporate the changes given in \cref{sec:improve-taping} that transform the algorithm into classical taping (because implementations of taping already abound), but it does include support for recursive functions, coproduct types, and user-defined data types as described in \cref{sec:source-language-extension}.

Template Haskell~\citep{fp-2002-template-haskell} is a built-in metaprogramming facility in GHC Haskell that (roughly) allows the programmer to write a Haskell function that takes a block of user-written Haskell code, do whatever it wants with the AST of that code, and finally splice the result back into the user's program.
The resulting code is still type-checked as usual.
The AST transformation that we implement is, of course, differentiation.

\myparagraph{Benchmarks}
To check that our implementation has reasonable performance in practice, we benchmark (in \texttt{bench/Main.hs}) against Kmett's Haskell \texttt{ad} library~\citep{ad-2021-kmett-hackage} (version 4.5.6) on a few basic functions.\footnote{
  We use \texttt{Numeric.AD.Double} instead of \texttt{Numeric.AD} to allow \texttt{ad} to specialise for \texttt{Double}, which we also do.
  We keep \texttt{ad}'s (non-default) \texttt{ffi} flag off for a fairer playing ground (we could do similar things, but do not).
}
These functions are the following (abbreviating \texttt{Double} as `\texttt{D}'):
\newcommand\rotatevecbyquat{rotate\textunderscore{}vec\textunderscore{}by\textunderscore{}quat\xspace}
\begin{itemize}
\item A single scalar multiplication of type \texttt{(D, D) -> D};
\item Dot product of type \texttt{([D], [D]) -> D};
\item Matrix--vector multiplication, then sum: of type \texttt{([[D]], [D]) -> D};
\item The \texttt{\rotatevecbyquat} example from \citep{ad-2021-krawiec-kmett-ad} of type \texttt{(Vec3 D, Quaternion D) -> Vec3 D}, with \texttt{data Vec3 s = Vec3 s s s} and \texttt{data Quaternion s = Quaternion s s s s};
\item A simple, dense neural network with 2 hidden layers, ReLU activations and (safe) softmax output processing.
  The result vector is summed to make the output a single scalar.
  The actual Haskell function is generic in the number of layers and has type \texttt{([([[D]], [D])], [D]) -> D}: the first list contains a matrix and a bias vector for each hidden layer, and the second tuple component is the input vector.
  In the benchmark, the input has length 50 and the two hidden layers have size 100 and 50.
\item Simulation of 4 particles in a simple force field with friction for 1000 time steps; this example has type \texttt{[((D, D), (D, D))] -> D}.
  The input is a list (of length 4) of initial positions and velocities; the output is $\sum_{(x,y)} x \cdot y$ ranging over the 4 result positions $p$, to ensure that the computation has a single scalar as output.
  The four particles are simulated in parallel using the $(\parCom)$ combinator from \cref{sec:parallelism}.
\end{itemize}
The fourth test case, \rotatevecbyquat, has a non-trivial return type; the benchmark executes its reverse pass three times (`3' being the number of scalars in the function output) to get the full Jacobian.
The fifth test case, `particles', is run on 1, 2 and 4 threads, where the ideal result would be perfect scaling due to the four paricles being independent.

\begin{table}
  \caption{\label{fig:bench-results}
    Benchmark results of \cref{sec:parallelism} + \cref{sec:improve-one-array,sec:improve-defunctionalisation} versus \texttt{ad-4.5.6}.
    The `TH' and `\texttt{ad}' columns indicate runtimes on one machine for our implementation and the \texttt{ad} library, respectively.
    The last column shows the ratio between the previous two columns.
    We give the size of the largest side of \texttt{criterion}'s 95\% bootstrapping confidence interval, rounded to 2 decimal digits.
    Setup: GHC 9.6.6 on Linux, Intel i9-10900K CPU, with Intel Turbo Boost disabled (i.e.\ running at a consistent 3.7 GHz).
  }
  \centering
  \begin{tabular}{l|r@{}l@{\ }lr@{}l@{\ }lc}
    & \multicolumn{3}{c}{TH} & \multicolumn{3}{c}{\texttt{ad}} & TH / \texttt{ad} \\ \hline
    scalar mult. & 0&.33 & $\mu$s $\pm0.00$ & 0&.80 & $\mu$s $\pm0.00$ & $\approx$0.4 \\
    dot product & 0&.95 & $\mu$s $\pm0.03$ & 2&.14 & $\mu$s $\pm0.07$ & $\approx$0.4 \\
    sum-mat-vec & 0&.59 & $\mu$s $\pm0.02$ & 1&.23 & $\mu$s $\pm0.03$ & $\approx$0.5 \\
    \rotatevecbyquat & 5&.62 & $\mu$s $\pm0.00$ & 6&.71 & $\mu$s $\pm0.01$ & $\approx$0.8 \\
    neural & 2&.4 & ms $\pm0.05$ & 8&.1 & ms $\pm0.01$ & $\approx$0.3 \\
    particles (1 thr.) & 7&.6 & ms $\pm0.05$ & 9&.1 & ms $\pm0.08$ & $\approx$0.8 \\
    particles (2 thr.) & 4&.6 & ms $\pm0.04$ & \multicolumn{3}{c}{\textcolor{gray}{---}} & \textcolor{gray}{---} \\
    particles (4 thr.) & 2&.4 & ms $\pm0.12$ & \multicolumn{3}{c}{\textcolor{gray}{---}} & \textcolor{gray}{---}
  \end{tabular}
\end{table}

The benchmark results are summarised in \cref{fig:bench-results} and shown in detail (including evidence of the desired linear complexity scaling) in \cref{sec:criterion-benchmark-details}.
The benchmarks are measured using the \texttt{criterion}\footnote{By Bryan O'Sullivan: \url{https://hackage.haskell.org/package/criterion}} library.
To get statistically significant results, we measure how the timings scale with increasing $n$:
\begin{itemize}
\item Scalar multiplication, \rotatevecbyquat{}, the neural network and the particle simulation are simply differentiated $n$ times;
\item Dot product is performed on lists of length $n$;
\item Matrix multiplication is done for a matrix and vector of size $\sqrt n$, to get linear scaling in $n$.
\end{itemize}
The reported time is the deduced actual runtime for $n = 1$.

By the results in \cref{fig:bench-results}, we see that on these simple benchmark programs, our implementation is faster than the existing \texttt{ad} library.
While this is encouraging, it is not overly surprising: because our algorithm is implemented as a compile-time code transformation, the compiler is able to optimise the derivative code somewhat before it gets executed.

Of course, performance results may well be different for other programs, and AD implementations that have native support for array operations can handle some of these programs much more efficiently.
Furthermore, there are various implementation choices for the code transformation that may seem relatively innocuous but have a large impact on performance (we give some more details below in \cref{sec:implementation-considerations}).

For these reasons, our goal here is merely to substantiate the claim that the implementation exhibits constant-factor performance in the right ballpark (in addition to it having the right asymptotic complexity, as we have argued).
Nevertheless, our benchmarks include key components of many AD applications, and seeing that we have not at all special-cased their implementation (the implementation knows only primitive arithmetic operations such as scalar addition, multiplication, etc.),\ we believe that they suffice to demonstrate our limited claim.

\subsection{Considerations for Implementation Performance}\label{sec:implementation-considerations}

The target language of the code transformation in our implementation is Haskell, which is a lazy, garbage-collected language.
This has various effects on the performance characteristics of our implementation.

\myparagraph{Garbage collection}
The graph of backpropagators (a normal data structure, `$\Contrib$', since \cref{sec:improve-defunctionalisation}) is a big data structure of size proportional to the number of scalar operations performed in the source program.
While this data structure grows during the forward pass, the nursery (zeroth generation) of GHC's generational garbage collector (GC) repeatedly fills up, triggering garbage collection of the heap.
Because a GC pass takes time on the order of the amount of live data on the heap, these passes end up very expensive: a naive taping reverse AD algorithm becomes \emph{quadratic} in the source program runtime.
Using a GHC runtime system flag (e.g.~\texttt{-A500m}) to set the nursery size of GHC's GC sufficiently large to prevent the GC from running during a benchmark (\texttt{criterion} runs the GC explicitly between each benchmark invocation), timings on some benchmarks above decrease significantly: on the largest benchmark (particles), this can save 10\% off \texttt{ad}'s runtime and 25\% off our runtime (though precise timings vary).
The times reported in \cref{fig:bench-results} are with GHC's default GC settings.

While this is technically a complexity problem in our implementation, we gloss over this because it is not fundamental to the algorithm: the backpropagator graph does not contain cycles, so it could be tracked with reference-counting GC to immediately eliminate the quadratic blowup.
Using a custom, manual allocator, one could also eliminate all tracking of the liveness of the graph because we know from the structure of the algorithm exactly when we can free a node in the graph (namely when we have visited it in the reverse pass).
Our reference implementation does not do these things to be able to focus on the workings of the algorithm.

\myparagraph{Laziness}
Because data types are lazy by default in Haskell, a naive encoding of the $\Contrib$ data type from \cref{sec:improve-defunctionalisation} would make the whole graph lazily evaluated (because it is never demanded during the forward pass).
This results in a significant constant-factor overhead (more than $2\times$), and furthermore means that part of the work of the forward pass happens when the reverse pass first touches the top-level backpropagator; this work then happens sequentially, even if the forward pass was meant to be parallel.
To achieve proper parallel scaling, it was necessary to make the $\Contrib$ graph strict, and furthermore to make the $\trans{}c[\R] = (\R, (\CID, \Contrib))$ triples strict, to ensure that the graph is fully evaluated as it is being \emph{constructed}, not when it is demanded in the reverse pass.

Using some well-chosen \verb|{-# UNPACK #-}| pragmas on some of these fields also had a significant positive effect on performance.

\myparagraph{Thread pool}
In \cref{sec:parallelism} we used an abstract \texttt{inParallel} operation for running two jobs in parallel, assuming some underlying thread pool for efficient evaluation of the resulting parallel job graph.
In a standard thread pool implementation, spawning tasks from within tasks can result in deadlocks.
Because nested tasks are essential to our model of task parallelism, the implementer has to take care to further augment $\fwdM$ from \cref{sec:parallelism} to be a continuation monad as well: this allows the continuation of the \texttt{inParallel} operation to be captured and scheduled separately in the thread pool, so that thread pool jobs are indeed individual \emph{jobs} as defined in \cref{sec:parallelism}.

The GHC runtime system has a thread scheduler that should handle this completely transparently, but in our tests it was not eager enough in assigning virtual Haskell threads to actual separate kernel threads, resulting in a sequential benchmark.
This motivated a (small) custom thread pool implementation that sufficed for our benchmarks, but has a significant amount of overhead that can be optimised with more engineering effort.

\myparagraph{Imperfect scaling}
Despite efforts to the contrary, it is evident from \cref{fig:bench-results} that even on an embarrassingly parallel task like `particles', the implementation does not scale perfectly.
From inspection of more granular timing of our implementation, we suspect that this is a combination of thread pool overhead and the fact that the forward pass simply allocates too quickly, exhausting the memory bandwidth of our system when run sufficiently parallel.
However, more research is needed here to uncover the true bottlenecks.

\section{Conclusions}

One may ask: if the final algorithm from \cref{sec:mutarrays} can be argued to be ``just taping'' (\cref{sec:improve-taping}), which is already widely used in practice---and the parallel extension is \emph{still} just taping, except on a non-linear tape---what was the point?
The point is the observation that optimisations are key to implementing efficient AD and that
multiple kinds of reverse AD algorithms proposed by the programming languages community (in particular the one from \cref{fig:algo-naive}, studied by~\cite{ad-2020-dualnum-revad-linear-factoring} and~\cite[Section 6]{ad-2020-sam-mathieu-matthijs}---for further examples, see below in \cref{sec:litrev-krawiec}) tend to all reduce to taping after optimisation.
We hope to have demonstrated that these techniques are quite flexible, allowing the differentiation of rich source languages with dynamic control flow, recursion and parallelism, and that the resulting algorithm can be relatively straightforward by starting from a naive differentiation algorithm and next optimising it to achieve the desired complexity.

The first of our optimisations (linear factoring) is quite specific to starting AD algorithms that need some kind of distributive law to become efficient (e.g.\ also~\citep{ad-2021-krawiec-kmett-ad}).
However, we think that the other optimisations are more widely applicable (and are, for example, also related to the optimisations that fix the time complexity of CHAD~\citep{chad-efficient-popl}): sparse vectors are probably needed in most functional reverse AD algorithms to efficiently represent the one-hot vectors resulting from projections (\texttt{fst}/\texttt{snd} as well as random access into arrays, through indexing), and mutable arrays are a standard solution to remove the ubiquitous log-factor in the complexity of purely functional algorithms.

If one desires to take the techniques in this paper further to an algorithm that is useful in practice, it will be necessary to add \emph{arrays} of scalars as a primitive type in the transformation.
This would allow us to significantly reduce the size of the allocated tape, and reuse much more of the structure of the original program in the reverse pass.
Since many AD applications tend to involve very large arrays of scalars, we expect to be able to gain significant constant factor performance by replacing an array of many scalar backpropagators with a single vector backpropagator.
We are currently exploring this direction of research.

\section{Origins of Dual-Numbers Reverse AD, Relationship With Vectorised Forward AD and Other Related Work}\label{sec:related-work}

The literature about automatic differentiation spans many decades and academic subcommunities (scientific computing, machine learning and---most recently---programming languages).
Important early references are \citep{ad-1964-ad,linnainmaa1970representation,ad-1980-ad}.
Good surveys can be found in \citep{ad-2018-survey-automatic-differentiation,ad-2018-survey-ad-implementation}.
In the rest of this section, we focus on the more recent literature that studies AD from a programming languages (PL) point of view, to extend the scope of our discussion in \cref{sec:improve}.

\subsection{Theoretical Foundations for Our Algorithm}
The first mention that we know of the naive dual-numbers reverse mode AD algorithm  that we analyse in this paper 
is \cite[page 12]{ad-2008-reverse-functional-ad}, where it is quickly dismissed before a different technique is pursued.
The algorithm is first thoroughly studied by~\cite{ad-2020-dualnum-revad-linear-factoring}
using operational semantics and in \cite[Section 6]{ad-2020-sam-mathieu-matthijs} using denotational semantics.
\cite{ad-2020-dualnum-revad-linear-factoring} introduce the key idea that underlies the optimisations in our paper: the linear factoring rule, stating that a term $f\ x + f\ y$, with $f$ a linear function, may be reduced to $f\ (x + y)$.
We build on their use of this rule as a tool in a complexity proof to make it a suitable basis for a performant implementation.

\cite{ad-2021-dual-revad-linear-factoring-pcf} extend the work of \cite{ad-2020-dualnum-revad-linear-factoring} to apply to a language with term recursion, showing that dual-numbers reverse AD on PCF is almost everywhere correct.
Similarly, \cite{nunes-2024-dual-numbers} extend the work of \cite{ad-2020-sam-mathieu-matthijs} to apply to partial programs involving iteration, recursion and recursive types, thus giving a correctness proof for the initial dual-numbers reverse AD transformation of \cref{fig:algo-naive} applied to, essentially, idealised Haskell98.

\subsection{Vectorised Forward AD}\label{sec:litrev-krawiec}
\newcommand\vfad{\textsc{vfad}\xspace}
\newcommand\dnrad{\textsc{dnrad}\xspace}
There are strong parallels between our optimisations to the sequential algorithm in \cref{sec:naive,sec:staging,sec:cayley,sec:just-scalars,sec:mutarrays} and the derivation by~\cite{ad-2021-krawiec-kmett-ad}.
Like the present paper, they give a sequence of steps that optimise a simple algorithm to an efficient implementation---but the starting algorithm is vectorised forward AD (\vfad) instead of backpropagator-based dual-numbers reverse AD (\dnrad).
In our notation, their initial type transformation does not have $\trans1c[\R] = (\R, \lR\lra c)$, but instead $\trans1c[\R] = (\R, c)$.
(As befits a dual-numbers algorithm, the rest of the type transformation is simply structurally recursive.)

Linear algebra tells us that the vector spaces $\lR\lra c$ and $c$ are isomorphic, and indeed inspection of the term transformations shows that both naive algorithms compute the same thing.
Their operational behaviour, on the other hand, is very different: the complexity problem with \dnrad is exponential blowup in the presence of sharing, whereas \vfad is ``simply'' $n$ times too slow, where $n$ is the number of scalars in the input.

But the first optimisation on \vfad, which defunctionalises the zero, one-hot, addition and scaling operations on the $c$ tangent vector, introduces the same sharing-induced complexity problem as we have in naive \dnrad as payment for fixing the factor-$n$ overhead.
The two algorithms are now on equal footing: we could defunctionalise the backpropagators in \dnrad just as easily (and indeed we do so, albeit later in \cref{sec:improve-defunctionalisation}).

Afterwards, \vfad is lifted to a combination (stack) of an ID generation monad and a Writer monad.
Each scalar result of a primitive operation gets an ID, and the Writer monad records for each ID (hence, scalar in the program) its defunctionalised tangent vector (i.e.\ an expression) in terms of other already-computed tangent vectors from the Writer record.
These expressions correspond to our primitive operation backpropagators with calls replaced with $\StagedCall$: where we replace calls with ID-tagged pairs of function and argument, \vfad replaces the usage of already-computed tangent vectors with scaled references to the IDs of those vectors.
The choice in our $\ResolveStaged$ of evaluation order from highest ID to lowest ID (\cref{sec:staging}) is encoded in \vfad's definitions of \textit{runDelta} and \textit{eval}, which process the record back-to-front.

Finally, our Cayley transform is encoded in the type of \vfad's \textit{eval} function, which interprets the defunctionalised operations on tangent vectors (including explicit sharing using the Writer log) into an actual tangent vector---the final gradient: its gradient return type is Cayley-transformed.
Our final optimisation to mutable arrays to eliminate log-factors in the complexity is also mirrored in \vfad.


\myparagraph{Distributive law}
Under the isomorphism $\lR\lra c\cong c$, the type $\Staged c$ can be thought of as a type $\Expr c$ of ASTs of expressions (with sharing) of type $c$.\footnote{In~\citep{ad-2021-krawiec-kmett-ad}, $\Expr c$ is called $\textrm{Delta}$.}
The linear factoring rule $f\ x+ f\ y\leadsto f\ (x+y)$ for a linear function $f:\lR\lra c$ that rescales a vector $v:c$ with a scalar  then corresponds to the distributive law $v\cdot x+ v\cdot y\leadsto v\cdot (x+y)$.
This highlights the relationship between our work and that of~\cite{ad-2019-fwd-ad-gradient-compiler-opts},
who try to (statically) optimise vectorised forward AD to reverse AD using precisely this distributive law.
A key distinction is that we apply this law (in the form of the linear factoring rule) at runtime rather than compile time, allowing us to always achieve the complexity of reverse AD, rather than merely on some specific programs with straightforward control and data flow.
The price we pay for this generality is a runtime overhead, similar to the usual overhead of taping.


\subsection{Other PL Literature About AD}\label{sec:related-work:pl}
\myparagraph{CHAD and category theory inspired AD}
Rather than interleaving backpropagators by pairing them with scalars in a type, we can also try to directly implement reverse AD as a structurally recursive code transformation that does not need a (de)interleaving wrapper: $\mathcal{R}_2$ from \cref{sec:rev-ad-type}.
This is the approach taken by~\cite{adfp-2018-categories-ad}.
It pairs vectors (and values of other composite types) with a single composite backpropagator, rather than decomposing to the point where each scalar is paired with a mini-backpropagator like in our dual-numbers approach.
The resulting algorithm is extended to source languages with function types in 
\citep{vakar-2021-higher-order-reverse-ad,vakar-2022-chad,vytiniotis2019differentiable} and to sum and (co)inductive types in 
\citep{DBLP:journals/mscs/NunesV23}.
\Citet{chad-efficient-popl} give a mechanised proof that the resulting algorithm attains the correct computational complexity, after a series of optimisations that are similar to the ones considered in this paper.
Like our dual-numbers reverse AD approach, the algorithm arises as a canonical structure-preserving functor on the syntax of a programming language.
However, due to a different choice in target category (a Grothendieck construction of a linear $\lambda$-calculus for CHAD rather than the syntax of a plain $\lambda$-calculus for dual-numbers AD), the resulting algorithm looks very different.

\myparagraph{Taping-like methods and non-local control flow}
Another family of approaches to AD recently taken by the PL community contains those that 
rely on forms of non-local control flow such as delimited continuations \citep{ad-2019-delimited-continuations}
or effect handlers \citep{sigal2021automatic,de2021verifying}.
These techniques are different in the sense that they generate code that is not purely functional. 
This use of non-local control flow makes it possible to achieve an efficient implementation of reverse AD that looks strikingly simple compared to alternative approaches.
Where the CHAD approaches and our dual-numbers reverse AD approach both have to manually invert control flow at compile time by passing around continuations, possibly combined with smart staging of those continuations like in this paper, this inversion of control can be deferred to run time by clever use of delimited control operators or effect handlers.
\Citet{de2021verifying} give a mechanised proof that the resulting (rather subtle) code transformation is correct.

Operationally, however, these effect-handler based techniques are essentially equivalent to taping: the mutable cells for cotangent accumulation scope over the full remainder of the program.
In this sense, they are operationally similar to the algorithm of \cref{sec:improve-taping}, which is also essentially taping.
The AD algorithm in Dex~\citep{dex-2021-ad} also achieves something like taping in a functional style, by conversion to an A-normal form.

\myparagraph{AD of parallel code}
\Citet{bischof1991exploiting,bischof1991issues} 
present some of the first work in parallel AD.
Rather than starting with a source program (with potential dynamic control flow) that has (fork-join) parallelism like we do and mirroring that in the reverse pass of the algorithm, they focus on code without dynamic control flow and analyse the dependency structure of the reverse pass at compile time to automatically parallelise it.
This approach is developed further by~\cite{bucker2002explicit}.

Building on the classic imperative AD tool Tapenade \citep{ad-2013-tapenade}, \cite{huckelheim2022source} discuss a method for differentiating parallel for-loops (with shared memory) in a parallelism-preserving way.
Industrial machine learning frameworks such as TensorFlow~\citep{ad-2016-tensorflow}, JAX~\citep{ad-2018-jax} and PyTorch~\citep{ad-2017-pytorch} focus on data parallelism through parallel array operations---typically first-order ones.

\cite{kaler2021parad} focus on AD of programs with similar forms of fork-join parallelism as we consider in this paper.
Their implementation builds on the Adept C++ library, which implements an AD algorithm that is very different from the one discussed in this paper.
A unique feature of their work is that they give a formal analysis of the complexity of the technique and give bounds for both the work and the span of the resulting derivative code; it is possible that ideas from the cited work can be used to improve and bound the span of our parallel algorithm.

Other recent work has focussed on developing data-parallel derivatives for data-parallel array programs \citep{paszke2021parallelism,ad-2022-futhark-partial-recompute,dex-2021-ad}.
These methods are orthogonal to the ideas focussed on task parallelism that we present in the present paper.

In recent work, we have shown that an optimised version of CHAD seems to preserve task and data parallelism \citep{chad-efficient-popl}.
We are working on giving formal guarantees about the extent (work and span) to which different forms of parallelism is preserved.
Compared to the dual numbers reverse AD method of the present paper, CHAD seems to accommodate higher order array combinators more easily, including their data parallel versions. 
This makes us hopeful for the use of CHAD as a candidate algorithm for parallel AD.

\myparagraph{Conflicts of interest}
None.

\myparagraph{Acknowledgements}
We gratefully acknowledge funding for this research from NWO Veni grant number VI.Veni.202.124 and the ERC project FoRECAST.

\clearpage 

\appendix
\section*{Appendix}

On the following page, we show benchmark details from our implementation on some sequential programs, as determined by the Haskell library \texttt{criterion}.
See \cref{sec:implementation}.




\section*{Supplementary material}

# 1 Criterion Benchmark Details

## criterion performance measurements

### overview

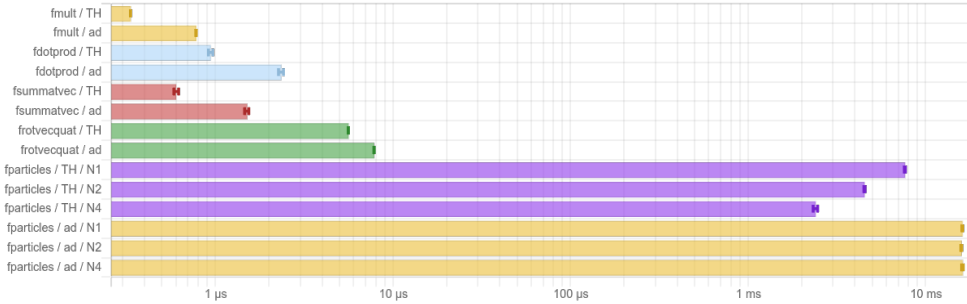

### fmult / TH

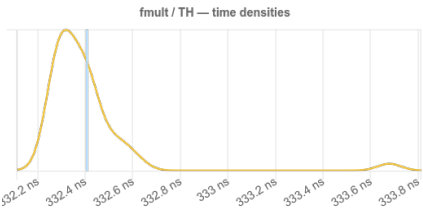
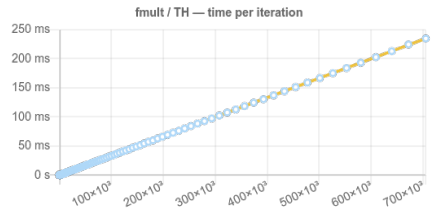

|  | lower bound | **estimate** | upper bound |
|---|---|---|---|
| OLS regression | 332 ns | 332 ns | 333 ns |
| $R^2$ goodness-of-fit | 1.00 | 1.00 | 1.00 |
| Mean execution time | 332 ns | 332 ns | 333 ns |
| Standard deviation | 85.9 ps | 218 ps | 462 ps |

Outlying measurements have no (0.426%) effect on estimated standard deviation.

### fmult / ad

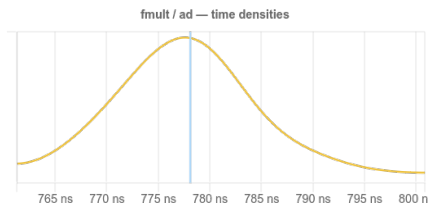
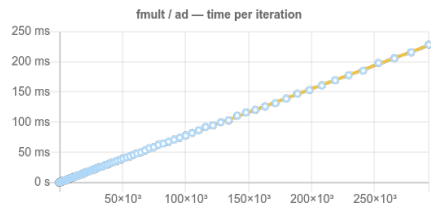

|  | lower bound | **estimate** | upper bound |
|---|---|---|---|
| OLS regression | 773 ns | 775 ns | 777 ns |
| $R^2$ goodness-of-fit | 1.00 | 1.00 | 1.00 |
| Mean execution time | 776 ns | 778 ns | 780 ns |
| Standard deviation | 5.25 ns | 6.72 ns | 8.60 ns |

Outlying measurements have a slight (3.32%) effect on estimated standard deviation.

## fdotprod / TH

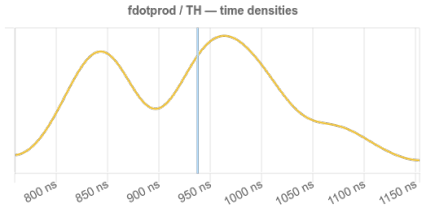 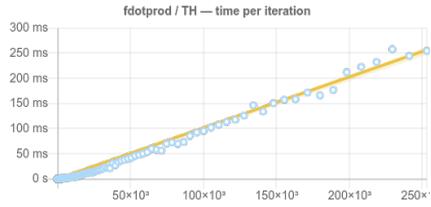

|                       | lower bound | estimate | upper bound |
|-----------------------|-------------|----------|-------------|
| OLS regression        | 980 ns      | 1.01 µs  | 1.04 µs     |
| R² goodness-of-fit    | 0.988       | 0.992    | 0.994       |
| Mean execution time   | 912 ns      | 938 ns   | 965 ns      |
| Standard deviation    | 73.7 ns     | 86.0 ns  | 102 ns      |

Outlying measurements have a severe (87.1%) effect on estimated standard deviation.

## fdotprod / ad

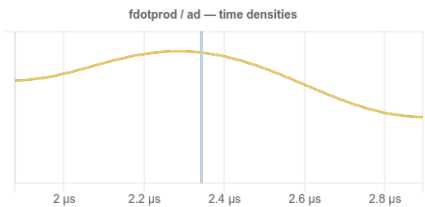 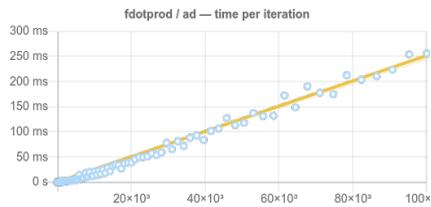

|                       | lower bound | estimate | upper bound |
|-----------------------|-------------|----------|-------------|
| OLS regression        | 2.43 µs     | 2.51 µs  | 2.57 µs     |
| R² goodness-of-fit    | 0.986       | 0.990    | 0.994       |
| Mean execution time   | 2.26 µs     | 2.34 µs  | 2.41 µs     |
| Standard deviation    | 200 ns      | 236 ns   | 277 ns      |

Outlying measurements have a severe (88.0%) effect on estimated standard deviation.

## fsummatvec / TH

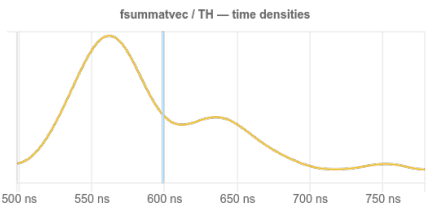 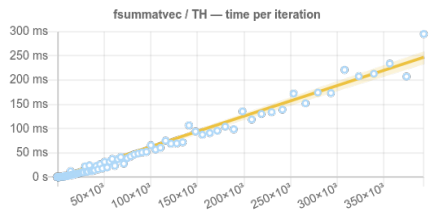

|                       | lower bound | estimate | upper bound |
|-----------------------|-------------|----------|-------------|
| OLS regression        | 595 ns      | 629 ns   | 662 ns      |
| R² goodness-of-fit    | 0.978       | 0.984    | 0.991       |
| Mean execution time   | 584 ns      | 599 ns   | 620 ns      |
| Standard deviation    | 46.6 ns     | 59.3 ns  | 75.9 ns     |

Outlying measurements have a severe (89.1%) effect on estimated standard deviation.

## fsummatvec / ad

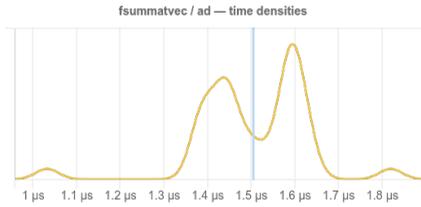 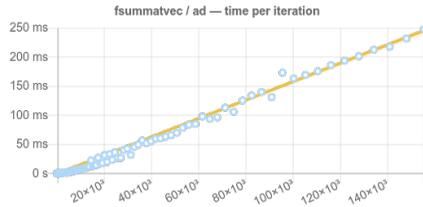

|                        | lower bound | estimate | upper bound |
|------------------------|-------------|----------|-------------|
| OLS regression         | 1.55 μs     | 1.58 μs  | 1.61 μs     |
| $R^2$ goodness-of-fit  | 0.989       | 0.993    | 0.996       |
| Mean execution time    | 1.46 μs     | 1.50 μs  | 1.54 μs     |
| Standard deviation     | 92.0 ns     | 125 ns   | 191 ns      |

Outlying measurements have a severe (84.4%) effect on estimated standard deviation.

## frotvecquat / TH

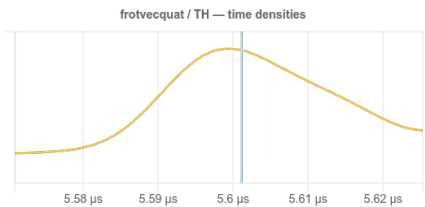 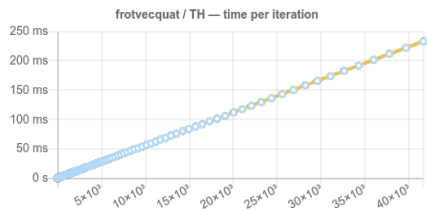

|                        | lower bound | estimate | upper bound |
|------------------------|-------------|----------|-------------|
| OLS regression         | 5.60 μs     | 5.60 μs  | 5.60 μs     |
| $R^2$ goodness-of-fit  | 1.00        | 1.00     | 1.00        |
| Mean execution time    | 5.60 μs     | 5.60 μs  | 5.60 μs     |
| Standard deviation     | 9.57 ns     | 11.5 ns  | 14.3 ns     |

Outlying measurements have no (0.565%) effect on estimated standard deviation.

## frotvecquat / ad

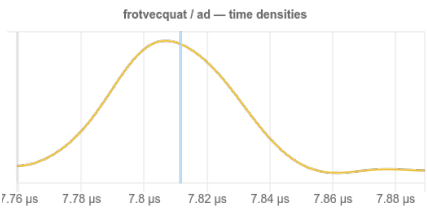 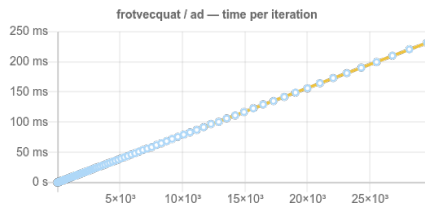

|                        | lower bound | estimate | upper bound |
|------------------------|-------------|----------|-------------|
| OLS regression         | 7.81 μs     | 7.81 μs  | 7.82 μs     |
| $R^2$ goodness-of-fit  | 1.00        | 1.00     | 1.00        |
| Mean execution time    | 7.81 μs     | 7.81 μs  | 7.82 μs     |
| Standard deviation     | 16.1 ns     | 21.7 ns  | 28.5 ns     |

Outlying measurements have no (0.588%) effect on estimated standard deviation.

## fparticles / TH / N1

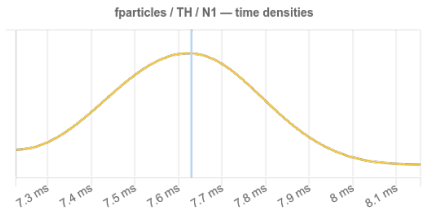
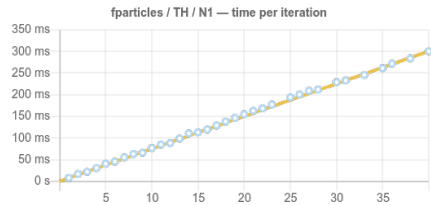

|  | lower bound | **estimate** | upper bound |
|---|---|---|---|
| OLS regression | 7.44 ms | 7.50 ms | 7.59 ms |
| R² goodness-of-fit | 0.999 | 0.999 | 1.00 |
| Mean execution time | 7.58 ms | 7.63 ms | 7.69 ms |
| Standard deviation | 130 µs | 165 µs | 221 µs |

Outlying measurements have a slight (4.68%) effect on estimated standard deviation.

## fparticles / TH / N2

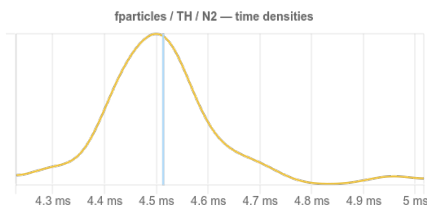
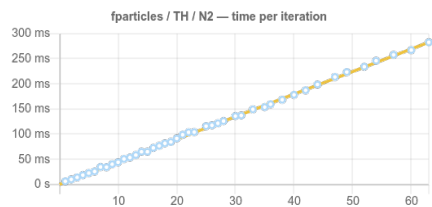

|  | lower bound | **estimate** | upper bound |
|---|---|---|---|
| OLS regression | 4.45 ms | 4.48 ms | 4.51 ms |
| R² goodness-of-fit | 0.999 | 0.999 | 1.00 |
| Mean execution time | 4.48 ms | 4.51 ms | 4.55 ms |
| Standard deviation | 78.5 µs | 113 µs | 176 µs |

Outlying measurements have a moderate (10.8%) effect on estimated standard deviation.

## fparticles / TH / N4

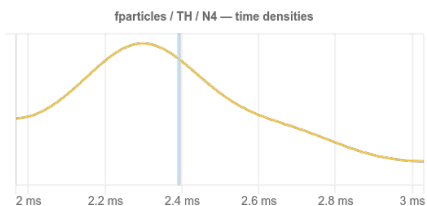
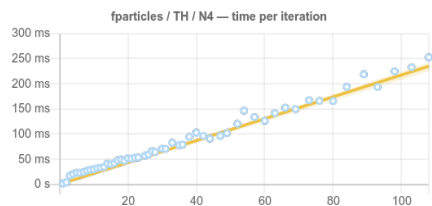

|  | lower bound | **estimate** | upper bound |
|---|---|---|---|
| OLS regression | 2.09 ms | 2.17 ms | 2.25 ms |
| R² goodness-of-fit | 0.980 | 0.988 | 0.994 |
| Mean execution time | 2.32 ms | 2.39 ms | 2.46 ms |
| Standard deviation | 191 µs | 232 µs | 282 µs |

Outlying measurements have a severe (65.3%) effect on estimated standard deviation.

## fparticles / ad / N1

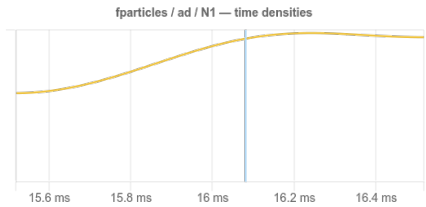 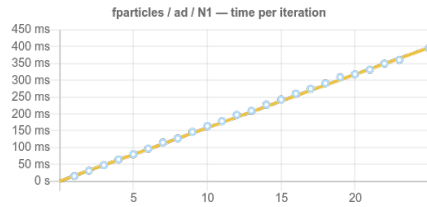

|  | lower bound | **estimate** | upper bound |
|---|---|---|---|
| OLS regression | 15.6 ms | **15.8 ms** | 16.1 ms |
| R² goodness-of-fit | 0.999 | **0.999** | 1.00 |
| Mean execution time | 16.0 ms | **16.1 ms** | 16.2 ms |
| Standard deviation | 196 µs | **239 µs** | 318 µs |

Outlying measurements have a slight (3.99%) effect on estimated standard deviation.

## fparticles / ad / N2

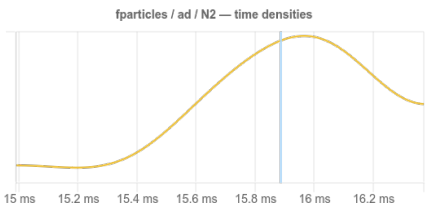 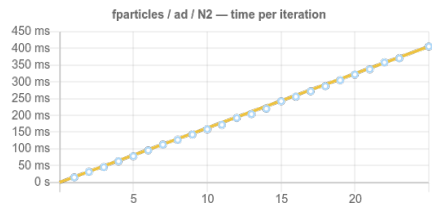

|  | lower bound | **estimate** | upper bound |
|---|---|---|---|
| OLS regression | 16.1 ms | **16.2 ms** | 16.3 ms |
| R² goodness-of-fit | 0.999 | **1.00** | 1.00 |
| Mean execution time | 15.8 ms | **15.9 ms** | 16.0 ms |
| Standard deviation | 185 µs | **259 µs** | 404 µs |

Outlying measurements have a slight (3.99%) effect on estimated standard deviation.

## fparticles / ad / N4

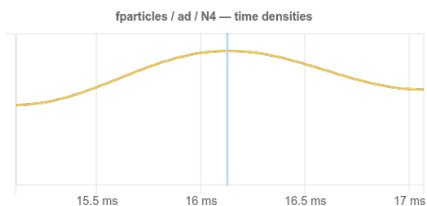 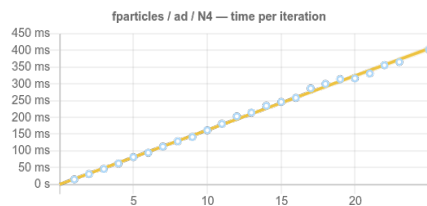

|  | lower bound | **estimate** | upper bound |
|---|---|---|---|
| OLS regression | 15.9 ms | **16.2 ms** | 16.6 ms |
| R² goodness-of-fit | 0.996 | **0.998** | 0.999 |
| Mean execution time | 15.9 ms | **16.1 ms** | 16.3 ms |
| Standard deviation | 355 µs | **442 µs** | 564 µs |

Outlying measurements have a slight (7.73%) effect on estimated standard deviation.

# understanding this report

In this report, each function benchmarked by criterion is assigned a section of its own.

- The chart on the left is a kernel density estimate (also known as a KDE) of time measurements. This graphs the probability of any given time measurement occurring. A spike indicates that a measurement of a particular time occurred; its height indicates how often that measurement was repeated.
- The chart on the right is the raw data from which the kernel density estimate is built. The *x*-axis indicates the number of loop iterations, while the *y*-axis shows measured execution time for the given number of loop iterations. The line behind the values is the linear regression estimate of execution time for a given number of iterations. Ideally, all measurements will be on (or very near) this line. The transparent area behind it shows the confidence interval for the execution time estimate.

Under the charts is a small table. The first two rows are the results of a linear regression run on the measurements displayed in the right-hand chart.

- *OLS regression* indicates the time estimated for a single loop iteration using an ordinary least-squares regression model. This number is more accurate than the *mean* estimate below it, as it more effectively eliminates measurement overhead and other constant factors.
- $R^2$; *goodness-of-fit* is a measure of how accurately the linear regression model fits the observed measurements. If the measurements are not too noisy, $R^2$; should lie between 0.99 and 1, indicating an excellent fit. If the number is below 0.99, something is confounding the accuracy of the linear model.
- *Mean execution time* and *standard deviation* are statistics calculated from execution time divided by number of iterations.

We use a statistical technique called the bootstrap to provide confidence intervals on our estimates. The bootstrap-derived upper and lower bounds on estimates let you see how accurate we believe those estimates to be.

A noisy benchmarking environment can cause some or many measurements to fall far from the mean. These outlying measurements can have a significant inflationary effect on the estimate of the standard deviation. We calculate and display an estimate of the extent to which the standard deviation has been inflated by outliers.

# colophon

This report was created using the criterion benchmark execution and performance analysis tool.

Criterion is developed and maintained by Bryan O'Sullivan.



\bibliographystyle{jfplike}

\bibliography{bibliography}

\begin{thebibliography}{48}

\bibitem[{Abadi et~al.}(2016){Abadi, Barham, Chen, Chen, Davis, Dean, Devin,
  Ghemawat, Irving, Isard, Kudlur, Levenberg, Monga, Moore, Murray, Steiner,
  Tucker, Vasudevan, Warden, Wicke, Yu, \& Zheng}]{ad-2016-tensorflow}
{Abadi, M., Barham, P., Chen, J., Chen, Z., Davis, A., Dean, J., Devin, M.,
  Ghemawat, S., Irving, G., Isard, M., Kudlur, M., Levenberg, J., Monga, R.,
  Moore, S., Murray, D.~G., Steiner, B., Tucker, P.~A., Vasudevan, V., Warden,
  P., Wicke, M., Yu, Y. \& Zheng, X.} (2016) Tensorflow: {A} system for
  large-scale machine learning. 12th {USENIX} Symposium on Operating Systems
  Design and Implementation, {OSDI} 2016, Savannah, GA, USA, November 2-4,
  2016. {USENIX} Association. pp. 265--283.

\bibitem[{Abadi and Plotkin}(2020){Abadi \& Plotkin}]{ad-2020-rev-ad-semantics}
{Abadi, M. \& Plotkin, G.~D.} (2020) A simple differentiable programming
  language. \textit{Proc. {ACM} Program. Lang.} {\bf 4}({POPL}), 38:1--38:28.

\bibitem[{Baydin et~al.}(2017){Baydin, Pearlmutter, Radul, \&
  Siskind}]{ad-2018-survey-automatic-differentiation}
{Baydin, A.~G., Pearlmutter, B.~A., Radul, A.~A. \& Siskind, J.~M.} (2017)
  Automatic differentiation in machine learning: a survey. \textit{J. Mach.
  Learn. Res.} {\bf 18}, 153:1--153:43.

\bibitem[{Bischof}(1991)]{bischof1991issues}
{Bischof, C.} (1991) {\em Issues in parallel automatic differentiation\/}.
  Argonne National Lab.

\bibitem[{Bischof et~al.}(1991){Bischof, Griewank, \&
  Juedes}]{bischof1991exploiting}
{Bischof, C., Griewank, A. \& Juedes, D.} (1991) Exploiting parallelism in
  automatic differentiation. Proceedings of the 5th international conference on
  Supercomputing. pp. 146--153.

\bibitem[{Boisseau and Gibbons}(2018){Boisseau \&
  Gibbons}]{2018-yoneda-profunctor}
{Boisseau, G. \& Gibbons, J.} (2018) What you needa know about yoneda:
  profunctor optics and the yoneda lemma (functional pearl). \textit{Proc.
  {ACM} Program. Lang.} {\bf 2}({ICFP}), 84:1--84:27.

\bibitem[{Bradbury et~al.}(2018){Bradbury, Frostig, Hawkins, Johnson, Leary,
  Maclaurin, Necula, Paszke, Vander{P}las, Wanderman-{M}ilne, \&
  Zhang}]{ad-2018-jax}
{Bradbury, J., Frostig, R., Hawkins, P., Johnson, M.~J., Leary, C., Maclaurin,
  D., Necula, G., Paszke, A., Vander{P}las, J., Wanderman-{M}ilne, S. \& Zhang,
  Q.} (2018) {JAX}: composable transformations of {P}ython+{N}um{P}y programs .

\bibitem[{Brunel et~al.}(2020){Brunel, Mazza, \&
  Pagani}]{ad-2020-dualnum-revad-linear-factoring}
{Brunel, A., Mazza, D. \& Pagani, M.} (2020) Backpropagation in the simply
  typed lambda-calculus with linear negation. \textit{Proc. {ACM} Program.
  Lang.} {\bf 4}({POPL}), 64:1--64:27.

\bibitem[{Bucker et~al.}(2002){Bucker, Lang, Rasch, Bischof, \&
  an~Mey}]{bucker2002explicit}
{Bucker, H., Lang, B., Rasch, A., Bischof, C.~H. \& an~Mey, D.} (2002) Explicit
  loop scheduling in openmp for parallel automatic differentiation. Proceedings
  16th Annual International Symposium on High Performance Computing Systems and
  Applications. IEEE. pp. 121--126.

\bibitem[{de~Vilhena and Pottier}(2023){de~Vilhena \&
  Pottier}]{de2021verifying}
{de~Vilhena, P.~E. \& Pottier, F.} (2023) Verifying an effect-handler-based
  define-by-run reverse-mode ad library. \textit{Logical Methods in Computer
  Science}. {\bf 19}.

\bibitem[{Elliott}(2018)]{adfp-2018-categories-ad}
{Elliott, C.} (2018) The simple essence of automatic differentiation.
  \textit{Proc. {ACM} Program. Lang.} {\bf 2}({ICFP}), 70:1--70:29.

\bibitem[{Griewank and Walther}(2008){Griewank \&
  Walther}]{adbook-2008-griewank-walther}
{Griewank, A. \& Walther, A.} (2008) {\em Evaluating derivatives - principles
  and techniques of algorithmic differentiation, Second Edition\/}. {SIAM}.

\bibitem[{Hasco{\"{e}}t et~al.}(2005){Hasco{\"{e}}t, Naumann, \&
  Pascual}]{ad-2005-to-be-recorded}
{Hasco{\"{e}}t, L., Naumann, U. \& Pascual, V.} (2005) "to be recorded"
  analysis in reverse-mode automatic differentiation. \textit{Future Gener.
  Comput. Syst.} {\bf 21}(8), 1401--1417.

\bibitem[{Hasco{\"{e}}t and Pascual}(2013){Hasco{\"{e}}t \&
  Pascual}]{ad-2013-tapenade}
{Hasco{\"{e}}t, L. \& Pascual, V.} (2013) The {Tapenade} automatic
  differentiation tool: Principles, model, and specification. \textit{{ACM}
  Trans. Math. Softw.} {\bf 39}(3), 20:1--20:43.

\bibitem[{H{\"u}ckelheim and Hasco{\"e}t}(2022){H{\"u}ckelheim \&
  Hasco{\"e}t}]{huckelheim2022source}
{H{\"u}ckelheim, J. \& Hasco{\"e}t, L.} (2022) Source-to-source automatic
  differentiation of openmp parallel loops. \textit{ACM Transactions on
  Mathematical Software (TOMS)}. {\bf 48}(1), 1--32.

\bibitem[{H{\"{u}}ckelheim et~al.}(2023){H{\"{u}}ckelheim, Menon, Moses,
  Christianson, Hovland, \& Hasco{\"{e}}t}]{ad-2023-pitfalls}
{H{\"{u}}ckelheim, J., Menon, H., Moses, W.~S., Christianson, B., Hovland,
  P.~D. \& Hasco{\"{e}}t, L.} (2023) Understanding automatic differentiation
  pitfalls. \textit{CoRR}. {\bf abs/2305.07546}.

\bibitem[{Hughes}(1986)]{fp-1986-difference-lists}
{Hughes, R. J.~M.} (1986) A novel representation of lists and its application
  to the function "reverse". \textit{Inf. Process. Lett.} {\bf 22}(3),
  141--144.

\bibitem[{Huot et~al.}(2020){Huot, Staton, \&
  V{\'{a}}k{\'{a}}r}]{ad-2020-sam-mathieu-matthijs}
{Huot, M., Staton, S. \& V{\'{a}}k{\'{a}}r, M.} (2020) Correctness of automatic
  differentiation via diffeologies and categorical gluing. Foundations of
  Software Science and Computation Structures - 23rd International Conference,
  {FOSSACS} 2020, Held as Part of the European Joint Conferences on Theory and
  Practice of Software, {ETAPS} 2020, Dublin, Ireland, April 25-30, 2020,
  Proceedings. Springer. pp. 319--338.

\bibitem[{Jacobs et~al.}(2022){Jacobs, Devriese, \&
  Timany}]{fp-2022-st-monad-proof}
{Jacobs, K., Devriese, D. \& Timany, A.} (2022) Purity of an {ST} monad: full
  abstraction by semantically typed back-translation. \textit{Proc. {ACM}
  Program. Lang.} {\bf 6}({OOPSLA1}), 1--27.

\bibitem[{Kaler et~al.}(2021){Kaler, Schardl, Xie, Leiserson, Chen, Pareja, \&
  Kollias}]{kaler2021parad}
{Kaler, T., Schardl, T.~B., Xie, B., Leiserson, C.~E., Chen, J., Pareja, A. \&
  Kollias, G.} (2021) Parad: A work-efficient parallel algorithm for
  reverse-mode automatic differentiation. Symposium on Algorithmic Principles
  of Computer Systems (APOCS). SIAM. pp. 144--158.

\bibitem[{Kmett and contributors}(2021){Kmett \&
  contributors}]{ad-2021-kmett-hackage}
{Kmett, E. \& contributors}. (2021) ad: Automatic differentiation (Haskell
  library on Hackage).

\bibitem[{Krawiec et~al.}(2022){Krawiec, Jones, Krishnaswami, Ellis, Eisenberg,
  \& Fitzgibbon}]{ad-2021-krawiec-kmett-ad}
{Krawiec, F., Jones, S.~P., Krishnaswami, N., Ellis, T., Eisenberg, R.~A. \&
  Fitzgibbon, A.~W.} (2022) Provably correct, asymptotically efficient,
  higher-order reverse-mode automatic differentiation. \textit{Proc. {ACM}
  Program. Lang.} {\bf 6}({POPL}), 1--30.

\bibitem[{Launchbury and Jones}(1994){Launchbury \& Jones}]{fp-1994-st-monad}
{Launchbury, J. \& Jones, S. L.~P.} (1994) Lazy functional state threads.
  Proceedings of the {ACM} SIGPLAN'94 Conference on Programming Language Design
  and Implementation (PLDI), Orlando, Florida, USA, June 20-24, 1994. {ACM}.
  pp. 24--35.

\bibitem[{Linnainmaa}(1970)]{linnainmaa1970representation}
{Linnainmaa, S.} (1970) The representation of the cumulative rounding error of
  an algorithm as a taylor expansion of the local rounding errors.
  \textit{Master’s Thesis (in Finnish), Univ. Helsinki}.

\bibitem[{Lucatelli~Nunes and Vákár}(2024){Lucatelli~Nunes \&
  Vákár}]{nunes-2024-dual-numbers}
{Lucatelli~Nunes, F. \& Vákár, M.} (2024) Automatic differentiation for
  {ML}-family languages: Correctness via logical relations.
  \textit{Mathematical Structures in Computer Science}. {\bf 34}(8), 747–806.

\bibitem[{Margossian}(2019)]{ad-2018-survey-ad-implementation}
{Margossian, C.~C.} (2019) A review of automatic differentiation and its
  efficient implementation. \textit{Wiley Interdiscip. Rev. Data Min. Knowl.
  Discov.} {\bf 9}(4).

\bibitem[{Mazza and Pagani}(2021){Mazza \&
  Pagani}]{ad-2021-dual-revad-linear-factoring-pcf}
{Mazza, D. \& Pagani, M.} (2021) Automatic differentiation in {PCF}.
  \textit{Proc. {ACM} Program. Lang.} {\bf 5}({POPL}), 1--27.

\bibitem[{Nunes and V{\'{a}}k{\'{a}}r}(2023){Nunes \&
  V{\'{a}}k{\'{a}}r}]{DBLP:journals/mscs/NunesV23}
{Nunes, F.~L. \& V{\'{a}}k{\'{a}}r, M.} (2023) {CHAD} for expressive total
  languages. \textit{Math. Struct. Comput. Sci.} {\bf 33}(4-5), 311--426.

\bibitem[{Paszke et~al.}(2017){Paszke, Gross, Chintala, Chanan, Yang, DeVito,
  Lin, Desmaison, Antiga, \& Lerer}]{ad-2017-pytorch}
{Paszke, A., Gross, S., Chintala, S., Chanan, G., Yang, E., DeVito, Z., Lin,
  Z., Desmaison, A., Antiga, L. \& Lerer, A.} (2017) Automatic differentiation
  in {PyTorch}. NIPS 2017 Autodiff Workshop: The future of gradient-based
  machine learning software and techniques. Red Hook, NY, USA. Curran
  Associates, Inc.

\bibitem[{Paszke et~al.}(2021a){Paszke, Johnson, Duvenaud, Vytiniotis, Radul,
  Johnson, Ragan{-}Kelley, \& Maclaurin}]{dex-2021-ad}
{Paszke, A., Johnson, D.~D., Duvenaud, D., Vytiniotis, D., Radul, A., Johnson,
  M.~J., Ragan{-}Kelley, J. \& Maclaurin, D.} (2021a) Getting to the point:
  index sets and parallelism-preserving autodiff for pointful array
  programming. \textit{Proc. {ACM} Program. Lang.} {\bf 5}({ICFP}), 1--29.

\bibitem[{Paszke et~al.}(2021b){Paszke, Johnson, Frostig, \&
  Maclaurin}]{paszke2021parallelism}
{Paszke, A., Johnson, M.~J., Frostig, R. \& Maclaurin, D.} (2021b)
  Parallelism-preserving automatic differentiation for second-order array
  languages. Proceedings of the 9th ACM SIGPLAN International Workshop on
  Functional High-Performance and Numerical Computing. pp. 13--23.

\bibitem[{Pearlmutter and Siskind}(2008){Pearlmutter \&
  Siskind}]{ad-2008-reverse-functional-ad}
{Pearlmutter, B.~A. \& Siskind, J.~M.} (2008) Reverse-mode {AD} in a functional
  framework: Lambda the ultimate backpropagator. \textit{{ACM} Trans. Program.
  Lang. Syst.} {\bf 30}(2), 7:1--7:36.

\bibitem[{Reynolds}(1998)]{fp-1998-defunctionalisation}
{Reynolds, J.~C.} (1998) Definitional interpreters for higher-order programming
  languages. \textit{High. Order Symb. Comput.} {\bf 11}(4), 363--397.

\bibitem[{Schenck et~al.}(2022){Schenck, R{\o}nning, Henriksen, \&
  Oancea}]{ad-2022-futhark-partial-recompute}
{Schenck, R., R{\o}nning, O., Henriksen, T. \& Oancea, C.~E.} (2022) {AD} for
  an array language with nested parallelism. {SC22:} International Conference
  for High Performance Computing, Networking, Storage and Analysis, Dallas, TX,
  USA, November 13-18, 2022. {IEEE}. pp. 58:1--58:15.

\bibitem[{Shaikhha et~al.}(2019){Shaikhha, Fitzgibbon, Vytiniotis, \&
  Jones}]{ad-2019-fwd-ad-gradient-compiler-opts}
{Shaikhha, A., Fitzgibbon, A., Vytiniotis, D. \& Jones, S.~P.} (2019) Efficient
  differentiable programming in a functional array-processing language.
  \textit{Proc. {ACM} Program. Lang.} {\bf 3}({ICFP}), 97:1--97:30.

\bibitem[{Sheard and Jones}(2002){Sheard \& Jones}]{fp-2002-template-haskell}
{Sheard, T. \& Jones, S. L.~P.} (2002) Template meta-programming for haskell.
  \textit{{ACM} {SIGPLAN} Notices}. {\bf 37}(12), 60--75.

\bibitem[{Sigal}(2021)]{sigal2021automatic}
{Sigal, J.} (2021) Automatic differentiation via effects and handlers: An
  implementation in frank. \textit{arXiv preprint arXiv:2101.08095}.

\bibitem[{Siskind and Pearlmutter}(2018){Siskind \&
  Pearlmutter}]{ad-2018-checkpointing-built-in}
{Siskind, J.~M. \& Pearlmutter, B.~A.} (2018) Divide-and-conquer checkpointing
  for arbitrary programs with no user annotation. \textit{Optimization Methods
  and Software}. {\bf 33}(4-6), 1288--1330.

\bibitem[{Smeding and V{\'{a}}k{\'{a}}r}(2022){Smeding \&
  V{\'{a}}k{\'{a}}r}]{ad-dualrev-th-arxivv2}
{Smeding, T. \& V{\'{a}}k{\'{a}}r, M.} (2022) Efficient dual-numbers reverse
  {AD} via well-known program transformations. \textit{CoRR}. {\bf
  abs/2207.03418v2}.

\bibitem[{Smeding and V{\'{a}}k{\'{a}}r}(2023){Smeding \&
  V{\'{a}}k{\'{a}}r}]{ad-dualrev-th}
{Smeding, T. \& V{\'{a}}k{\'{a}}r, M.} (2023) Efficient dual-numbers reverse
  {AD} via well-known program transformations. \textit{Proc. {ACM} Program.
  Lang.} {\bf 7}({POPL}), 1573--1600.

\bibitem[{Smeding and V{\'{a}}k{\'{a}}r}(2024){Smeding \&
  V{\'{a}}k{\'{a}}r}]{chad-efficient-popl}
{Smeding, T. \& V{\'{a}}k{\'{a}}r, M.} (2024) Efficient {CHAD}. \textit{Proc.
  {ACM} Program. Lang.} {\bf 8}({POPL}), 1060--1088.

\bibitem[{Speelpenning}(1980)]{ad-1980-ad}
{Speelpenning, B.} (1980) Compiling fast partial derivatives of functions given
  by algorithms. Technical report. Illinois University.

\bibitem[{V{\'{a}}k{\'{a}}r}(2021)]{vakar-2021-higher-order-reverse-ad}
{V{\'{a}}k{\'{a}}r, M.} (2021) Reverse {AD} at higher types: Pure, principled
  and denotationally correct. Programming Languages and Systems. Springer. pp.
  607--634.

\bibitem[{V{\'{a}}k{\'{a}}r and Smeding}(2022){V{\'{a}}k{\'{a}}r \&
  Smeding}]{vakar-2022-chad}
{V{\'{a}}k{\'{a}}r, M. \& Smeding, T.} (2022) {CHAD:} combinatory homomorphic
  automatic differentiation. pp. 20:1--20:49.

\bibitem[{Vytiniotis et~al.}(2019){Vytiniotis, Belov, Wei, Plotkin, \&
  Abadi}]{vytiniotis2019differentiable}
{Vytiniotis, D., Belov, D., Wei, R., Plotkin, G. \& Abadi, M.} (2019) The
  differentiable curry. \textit{NeurIPS Workshop on Program Transformations}.

\bibitem[{Wang et~al.}(2019){Wang, Zheng, Decker, Wu, Essertel, \&
  Rompf}]{ad-2019-delimited-continuations}
{Wang, F., Zheng, D., Decker, J.~M., Wu, X., Essertel, G.~M. \& Rompf, T.}
  (2019) Demystifying differentiable programming: shift/reset the penultimate
  backpropagator. \textit{Proc. {ACM} Program. Lang.} {\bf 3}({ICFP}),
  96:1--96:31.

\bibitem[{Wengert}(1964)]{ad-1964-ad}
{Wengert, R.~E.} (1964) A simple automatic derivative evaluation program.
  \textit{Commun. {ACM}}. {\bf 7}(8), 463--464.

\bibitem[{Westrick et~al.}(2024){Westrick, Fluet, Rainey, \&
  Acar}]{westrick2024parallelism}
{Westrick, S., Fluet, M., Rainey, M. \& Acar, U.~A.} (2024) Automatic
  parallelism management. \textit{Proc. {ACM} Program. Lang.} {\bf 8}({POPL}),
  1118--1149.

\end{thebibliography}
 
\label{lastpage}

\end{document}